\documentclass[aps,prd,onecolumn,nofootinbib,superscriptaddress]{revtex4}
\pdfoutput=1
\usepackage[colorlinks=true,linkcolor=blue,urlcolor=blue,filecolor=black,citecolor=red,pdfstartview=FitV,pdftitle={},pdfsubject={},pdfkeywords={},pdfpagemode=None,bookmarksopen=true]{hyperref}
\usepackage{graphicx}
\usepackage{amsmath}
\usepackage{amsfonts}
\usepackage{amssymb,ulem}
\usepackage{color}%
\usepackage{dcolumn}
\usepackage{braket}
\usepackage{eurosym}
\usepackage[usenames,dvipsnames,svgnames]{xcolor}

\newcommand{\RNum}[1]{\uppercase\expandafter{\romannumeral #1\relax}}

\usepackage[title]{appendix}
\usepackage{wrapfig,boxedminipage,setspace,subfigure,epsfig}

\begin{document}
\baselineskip=0.5 cm

\title{Rings and images of Horndeski hairy black hole illuminated by various thin accretions}

\author{Xi-Jing Wang}
\email{xijingwang@yzu.edu.cn}
\affiliation{Center for Gravitation and Cosmology, College of Physical Science and Technology, Yangzhou University, Yangzhou, 225009, China}

\author{Xiao-Mei Kuang}
\email{xmeikuang@yzu.edu.cn (corresponding author)}
\affiliation{Center for Gravitation and Cosmology, College of Physical Science and Technology, Yangzhou University, Yangzhou, 225009, China}

\author{Yuan Meng}
\email{mengyuanphy@163.com}
\affiliation{Center for Gravitation and Cosmology, College of Physical Science and Technology, Yangzhou University, Yangzhou, 225009, China}

\author{Bin Wang}
\email{wang$_$b@sjtu.edu.cn}
\affiliation{Center for Gravitation and Cosmology, College of Physical Science and Technology, Yangzhou University, Yangzhou, 225009, China}
\affiliation{Shanghai Frontier Science Center for Gravitational Wave Detection, Shanghai Jiao Tong University, Shanghai 200240, China}
\author{Jian-Pin Wu}
\email{jianpinwu@yzu.edu.cn}
\affiliation{Center for Gravitation and Cosmology, College of Physical Science and Technology, Yangzhou University, Yangzhou, 225009, China}

\begin{abstract}
\baselineskip=0.4 cm
{In this paper, we analyze the light rays around a static hairy black hole in Horndeski gravity with the use of ray-tracing procedure. We find that a stronger Horndeski hairy parameter corresponds to larger photon sphere as well as critical impact parameter, and wider ranges of photon ring and lensed ring emissions. These influences can be robustly interpreted from the shape of the  effective potential of the photon's radial motion. Based on the distribution of the light rays, we then investigate the optical appearances of the Horndeski hairy black hole surrounded by various thin accretions.} Firstly, we consider that the Horndeski hairy black hole is illuminated by the optically and geometrically thin accretion disk. We carefully clarify the contributions from the direct, lensed ring and photon ring intensities to the total observed intensity via the transfer function, which is rarely discussed in this scenario. We find that the Horndeski hair has significant influences on both shadow size and distributions of direct, lensed ring and photon ring brightness in three standard emission profiles. As a result, the rings and images of Horndeski hairy black hole and the origination of their brightness differentiate from those of Schwarzschild black hole (SBH) when viewed in face-on orientation. Then, when the Horndeski hairy black hole is illuminated by thin spherical accretions, the hairy black hole's shadow surrounded by a bright ring is larger than that of SBH, but the brightness of ring is fainter. Similar to that of SBH, the size of hairy black hole shadow does not change as the radial moving of the spherical accretion, and the brightness for the infalling accretion is fainter than that for the static accretion due to the Doppler effect. Therefore, we argue that the black hole image consisting of the shadow and accretion construction could, in theory, reflect the observational differences between the Horndeski hairy black hole and SBH.
\end{abstract}

\maketitle

\newpage
\tableofcontents

\section{Introduction}
It is generally accepted that our current understanding of gravity may be incomplete, due to fundamental hints (e.g., the quantum theory of gravity) and observational hints (e.g., the accelerated expansion of our Universe).  Moreover, most observations fall into the prediction of classical general relativity (GR), but they allow the space for the modified theories of gravity. In particular, the no-hair theorem, stating that black holes are only described by mass, electric charge and spin, is one of the main characteristics in classical GR, however, whether it can be evaded is still pending. In fact, there are considerable affords to modify GR by means of additional matter fields so that the theories admit black holes with hair.  Among them, the scalar-tensor gravity theory, which contains a scalar field $\varphi$ as well as a metric tensor $g_{\mu\nu}$,  are known as a simplest nontrivial extension of GR \cite{Damour:1992we}. One of the most famous four-dimensional scalar-tensor theories is the Horndeski gravity proposed in 1974 \cite{Horndeski:1974wa},  which contains higher derivatives of $\varphi$ and $g_{\mu\nu}$ and is free of Ostrogradski instabilities because it posses at most second-order differential field equations.
Various observational constraints or bounds on Horndeski gravity have been explored in \cite{Bellini:2015xja,Bhattacharya:2016naa,Kreisch:2017uet,Hou:2017cjy,SpurioMancini:2019rxy,Allahyari:2020jkn}. In fact, Horndeski gravity attracts lots of attentions in cosmological and astrophysical communities because it has significant consequences  in describing the accelerated expansion and other interesting features (please see \cite{Kobayashi:2019hrl} for a review), as well as holographic applications \cite{Feng:2015oea,Kuang:2016edj,Jiang:2017imk,Baggioli:2017ojd,Feng:2018sqm,Wang:2019jyw,Zhang:2022hxl,Bravo-Gaete:2020lzs,Bravo-Gaete:2022lno} and references therein.

Hairy black holes in Horndeski gravity have also been widely constructed and analyzed, including the radially
dependent scalar field \cite{Rinaldi:2012vy,Cisterna:2014nua,Feng:2015oea,Sotiriou:2013qea, Miao:2016aol,Kuang:2016edj,Babichev:2016rlq,Benkel:2016rlz,Filios:2018xvy,Cisterna:2018hzf,Giusti:2021sku} and time-dependent scalar field \cite{Babichev:2013cya,Babichev:2017lmw,BenAchour:2018dap,Takahashi:2019oxz,Minamitsuji:2019shy,Arkani-Hamed:2003juy}. However, the hairy solution with scalar hair in linear time dependence was found to be unstable, and so this type of hairy solution was ruled out in Horndeski gravity \cite{Khoury:2020aya}.  Since Horndeski theory has diffeomorphism invariance and second order field equations, which is similar to GR, so it is of importance to test the no-hair theorem in Horndeski framework.  In fact, in \cite{Hui:2012qt}, the no-hair theorem was  demonstrated not be held when a Galileon field is coupled to gravity, but the static spherical black hole only  admits trivial Galileon profiles.  Then, the authors of \cite{Babichev:2017guv} further examined the no-hair theorem in Horndeski theory and beyond. They demonstrated that the shift-symmetric Horndeski theory and beyond allow for static and asymptotically flat black holes with a nontrivial static scalar field,  and the action they considered is dubbed  quartic Horndeski gravity
\begin{eqnarray}\label{eq:action}
S=\int d^4x \sqrt{-g}\big[Q_2+Q_3\Box\varphi+Q_4R+Q_{4,\chi}\left((\Box\varphi)^2-(\nabla^\mu\nabla^\nu\varphi)(\nabla_\mu\nabla_\nu\varphi)\right)
+Q_5G_{\mu\nu}\nabla^\mu\nabla^\nu\varphi\nonumber\\
-\frac{1}{6}Q_{5,\chi}\left((\Box\varphi)^3-3(\Box\varphi)(\nabla^\mu\nabla^\nu\varphi)(\nabla_\mu\nabla_\nu\varphi)
+2(\nabla_\mu\nabla_\nu\varphi)(\nabla^\nu\nabla^\gamma\varphi)(\nabla_\gamma\nabla^\mu\varphi)\right)\big],
\end{eqnarray}
where $\chi=-\partial^\mu\varphi\partial_\mu\varphi/2$ is the canonical kinetic term, $Q_i (i=2,3,4,5)$ are arbitrary functions of $\chi$ and $Q_{i,\chi} \equiv \partial Q_{i}/\partial \chi $, $R$ is the Ricci scalar and $G_{\mu\nu}$ is the Einstein tensor. In particular, very recently a static hairy black hole in a specific quartic Horndeski theory, saying $Q_5$ in the above action vanishes, has been constructed in \cite{Bergliaffa:2021diw}
\begin{eqnarray}\label{eq:metric}
ds^2=-f(r)dt^2+\frac{dr^2}{f(r)}+r^2(d\theta^2+\sin^2\theta d\phi^2)\nonumber\\
\mathrm{with}~~~f(r)=1-\frac{2M}{r}+\frac{h}{r}\ln\left(\frac{r}{2M}\right).
\end{eqnarray}
Here, $M$ and $h$ are the parameters related with the black hole mass and scalar hair, respectively.
The metric reduces to Schwarzschild case as $h\to 0$, and it is asymptotically flat. Its thermodynamical properties were explored in \cite{Walia:2021emv} and  the superradiant energy extraction was carried on in  \cite{Jha:2022tdl}.

On the other hand, recently remarkable breakthroughs in the observation of black holes have been made, which could promote us to investigate the physics in strong gravity field regime. One of the most important
achievements is that the Event Horizon Telescope (EHT) collaboration has released images of the supermassive black holes in M87* \cite{EventHorizonTelescope:2019dse,EventHorizonTelescope:2019uob,EventHorizonTelescope:2019jan,EventHorizonTelescope:2019ths,EventHorizonTelescope:2019pgp,EventHorizonTelescope:2019ggy}, and further in SgrA* at the center of the Milky Way system \cite{EventHorizonTelescope:2022wkp,EventHorizonTelescope:2022apq,EventHorizonTelescope:2022wok,EventHorizonTelescope:2022exc,EventHorizonTelescope:2022urf,EventHorizonTelescope:2022xqj}. Those images  provide direct evidences of the existence of black hole, and also disclose that there exists the central dark region usually known as black hole shadow surrounded by a bright ring. Nowadays, the black hole shadow and the observation data of EHT, which are believed to contain the information of spacetime, are extensively used  to further explore  black hole physics, GR and beyond,  for example, parameter estimations of black holes \cite{Kumar:2018ple,Ghosh:2020spb,Afrin:2021imp,Ghosh:2022kit}, extra dimension size constraint \cite{Vagnozzi:2019apd,Banerjee:2019nnj,Tang:2022hsu} and testing general relativity or alternative theories of gravity and fundamental physics \cite{Mizuno:2018lxz,Psaltis:2018xkc,Stepanian:2021vvk,Younsi:2021dxe,KumarWalia:2022aop,Vagnozzi:2022moj,Meng:2022kjs,Kuang:2022ojj,Gussmann:2021mjj,Khodadi:2022pqh,Khodadi:2021gbc}, and so on.

The black hole images published by EHT reflect the optical features of black holes, which result from the deflection of photon or the effect of gravitational lensing by the black holes. In particular,  the bright sector around the dark region is radiated from the  accretion matters surrounding the real astrophysical black holes,  which definitely affect the black hole shape we observe due to the different distributions of accretion matters. Though it is difficult to mimic the realistic accretion disk from theoretical study, the first image of black hole with a thin accretion disk was calculated analytically in \cite{Luminet:1979nyg}, which shows that there are primary and secondary images appeared outside black hole shadow.  Then  the author of  \cite{Bambi:2013nla}  pointed out that it is relatively easy to distinguish a Schwarzschild black hole from the static wormhole according to shadow images.  More recent investigations of a Schwarzschild black hole with thin and thick accretion disks show that the lensed ring together with photon ring contribute additional observed flux to the image \cite{Gralla:2019xty}. As another kind of accretion, the spherical accretion has been applied to analyze the image of a Schwarzschild black hole \cite{Falcke:1999pj,Narayan:2019imo}. It is shown that the shadow is a robust feature, and its size and shape are primarily influenced by rather the spacetime geometry than the  details of the accretion. Currently the studies on photon ring and observational appearances of black holes have attracted much more attentions \cite{Zeng:2020dco,Zeng:2020vsj,Peng:2020wun,Saurabh:2020zqg,Qin:2020xzu,Gan:2021pwu,Okyay:2021nnh,Li:2021ypw,Li:2021riw,Guo:2021bhr,Hu:2022lek,Guo:2021bwr,Wen:2022hkv,Chakhchi:2022fls,Hou:2022eev,Kuang:2022xjp,Uniyal:2022vdu,Uniyal:2023inx}, which disclosed the images of static black holes surrounded by various accretions beyond GR.

The aim of this paper is to disclose the effect of the Horndeski hair on the image of static black hole by investigating the observational appearances of Horndeski hairy black hole \eqref{eq:metric} illuminated by various accretions. In fact, some observational investigations on this hairy black hole and its rotating counterpart have been carried out, for example  the strong gravitational lensing \cite{Kumar:2021cyl}, weak gravitational lensing \cite{Atamurotov:2022slw}, shadow constraint from EHT observations \cite{Afrin:2021wlj}, which indicated that the non-trivial scalar has significant influences on
the optical phenomena of the Horndeski black hole. Here, we will explore the light rays around the Horndeski hairy black hole \eqref{eq:metric} using the ray-tracing method, and utilize this to figure out the optical appearance of black hole surrounded by thin accretion disk and thin spherical accretion flow, respectively. Especially, we will peer the contributions from the direct, lensed ring and photon ring brightness to the total brightness distribution of the optical appearance under the thin disk accretion, and analyze the effect of the Horndeski hair in this process by comparing to the Schwarzschild case. It is worthwhile to emphasize that to our knowledge, all the existed literatures in this scenario mainly concentrated on the distribution of the total brightness rather than its sources of contribution, except the reference \cite{Wang:2020jek} in which the authors traced back the sources of forming extra rings in the image of gravastars. Before the main-body study, we should first clarify the range of hairy parameter we will consider. Since the hairy parameter $h$ has the same dimension as $M$ and $r$, we shall use the dimensionless quantity $h/M$. From Eq.\eqref{eq:metric}, it is not difficult to induce that for arbitrary $h$, $r=0$ is an intrinsic singularity as the curvature scalar is singular and $f(r)=0$ always admits a solution $r=2M$ which indicates a horizon at $r=2M$.  In addition, when $h/M>0$, $r=2M$ is the unique root of $f(r)=0$ , so it is the unique horizon, i.e., the event horizon for the hairy black hole as for the Schwarzschild black hole. While for $-2<h/M<0$, $f(r)=0$ has two roots: one is $r_+=2M$ indicating the event horizon, and the other $r_-$ denotes the Cauchy horizon which is smaller than the event horizon. $r_-$ increase as $h$ decreases, and finally approaches $r_+$ as $h/M\to -2$, meaning the extremal case. Thus, here we will focus on the hairy parameter in the range $-2\leq h/M \leq 0$ as considered in \cite{Kumar:2021cyl,Atamurotov:2022slw,Afrin:2021wlj} to explore the effect of Horndeski hair on the black hole's optical appearance.

This paper is organized as follows. In section \ref{sec-Trajectories}, we calculate the trajectories of photons deflected by the Horndeski hairy black hole with the help of Lagrangian formalism and ray tracing method. In section \ref{thin}, we investigate the total observed intensity and the possible sources of its contribution for three standard disk accretion profiles, and then compare the optical appearances of the hairy black hole to that of Schwarzschild black hole. In section \ref{spherical}, we consider the hairy black hole surrounded by the static and radially infalling spherical accretions, respectively, and explore the optical appearances. Finally, we conclude in section \ref{conclusion}.

\section{Trajectories of photons around Horndeski hairy black hole}\label{sec-Trajectories}

In this section, we will analyze the light rays near the Horndeski hairy black hole (\ref{eq:metric}) from the null geodesics.
The photons in the spacetime take the Lagrangian
\begin{equation}
\mathcal{L}=\frac{1}{2}g_{\mu\nu}\dot{x}^\mu\dot{x}^\nu=\frac{1}{2}\left(-f(r)\dot{t}^2+\frac{1}{f(r)}\dot{r}^2+r^2\left(\dot{\theta}^2+\text{sin}^2\theta \dot{\phi}^2\right)\right),\label{lag}
\end{equation}
and their motions are determined by the Euler-Lagrange equation
\begin{equation}\label{eq-ELeq}
\frac{d}{d\lambda}\left(\frac{\partial \mathcal{L}}{\partial \dot{x}^\mu}\right)=\frac{\partial \mathcal{L}}{\partial x^\mu},
\end{equation}
where $\lambda$ is the affine parameter and $\dot{x}^\mu=d{x}^\mu/d\lambda$ represents the four-velocity of the photon.
Due to the spherical symmetry of the spacetime, we are safe to focus on the photons moving on the equatorial plane ($\theta=\pi/2$) for convenience, which leads to $\dot{\theta}=0$. Then, from the Lagrangian, we can obtain two conserved quantities for the photons which are the energy $E$ and angular momentum $L_z$ respectively given by
\begin{equation}
E\equiv -\frac{\partial \mathcal{L}}{\partial \dot{t}}=f(r)\dot{t},   \qquad  L_z\equiv \frac{\partial \mathcal{L}}{\partial \dot{\phi}}=r^2\dot{\phi}.\label{conserved}
\end{equation}
Subsequently, by recalling $\mathcal{L}=0$ for photons and defining the impact parameter $b\equiv L_z/E$, we can finally reduce \eqref{eq-ELeq} into three first-order  differential equations that control the photons' motion in the spacetime
\begin{align}
&\dot{t}=\frac{1}{b f(r)},\label{eq1}\\
&\dot{\phi}=\pm\frac{1}{r^2},\label{eq2}\\
&\dot{r}^2=\frac{1}{b^2}-V_{\text{eff}}(r),\label{eq3}
\end{align}
where the affine parameter $\lambda$ is redefined as $\lambda/L_z$ and the effective potential takes the form
\begin{equation}
V_{\text{eff}}(r)=\frac{f(r)}{r^2}. \label{eqveff}
\end{equation}
The sign $``+"$ and $``-"$ denote that the photon travels in the counterclockwise and clockwise direction, respectively.

\begin{figure}[htbp]
\centering
\includegraphics[width=6cm]{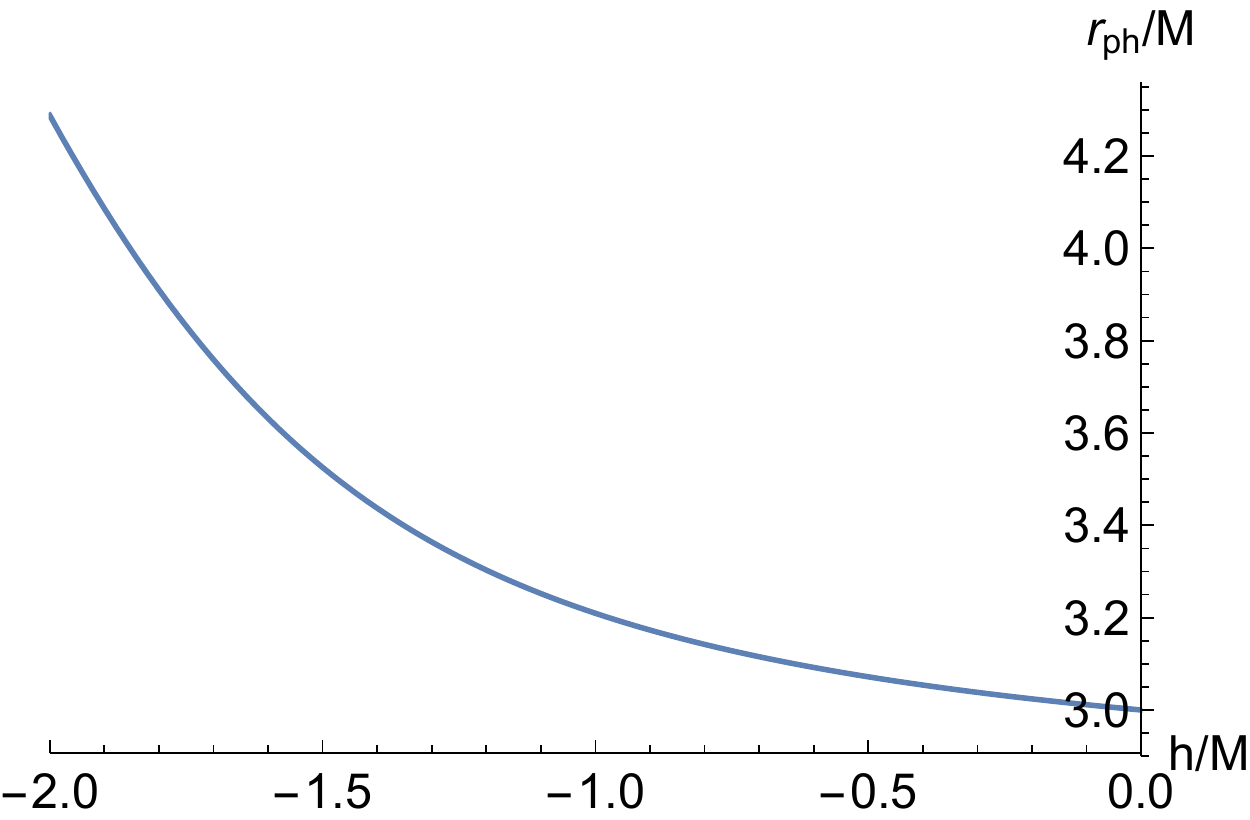}\hspace{1cm}
\includegraphics[width=6cm]{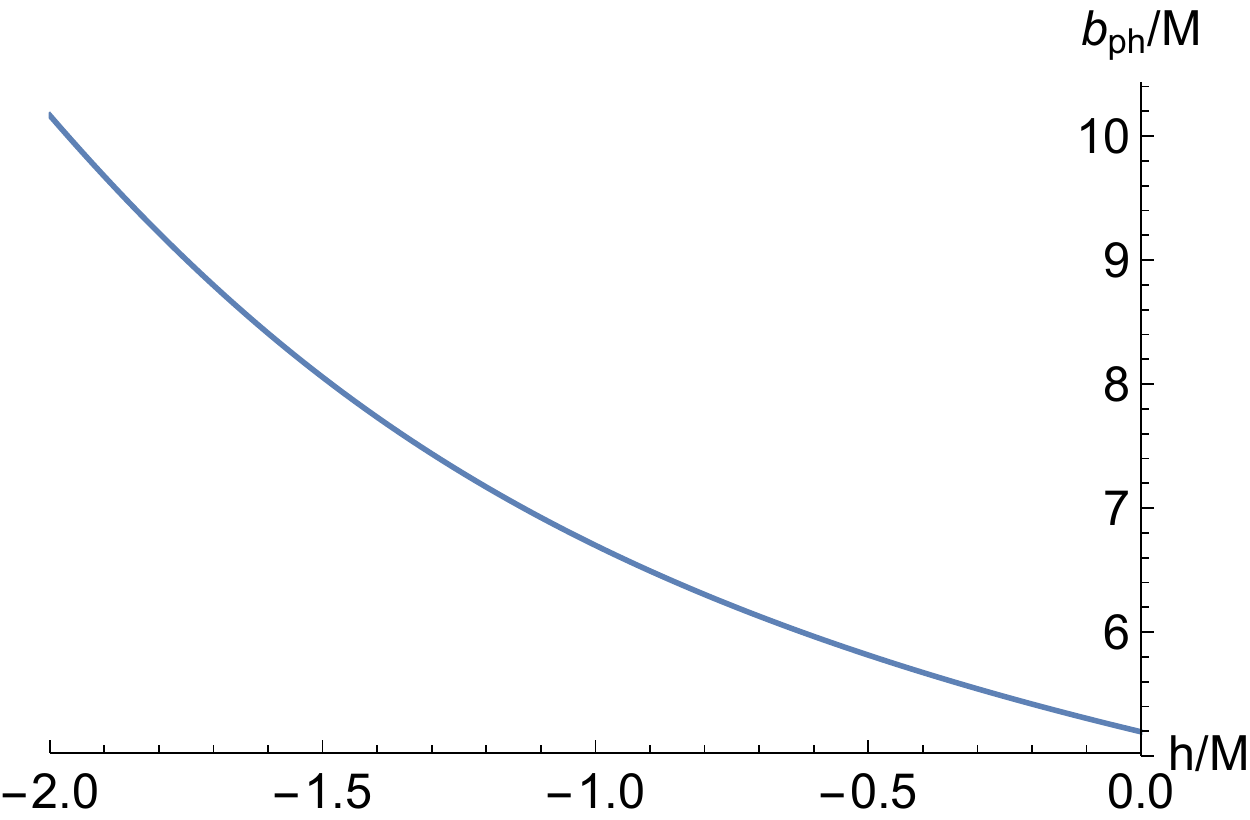}
\caption{The radius of  photon sphere $r_{ph}$ and critical impact parameter $b_{ph}$, as the function of the Horndeski hairy parameter $h/M$.}
\label{figBHquantity}
\end{figure}

Before studying the complete trajectories of photon, we shall analyze the circular orbit or photon sphere, which satisfies $\dot{r}=0$ and $\ddot{r}=0$. According to Eq.\eqref{eq3}, the conditions of photon sphere will be translated to
\begin{equation}
V_{\text{eff}}(r)=\frac{1}{b^2}, \qquad V_{\text{eff}}'(r)=0,  \label{formulabrph}
\end{equation}
where the prime represents the derivative with respect to $r$. Based on the above equations, one should numerically determine the radius of photon sphere, $r_{ph}$,  and the critical impact parameter $b_{ph}$, since analytical expressions are difficult due to the logarithmic term in the metric. {Note that in this system we have the quantities ($r$, $b$, $h$, $M$), which will be rescaled by $M$ into the dimensionless ones ($r/M$, $b/M$, $h/M$, $1$). Therefore, in the following study, all the quantities we evaluate are dimensionless whose physics is independent of $M$, such that we are safe to set $M=1$. } Thus their explicit dependence on the hairy parameter are shown in Fig.\ref{figBHquantity}, and samples of numerical results are listed in Table \ref{BHquantity} . We find that though the radius of event horizon is always $r=2M$, independent of the hairy parameter, both the radius $r_{ph}$ and impact parameter $b_{ph}$ of the photon sphere for the hairy black hole are larger than those of Schwarzschild black hole ($r_{ph}=3M$ and $b_{ph}=3\sqrt{3}M$), and they monotonously increase as $h$ decreases till the maximum in extremal case. {These effects of the Horndeski hair  will soon be understood from the behavior of the  potential function of the photon's radial motion.}

\begin{table}[htbp]
\begin{center}
\begin{tabular}{|c|c|c|c|c|c|c|c|c|c|}
\hline
{$h/M$}&{$0$}   &{$-0.5$}   &  {$-1$}   &  {$-1.5$}   &  {$-2$}\\
\hline
{$r_{ph}/M$}  &  {3}   &{3.07185}   &{3.20941}   &{3.52541}    &{4.28807}  \\
\hline
{$b_{ph}/M$}  &  {5.19615}   &{5.81484}    &{6.69978}   &{8.05599}    &{10.1676}\\

\hline
\end{tabular} \\
\caption{The numerical results of radius of photon sphere $r_{ph}$ and impact parameter $b_{ph}$ of the photon sphere for different $h$ with $M=1$.}\label{BHquantity}
\end{center}
\end{table}

\begin{figure}[htbp]
\centering {\includegraphics[width=6.2cm]{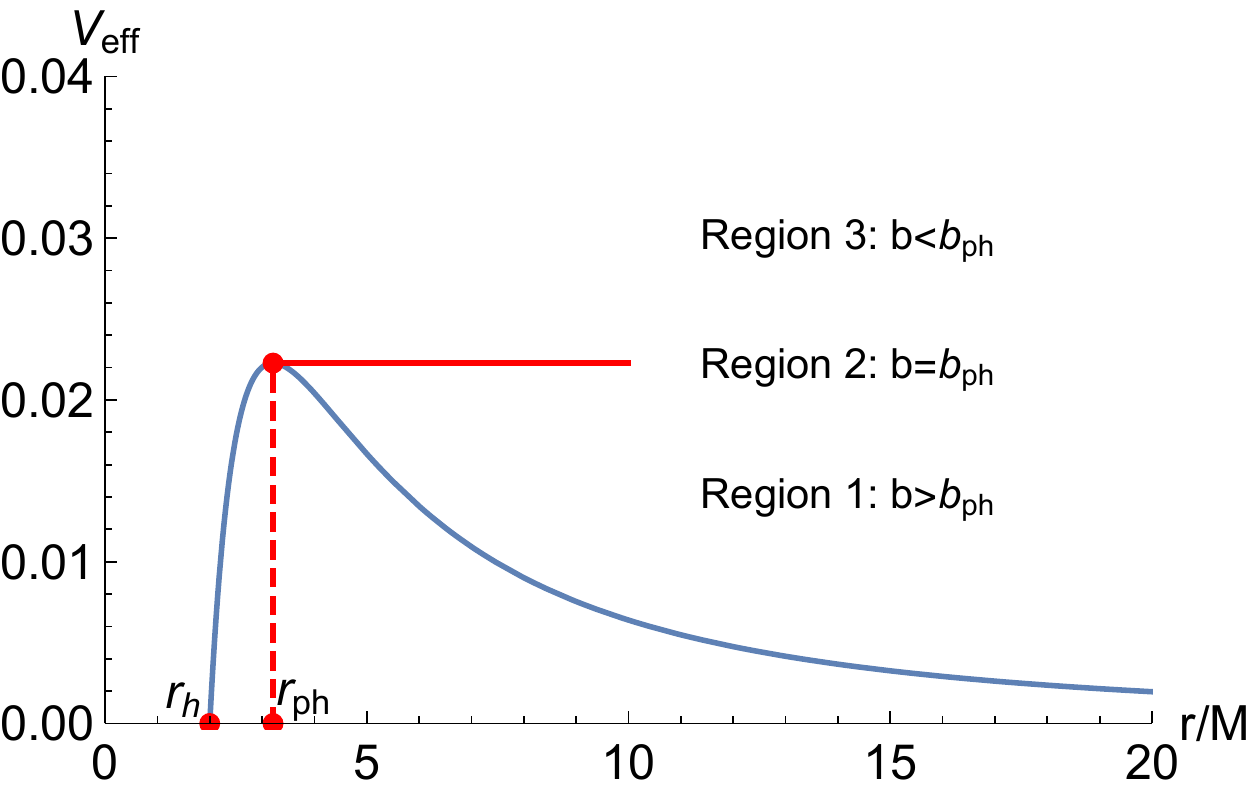}}\hspace{0.5cm}
{\includegraphics[width=4.3cm]{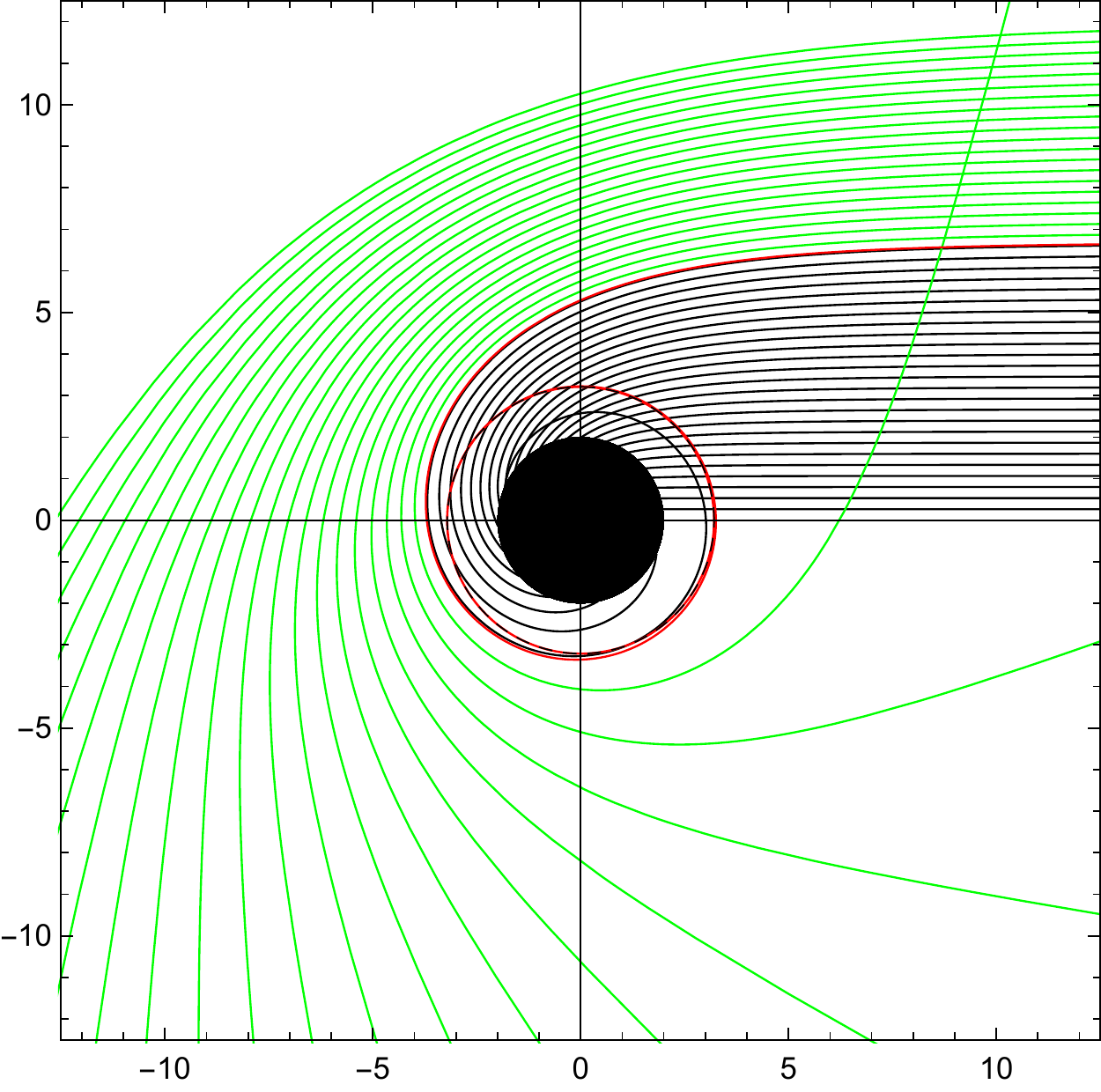}}\hspace{0.5cm}
{\includegraphics[width=6.2cm]{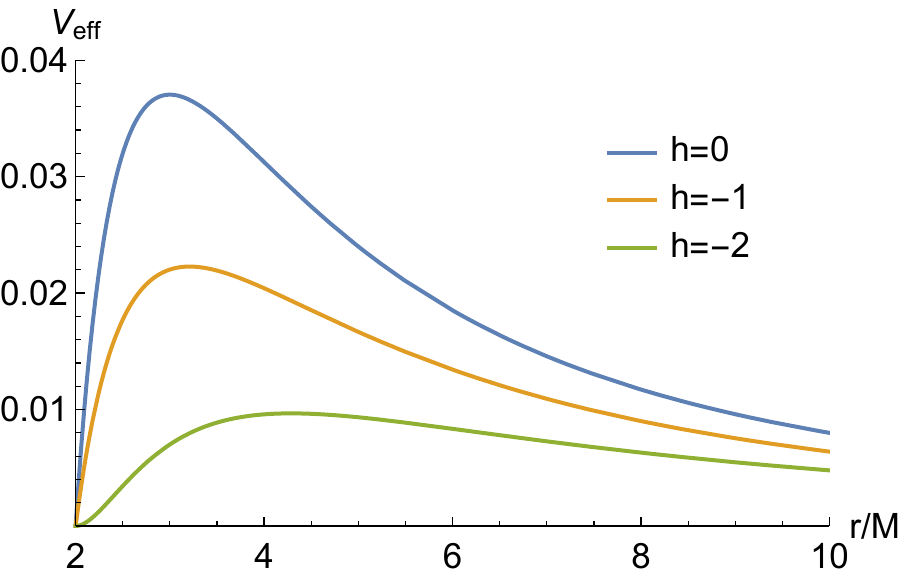}}
\caption{\textbf{Left:} The effective potential $V_{\text{eff}}$ as a function of radial coordinate $r$ for $h=-1$ as an example with $M=1$. \textbf{Middle:} The trajectories of photons for $h=-1$ as an example with $M=1$  in the polar coordinates. The red, black and green curves correspond to trajectories of photon emitted at $b=b_{ph}$, $b<b_{ph}$ and $b>b_{ph}$, respectively. The black hole is shown as the solid black disk and the photon sphere is denoted by the black dashed circle. \textbf{Right:} The effective potential as a function of radial coordinate $r$ for different $h$.}\label{figveff}
\end{figure}

We move on to study the trajectory of photon which can be depicted by solving the equation of motion
\begin{equation}
\frac{dr}{d\phi}=\pm r^2\sqrt{\frac{1}{b^2}-V_{\text{eff}}(r)},\label{eqphoton}
\end{equation}
which is obtained from combining Eq.({\ref{eq2}}) and Eq.({\ref{eq3}}). The trajectory of photon is somehow determined by effective potential $V_{\text{eff}}$ and the impact parameter $b$ (due to the symmetry, we shall focus on the positive impact parameter.).  So we scan the hairy parameter space and find that the effective potential always has one-single peak indicating the existence of single photon sphere. In details, the effective potential firstly vanishes at the event horizon, then as increasing  $r$ it increases to a maximum at the photon sphere, and finally decreases monotonously. This process is depicted in the left panel of Fig.\ref{figveff} where we have fixed $h/M=-1$ as an example. Now let us discuss the motion of photon in combination with the effective potential. In Region 1 ($b>b_{ph}$), the photon will encounter the potential barrier and then be scattered into infinity after passing through the turning point. In Region 2 ($b=b_{ph}$), the photon asymptotically approaches the photon sphere and then revolve around the black hole infinitely times. In Region 3 ($b<b_{ph}$), the photon which does not encounter the  potential barrier will fall into the black hole. These analyses match well with the trajectories of light rays we draw from solving \eqref{eqphoton}. The distribution of light rays around the hairy black hole are shown in the middle panel of Fig.\ref{figveff}, in which the solid black disk  stands for the hairy black hole and the black dashed circle denotes the photon sphere, while the red, black and green curves correspond to trajectory of photon emitted or ended at $b=b_{ph}$, $b<b_{ph}$ and $b>b_{ph}$, respectively. {Moreover, in order to further understand the influence of $h$ on the photon's fate, we show the effective potential for different $h$ in the right panel of Fig.\ref{figveff}. We find that when decreasing $h$, the position of the  effective potential's peak moves right and the value of peak is suppressed, which indeed can interpret the increasing of $r_{ph}$ and $b_{ph}$ shown in Fig.\ref{figBHquantity}.}

{Now, we are ready to study the optical appearance of the Horndeski hairy black hole under various illumination conditions. It was addressed in \cite{Yuan:2014gma} that  the hot, optically thin accretion flows are surrounding M87*, Sgr A* and many other supermassive black holes in our Universe. Here we will consider the hairy black hole illuminated by thin accretion disk and thin spherical accretions flow, respectively,  which are simple but enough for the purpose of this paper.}

\section{Rings and images of hairy black hole illuminated by thin disk accretions}\label{thin}
In this section, we shall explore the classification of light rays, and the images of the Horndeski hairy black hole illuminated by optically and geometrically thin accretion disks, which are located at rest on the equatorial plane around the black hole, viewed face-on. In this case, the light rays may intersect with the accretion disk for arbitrary times, which should contribute differently to the total observed  intensity. So we shall first classify  the light rays, then disclose the black hole images by studying the total observed  intensity for three standard accretion profiles.

\subsection{Classification of rays: direct, lensed ring and photon ring emissions}
 As illustrated in \cite{Gralla:2019xty}, the photons are classified into three classes in terms of the orbit numbers $n=\phi/(2\pi)$, where $\phi$ is the total change of azimuthal angle that the photon orbits the black hole, depending on the impact parameter via \eqref{eqphoton}. Concretely, the first class is defined as the direct emission with $n<3/4$, where light ray intersects the accretion disk only once. The second class with $3/4<n<5/4$, where light ray intersects the accretion disk twice, corresponds to the lensed ring emission. The final class is photon ring emission where the light ray with $n>5/4$ crosses the accretion disk at least three times. For the above descriptions of three classes of trajectories, one refers to \cite{Wielgus:2021peu,Bisnovatyi-Kogan:2022ujt,Hu:2022lek} to see the schematic diagrams to help understand well.
\begin{figure}[htbp]
\centering
{\includegraphics[width=5cm]{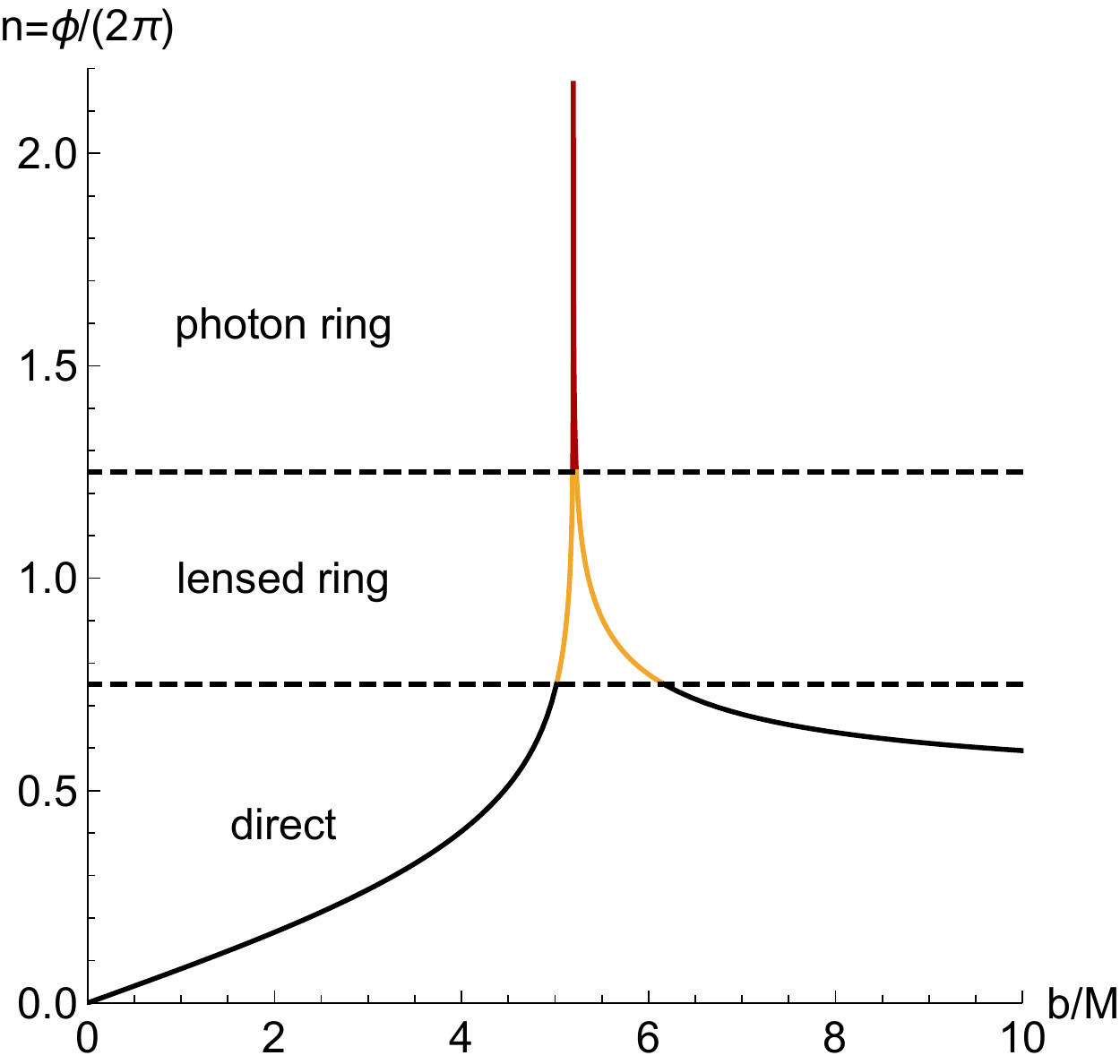}\hspace{3mm}  \includegraphics[width=5cm]{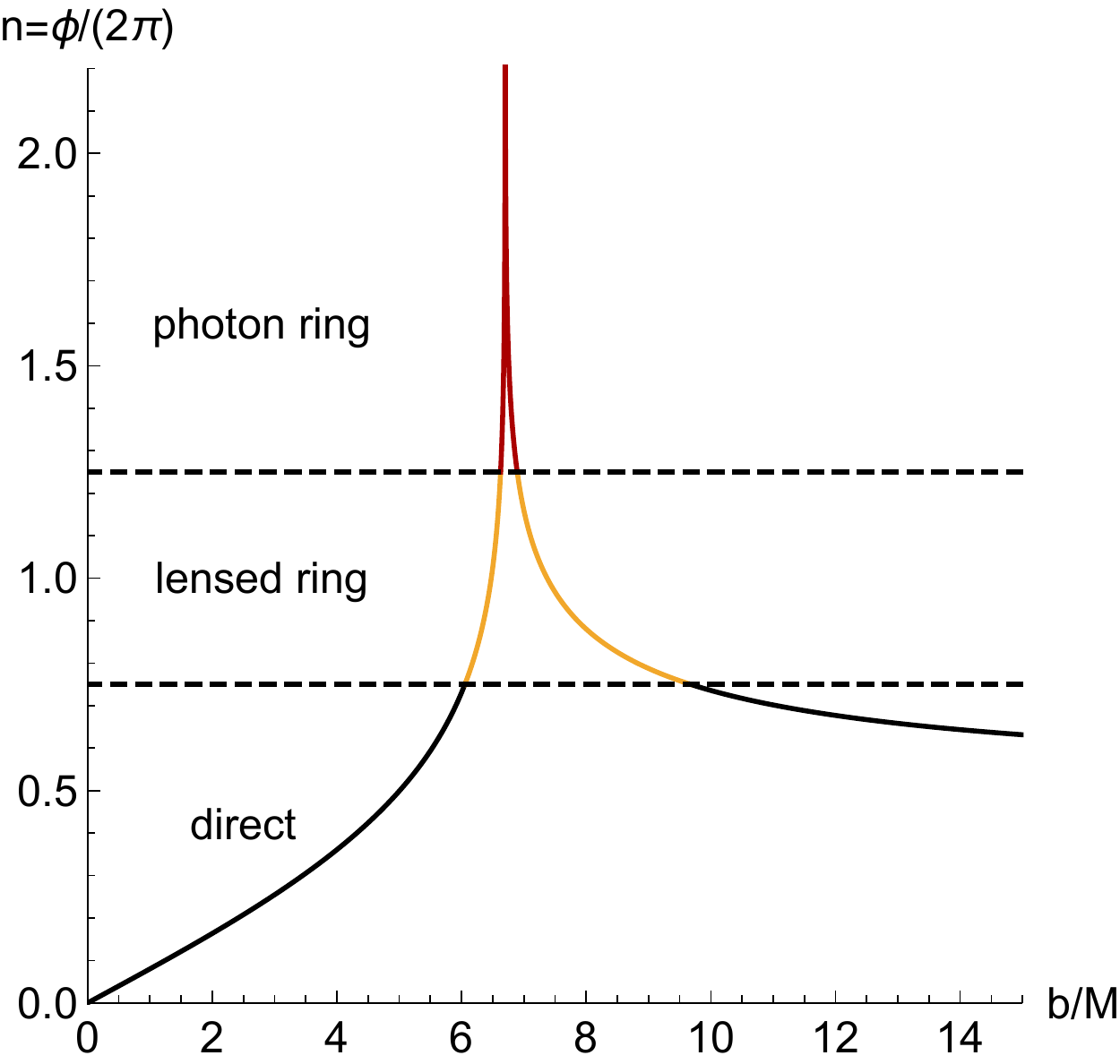}\hspace{3mm}
\includegraphics[width=5cm]{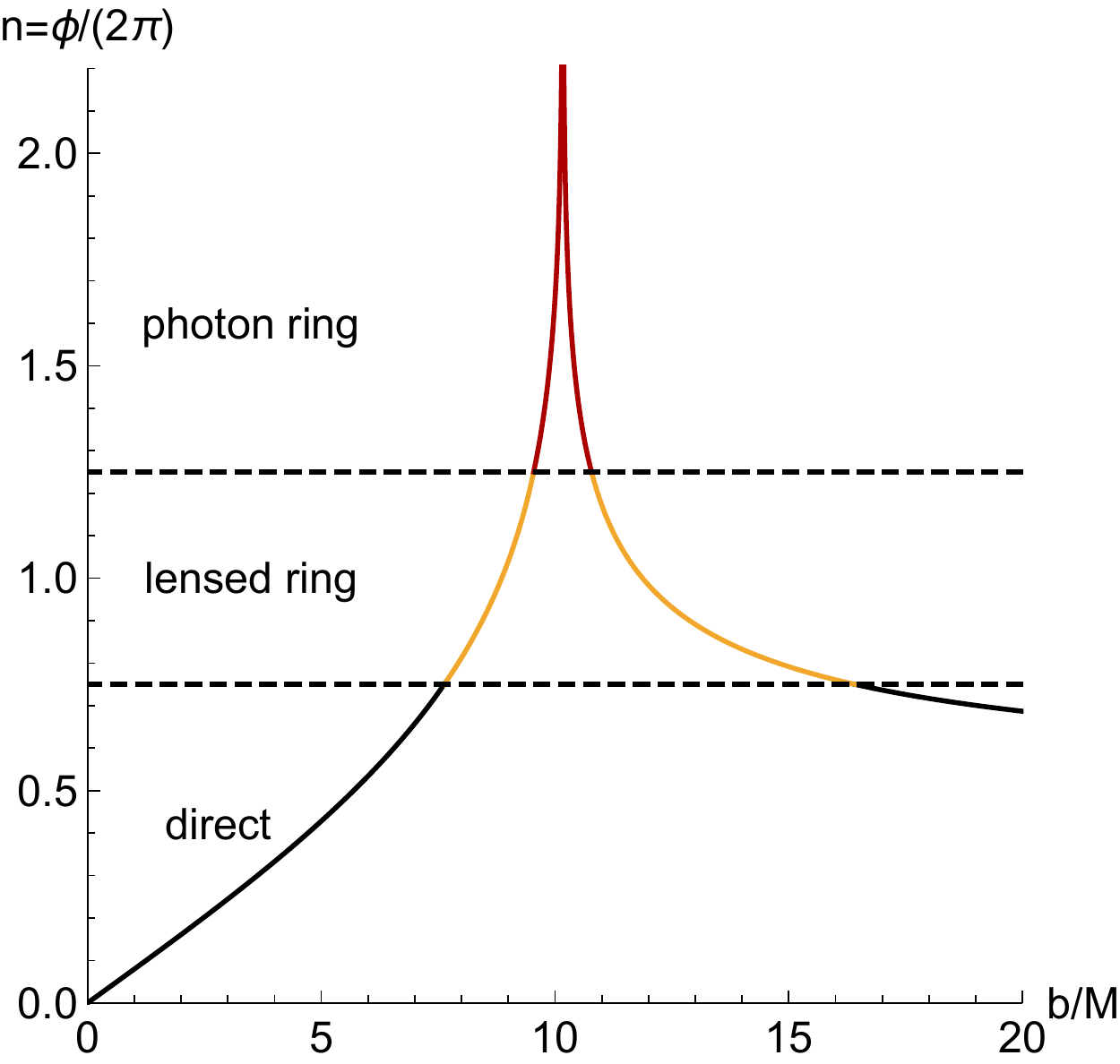}}\\
\subfigure[\, $h=0$]
{\includegraphics[width=5cm]{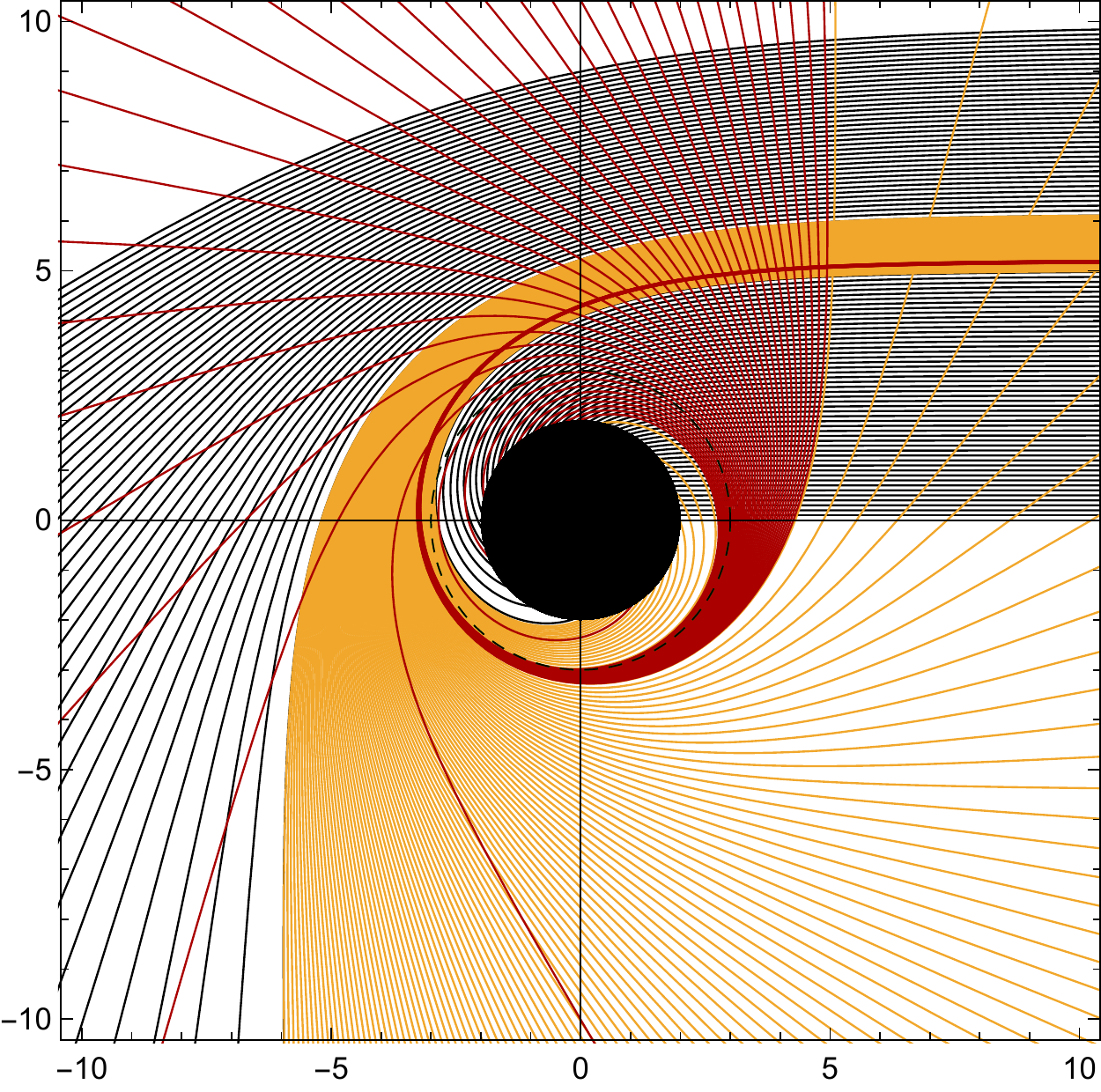}}\hspace{3mm}
\subfigure[\, $h=-1$]
{\includegraphics[width=5cm]{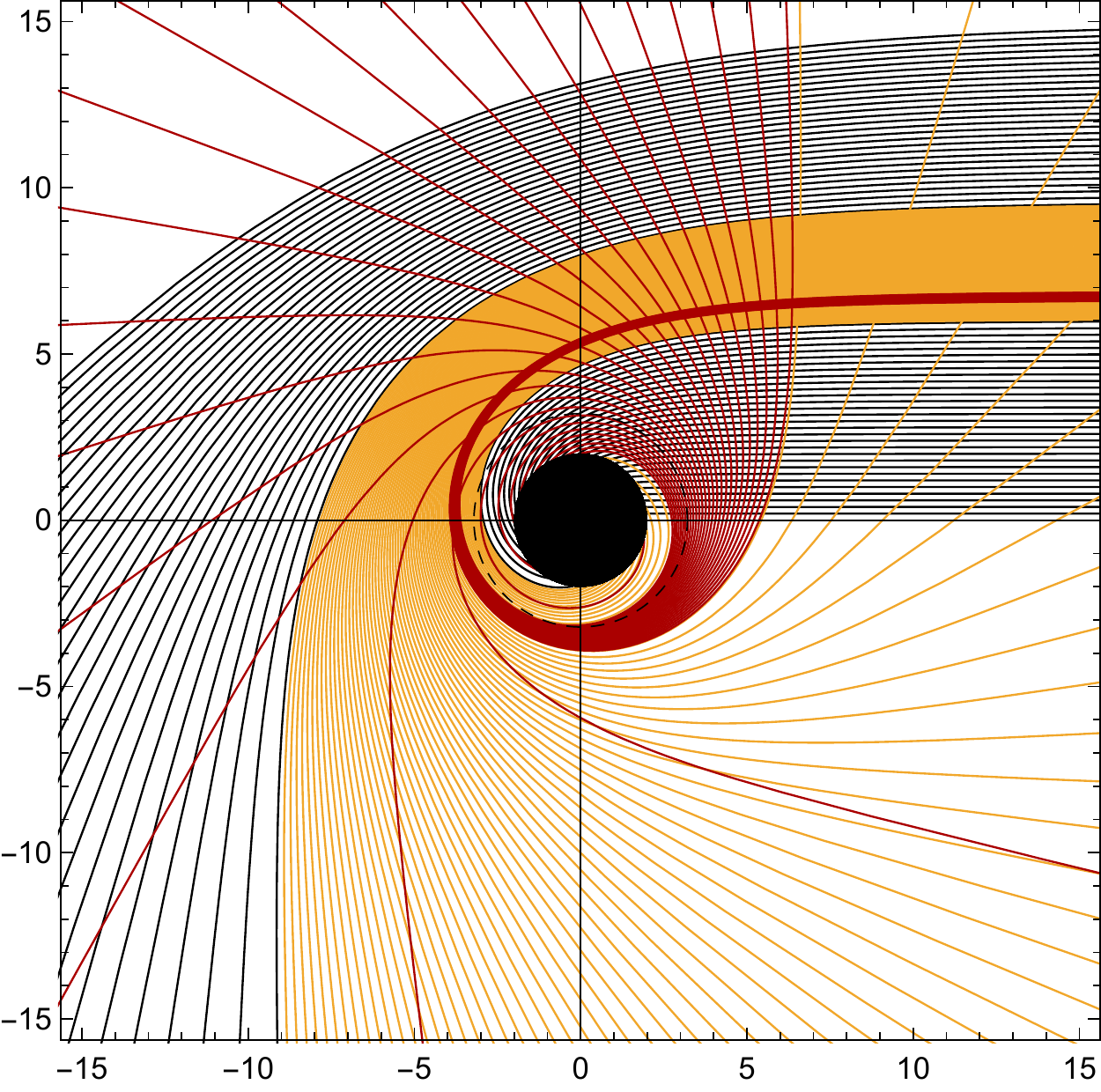}}\hspace{3mm}
\subfigure[\, $h=-2$]
{\includegraphics[width=5cm]{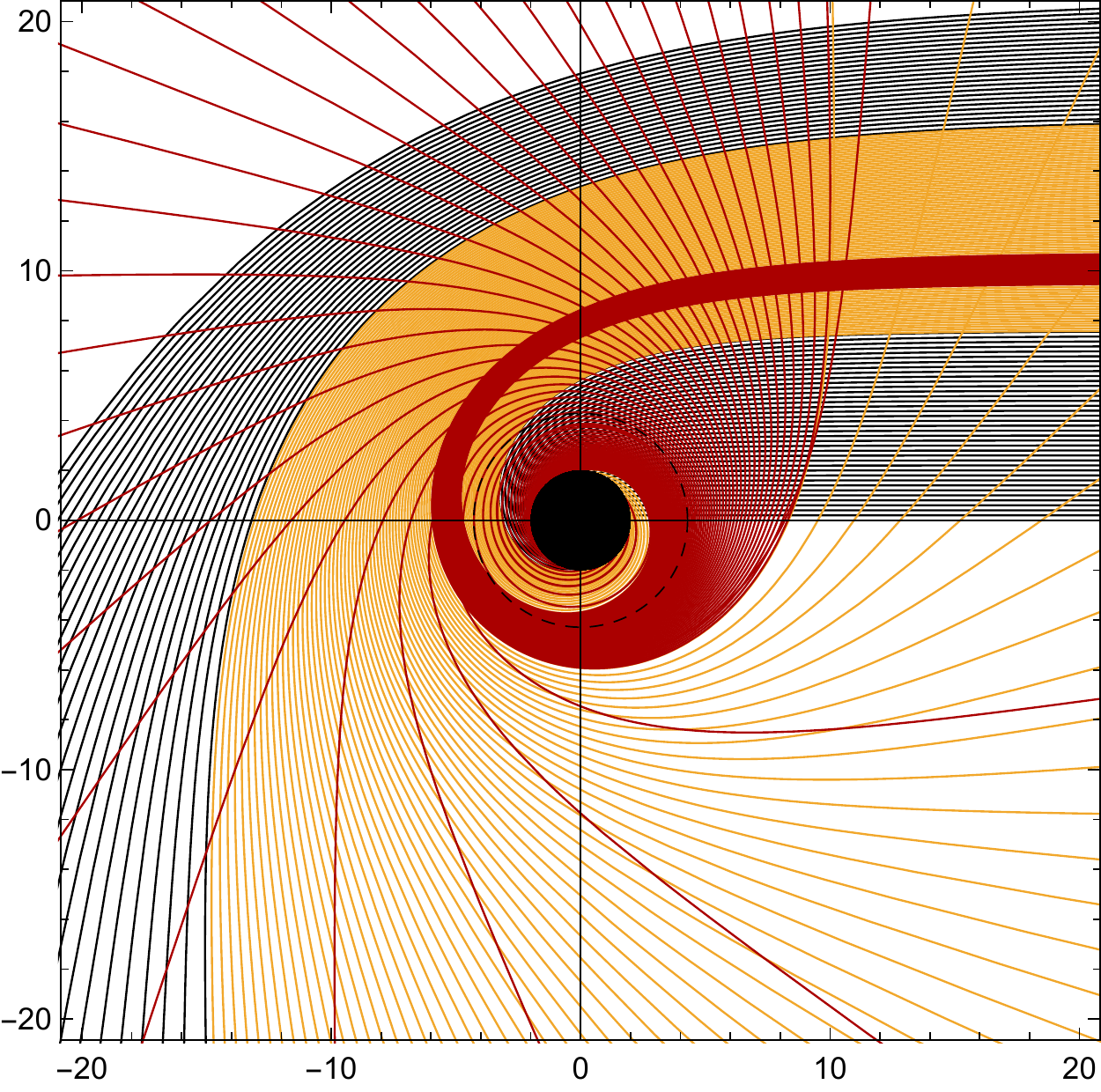}}
\caption{The orbit numbers $n$ (\textbf{top}) and trajectories of photons (\textbf{bottom}) as the functions of impact parameter $b$ for different $h$ with $M=1$. The black, gold, and red curves correspond to the direct ($n<3/4$), lensed ring ($3/4<n<5/4$), and photon ring ($n>5/4$) emissions respectively. The black hole is shown as the solid black disk and the photon sphere is denoted by a black dashed curve.}
\label{orbitNo}
\end{figure}
\begin{table}[htbp]
\begin{center}
\begin{tabular}{|c|c|c|c|c|}
\hline
{$h$}   &{Direct ($n<3/4$)}   &  {Lensed ring ($3/4<n<5/4$)}   &  {Photon ring ($n>5/4$)}\\
\hline
{$0$}   &{$b<5.01514$ and $b>6.16757$}   &  {$5.01514<b<5.18781$ and $5.22794<b<6.16757$}   &  {$5.18781<b<5.22794$} \\
\hline
{$-1$}  &{$b<6.05358$ and $b>9.67298$}   &  {$6.05358<b<6.62643$ and $6.89312<b<9.67298$}   &  {$6.62643<b<6.89312$} \\
\hline
{$-2$}  &{$b<7.62893$ and $b>16.415$}   &  {$7.62893<b<9.54309$ and $10.7805<b<16.415$}   & {$9.54309<b<10.7805$} \\
\hline
\end{tabular} \\
\caption{The ranges of impact parameter $b$ correspond to the direct, lensed ring, and photon ring emissions of the Horndeski black hole for different $h$ with $M=1$.}\label{tableb}
\end{center}
\end{table}

The orbit numbers with respect to impact parameter $b$ for different hairy parameter $h$ are shown in the top panel of Fig.\ref{orbitNo}. Similar to that in Schwarzschild black hole,  the ranges of impact parameter for  both photon ring (red curves) and lensed ring (gold curves) emission are narrow, and the direct emission (black curves) consist of two separate parts; but the Horndeski hairy black hole corresponds to  wider ranges of impact parameter for both photon and lensed rings emissions, and larger photon sphere indicated by the position of the peak. This phenomena can be simultaneously seen from the exact range of  $b$ for the direct, lensed ring, and photon ring listed in Table.\ref{tableb}, where the borders in each range are related with the intersections of each type of emission in the top panel of Fig.\ref{orbitNo}.
{The finding that the smaller hairy parameter gives wider range of photon and lensed rings emissions may attribute to the smoother peak of the potential function shown in the right panel of Fig.\ref{figveff}; and this smoother vicinity corresponds to a wider range that the photon could prefer to travel around the black hole rather than scatter directly.} Moreover, we also draw the photon trajectories in the polar coordinates in the bottom panel of Fig.\ref{orbitNo}. The Horndeski hairy black hole is denoted by the solid black disk and the black dashed circle indicates its photon sphere. Again, the red, gold, and black curves indicate the light rays emitted from photon ring, lensed ring and direct emissions, respectively.

Since the light ray will extract energy from the thin accretion disk each time when passing through it, different types of light rays will contribute differently to the observed light intensity. The above analysis indicates that the Horndeski hair has a significant effect on the type of light rays in thin accretion disk environment, which makes it interesting to further study the observed intensities and see the hairy black hole's observational appearances.

\subsection{Observed intensities and optical appearances }
Considering that the thin accretion disk emits isotropically in the rest frame of static worldlines, the specific intensity received by the observer with emission frequency $\nu_e$ is
\begin{equation}
I_{o}(r, \nu_o)=g^3 I_{e}(r,\nu_e),
\end{equation}
where $g=\nu_o/\nu_e=\sqrt{f(r)}$ is the redshift factor, and $I_{e}(r,\nu_e)$ is the specific intensity of the accretion disk.
The total observed intensity $I_{obs}(r)$ can be obtained by integrating all observed frequencies of $I_{o}(r, \nu_o)$ written as
\begin{equation}
I_{obs}(r)=\int I_{o}(r, \nu_o) d\nu_o=\int g^4 I_{e}(r,\nu_e) d\nu_e=f(r)^2 I_{em}(r),
\end{equation}
where we denote $I_{em}(r)=\int I_{e}(r,\nu_e) d\nu_e$ as the total emitted intensity. {We note that if the light ray followed backward from the observer intersects the disk, it will extract energy from the accretion disk such as to contribute  brightness \cite{Gralla:2019xty}. So, as we analyzed in previous subsection, depending on the light ray's impact parameter, it could pick up brightness once, twice or more determined by the class of the emission. Therefore, the total observed intensity should be the sum of the intensities from each intersection, yielding
\begin{equation}
I_{obs}(b)=\sum_{m}f(r)^2I_{em}(r)|_{r=r_m (b)}, \label{eqtransfer}
\end{equation} 
which is now rewritten as a function of $b$ instead of $r$ due to the introduction of the  transfer function $r_m(b)$. In fact, $r_m(b)$ is named as transfer function because it describes the mapping or transferring  from the light ray's impact parameter $b$ to 
the radial coordinate of the $m-th$ intersection with the accretion disk. In addition, its slope $dr/db$ describes the demagnification factor at each $b$ \cite{Gralla:2019xty}.
We plot the first three transfer functions $r_m (b)$ for different hairy parameter in Fig.\ref{figtransfer}, from which we can read off the following properties. (i) For the first transfer function (black curves) which corresponds to the direct image originating from direct, lensed and photon rings emission, the hairy parameter slightly enhances the value of $b$ for the emergence of transfer function, indicating the central dark region seen by the distant observer. In addition, the slope is almost $1$, which means that this direct image can be seen as the source profile after redshift. (ii) For the second transfer function (gold curves) which can origin from lensed ring and photon ring emission, the slope is large, and so this will contribute a bit into the total photon flux. Therefore, a highly demagnified image will be observed. We see that for  the Horndeski hairy black hole, the demagnification factor is suppressed by the hair, meaning that the lensed ring emission could be more visible in the hairy black hole image. (iii) For the third transfer function (red curves) which can only origin from photon ring emission, it is obvious that this image is extremely demagnified, so it will contribute little to the total flux and hardly visible. Again, the Horndeski hair slightly suppresses the demagnification factor and makes it easier to be seen than in Schwarzschild case.
\begin{figure}[htbp]
\centering
\subfigure[\, $h=0$]
{\includegraphics[width=5cm]{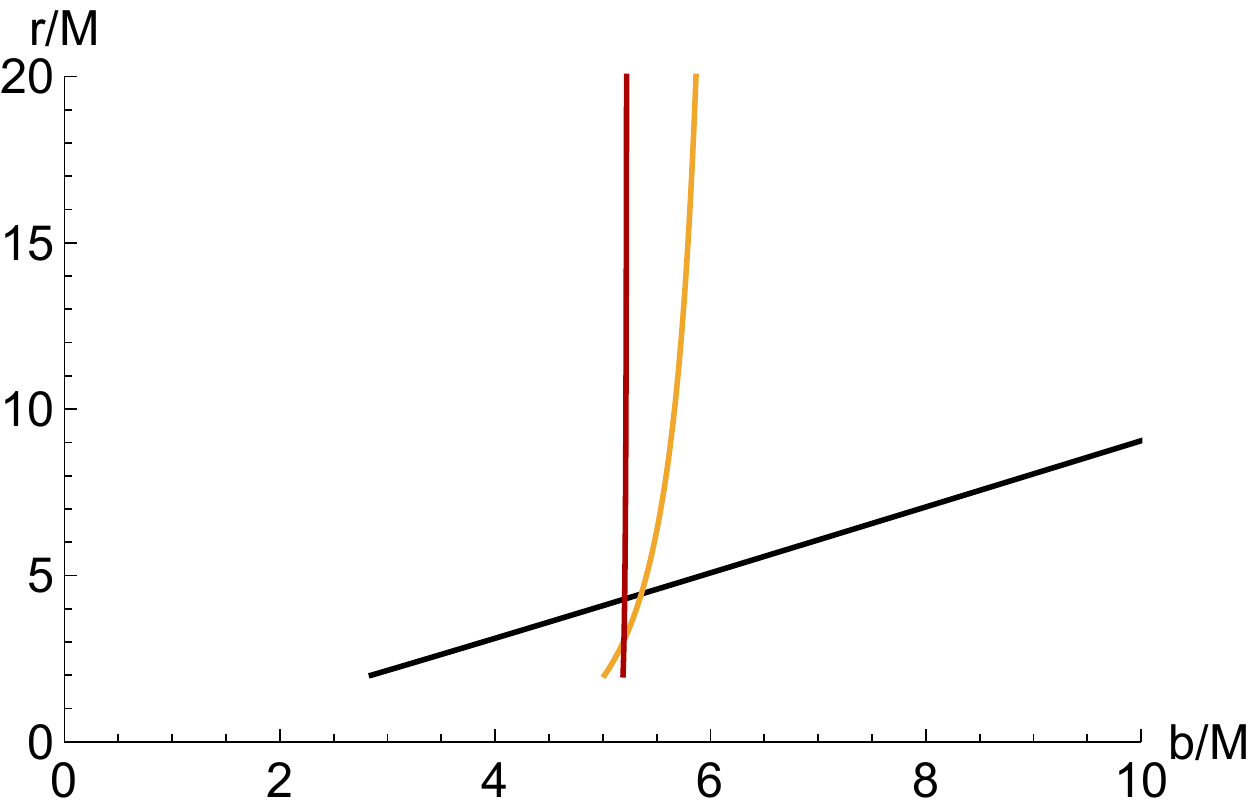}}\hspace{5mm}
\subfigure[\, $h=-1$]
{\includegraphics[width=5cm]{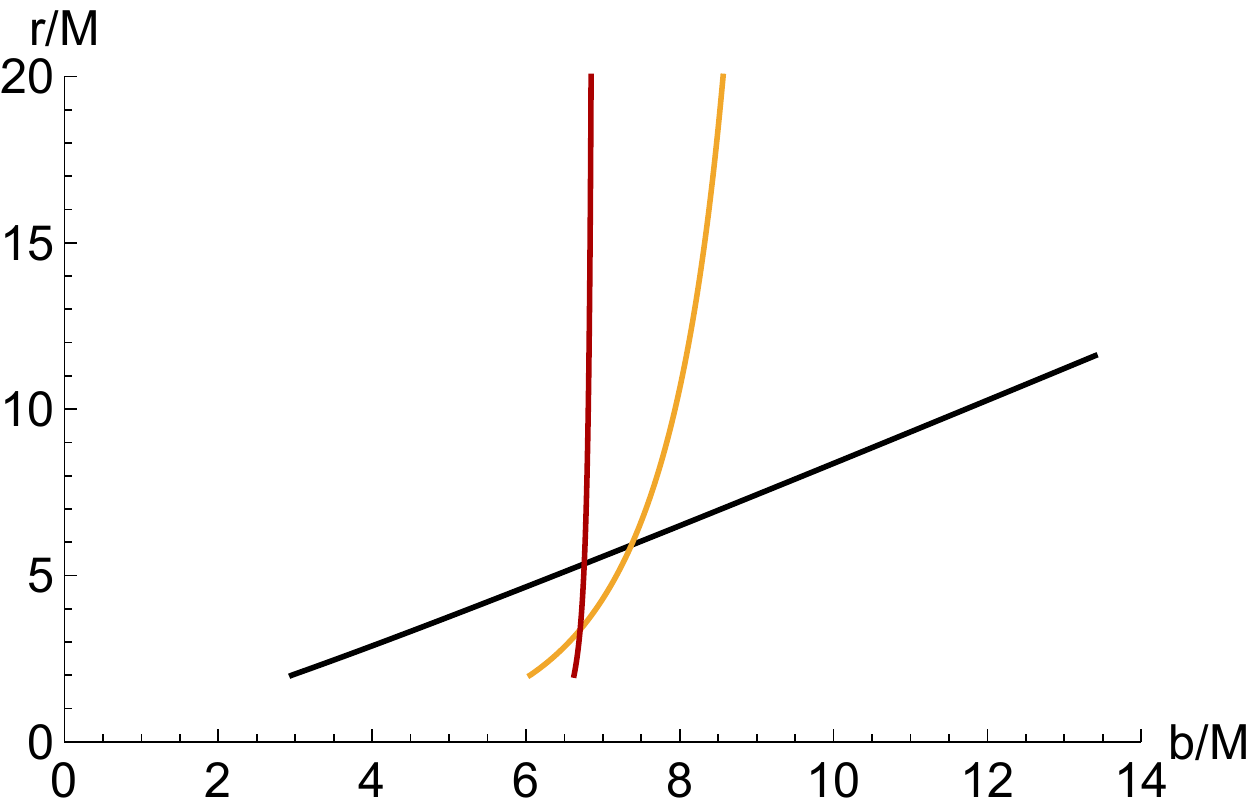}}\hspace{5mm}
\subfigure[\, $h=-2$]
{\includegraphics[width=5cm]{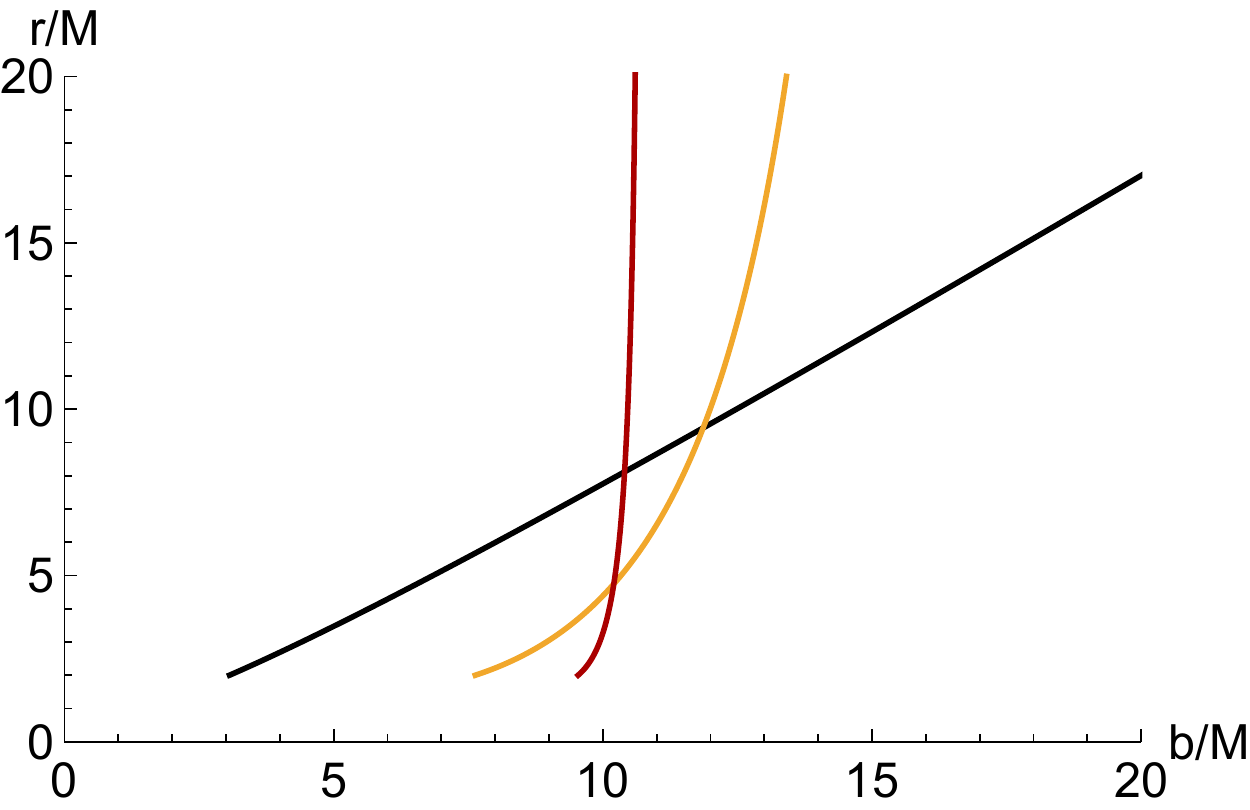}}\hspace{5mm}
\caption{The first three transfer functions in Horndeski black hole for different $h$ with $M=1$. They represent the radial coordinate of the first (black), second (gold), and third (red) intersections with the emission disk.}
    \label{figtransfer}
\end{figure}

\begin{figure}[htbp]
\centering
\subfigure[\, profile 1]
{\includegraphics[width=4cm]{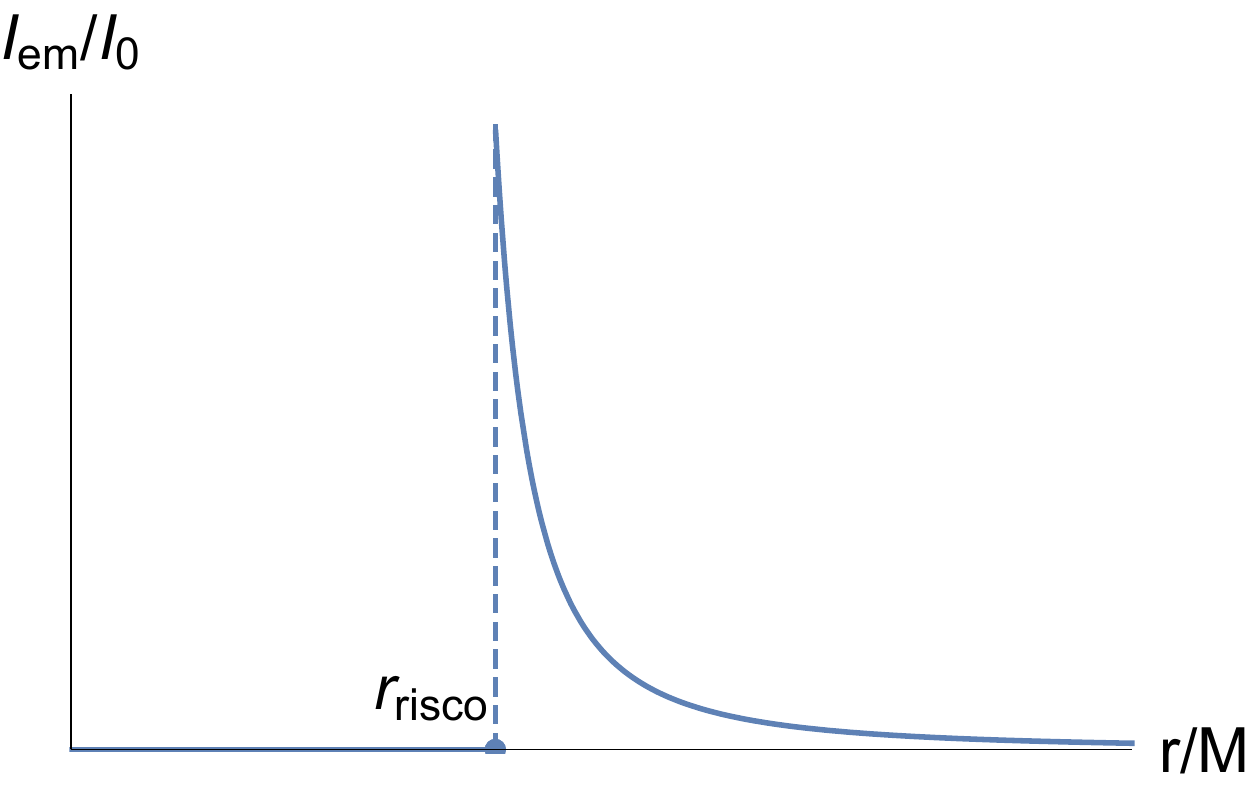}}\hspace{6mm}
\subfigure[\, profile 2]
{\includegraphics[width=4cm]{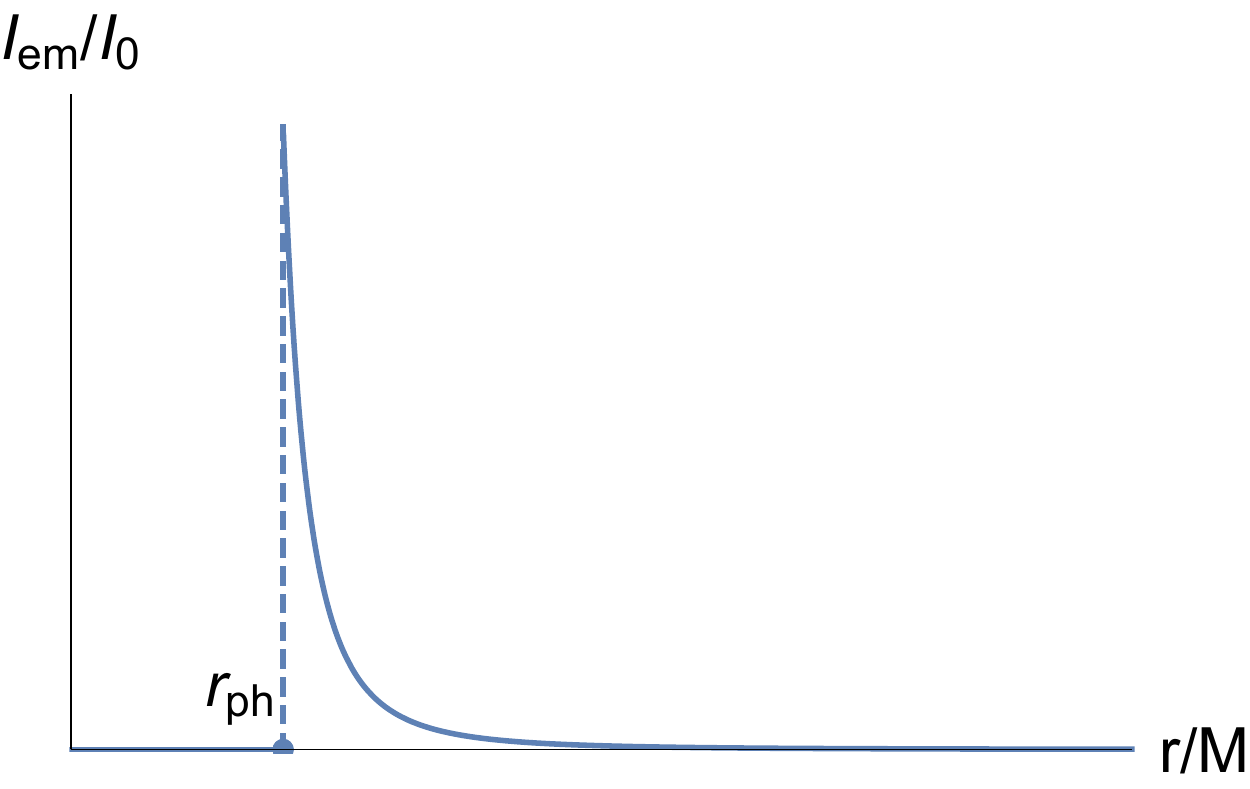}}\hspace{6mm}
\subfigure[\, profile 3]
{\includegraphics[width=4cm]{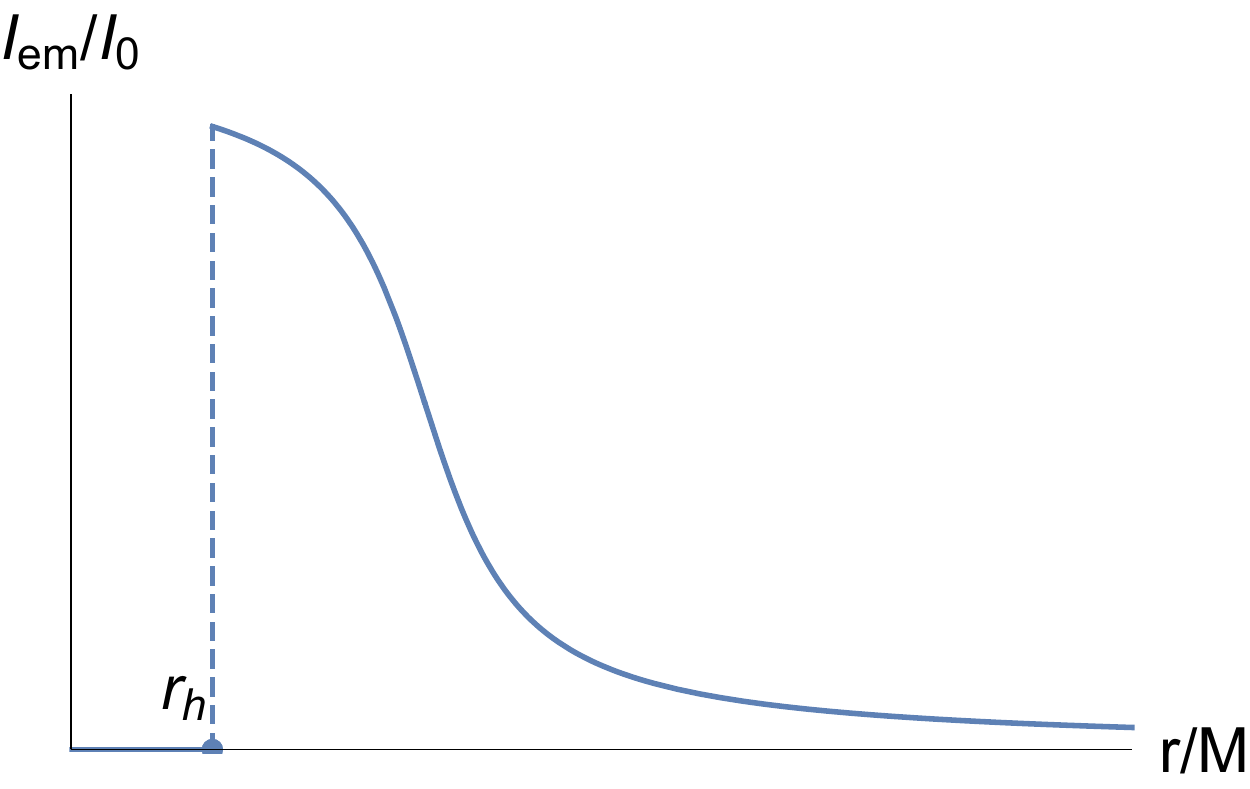}}\hspace{6mm}
\caption{The sketches of three accretion disk emission profiles plotted by Eq.\eqref{diskprofile1}, Eq.\eqref{diskprofile2}, and Eq.\eqref{diskprofile3} respectively.}
    \label{figprofile}
\end{figure}
To testify the prediction from the transfer function, we will consider some specific emission profile of accretion disk, and then evaluate the total observed  intensity from each kind of emission via \eqref{eqtransfer}. We expect to see how the direct, lensed and photon rings emissions contribute  to the total intensity and the image of black hole. Here we consider the following three toy-model emission functions \cite{Wang:2022yvi,Yang:2022btw}. Firstly, we assume that the emission of accretion disk is the power of second-order decay function that starts from the innermost stable circular orbit $r_{isco}$
\begin{align}
\text{profile 1:}\quad I_{em}(r)&:=
\begin{cases}
    I_0\left[\frac{1}{r-(r_{ isco}-1)}\right]^2, &\hspace{1.1cm}  r>r_{ isco}\\
    0,&\hspace{1.1cm} r \leqslant r_{ isco}
\end{cases},\label{diskprofile1}
\end{align}
where $I_0$ is the maximum intensity (the same below) and the $r_{isco}$ of the Horndeski hairy black hole is calculated in the Appendix A.  Secondly, the emission function starts from the photon sphere $r_{ph}$ and exhibit a cubic decay behaviour
\begin{align}
\text{profile 2:}\quad I_{ em}(r)&:=
\begin{cases}
I_0\left[\frac{1}{r-(r_{ ph}-1)}\right]^3, &\hspace{1.15cm}  r>r_{ ph}\\
0,&\hspace{1.15cm} r\leqslant r_{\rm ph}
 \end{cases}.\label{diskprofile2}
\end{align}
Thirdly, we consider a more moderate decay emission function starting from the event horizon $r_h$
\begin{align}
\text{profile 3:}\quad I_{ em}(r)&:=
\begin{cases}
I_0\frac{\frac{\pi}{2}-\arctan[r-(r_{ isco}-1)]}{\frac{\pi}{2}-\arctan[r_{ h}-(r_{ isco}-1)]}, &\quad r>r_{ h}\\
0,&\quad r\leqslant r_{ h}
\end{cases}.\label{diskprofile3}
\end{align}
The sketches of three accretion disk emission profiles are explicitly drawn in Fig.\ref{figprofile}.

Then we move on to study the optical appearance of the Horndeski hairy black hole with thin accretion disks, and the results for the three accretion disks are shown  in Fig.\ref{figprofile1}, Fig.\ref{figprofile2} and Fig.\ref{figprofile3}, respectively. We specifically show the different observed intensities originated from the first (black), second (gold) and third (red) transfer function in Eq.\eqref{eqtransfer} respectively, shown in the first columns of the figures. Note that we regard the brightness contributed by the first, second and third transfer function as the direct, lensed ring and photon ring intensity respectively. The second columns  correspond to the total observed intensities, which is the sum of the contributions of direct, lensed ring and photon ring intensity. Meanwhile, the third and fourth columns translate these total observed intensities into the corresponding two-dimensional images. Next, we will discuss the observational appearances of hairy  black hole with a thin accretion disk corresponding to three accretion profiles, respectively.
\begin{figure}[htbp]
\centering
\subfigure[\, $h=0$]
{\includegraphics[width=4.cm]{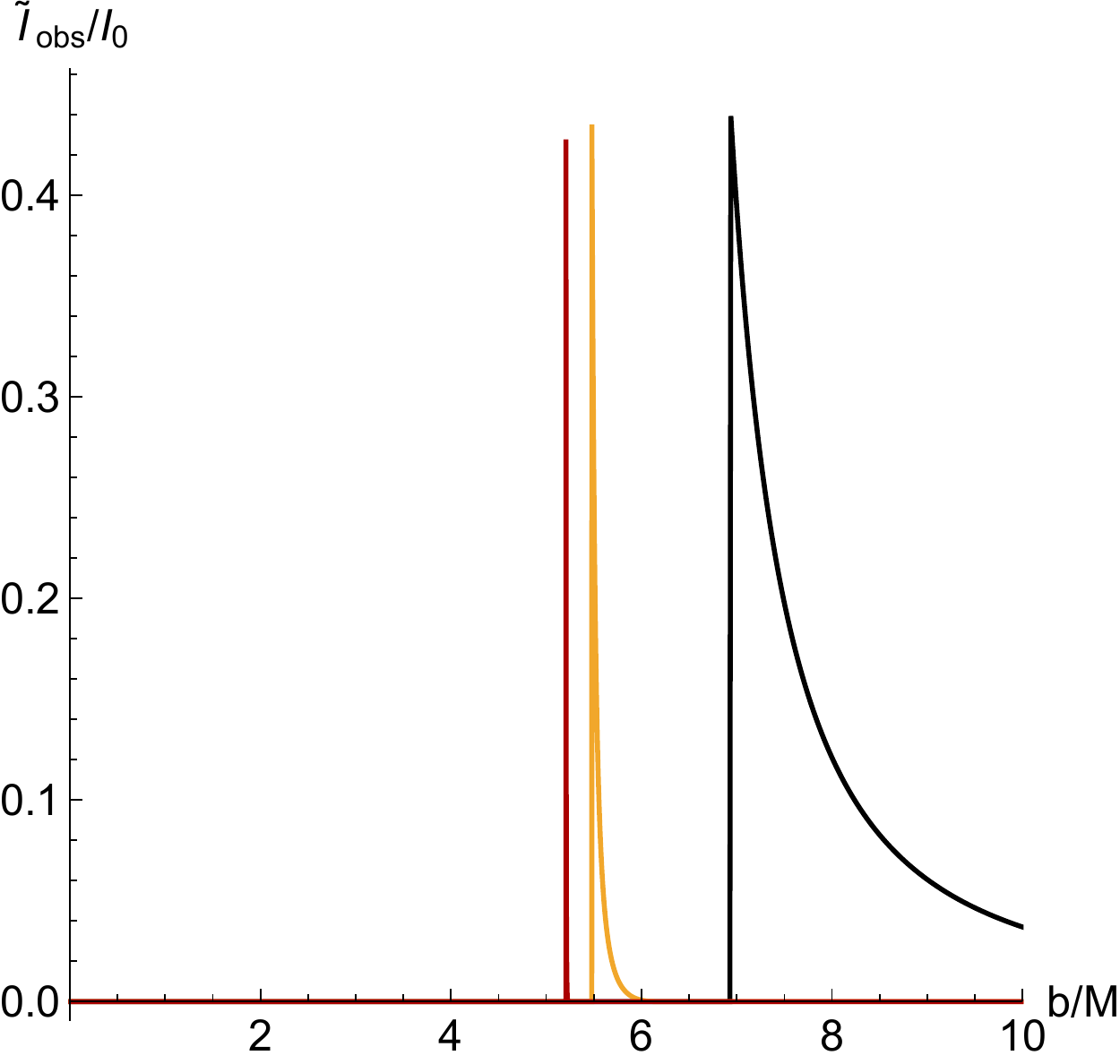} \label{}\hspace{2mm} \includegraphics[width=4.cm]{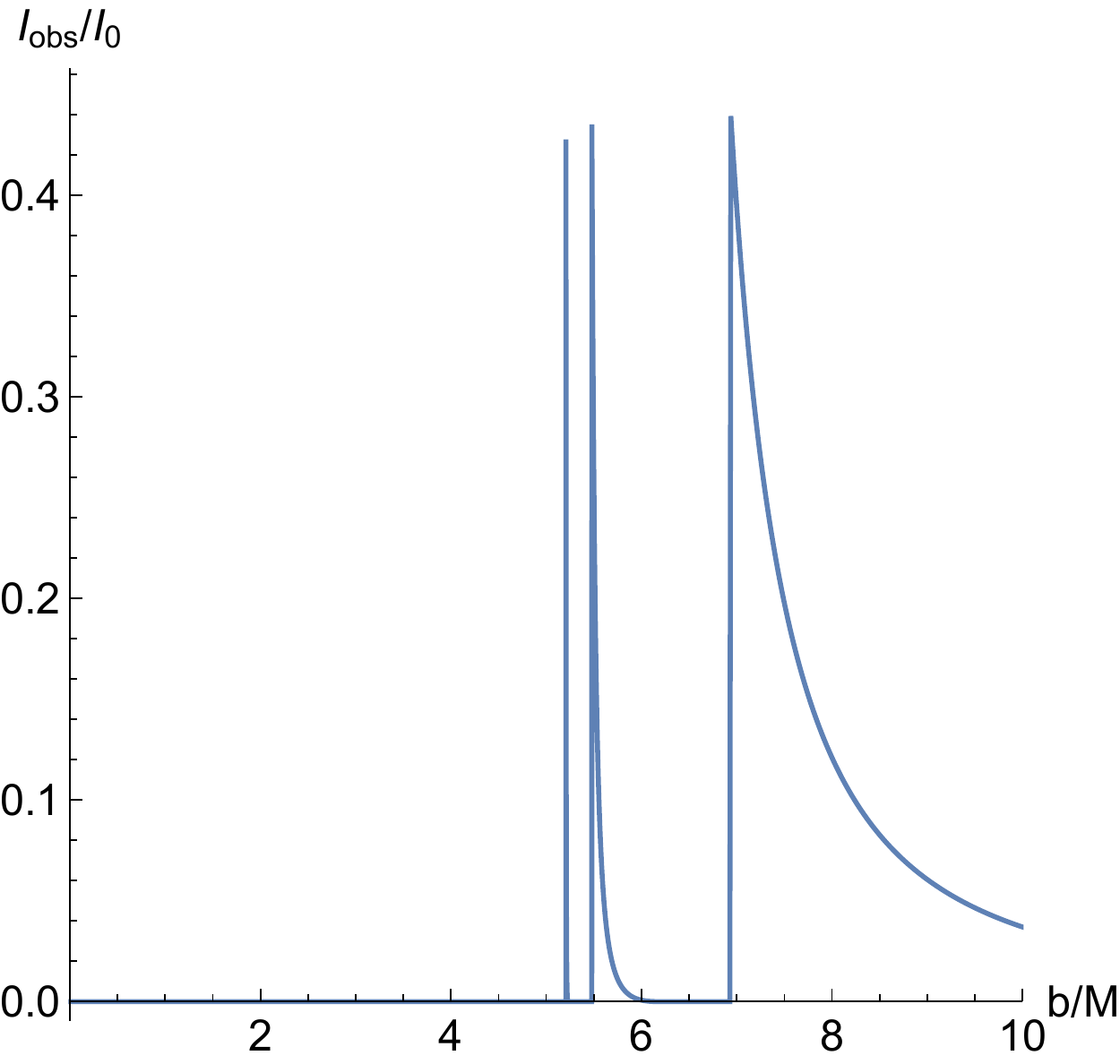}\hspace{2mm} \includegraphics[width=5cm]{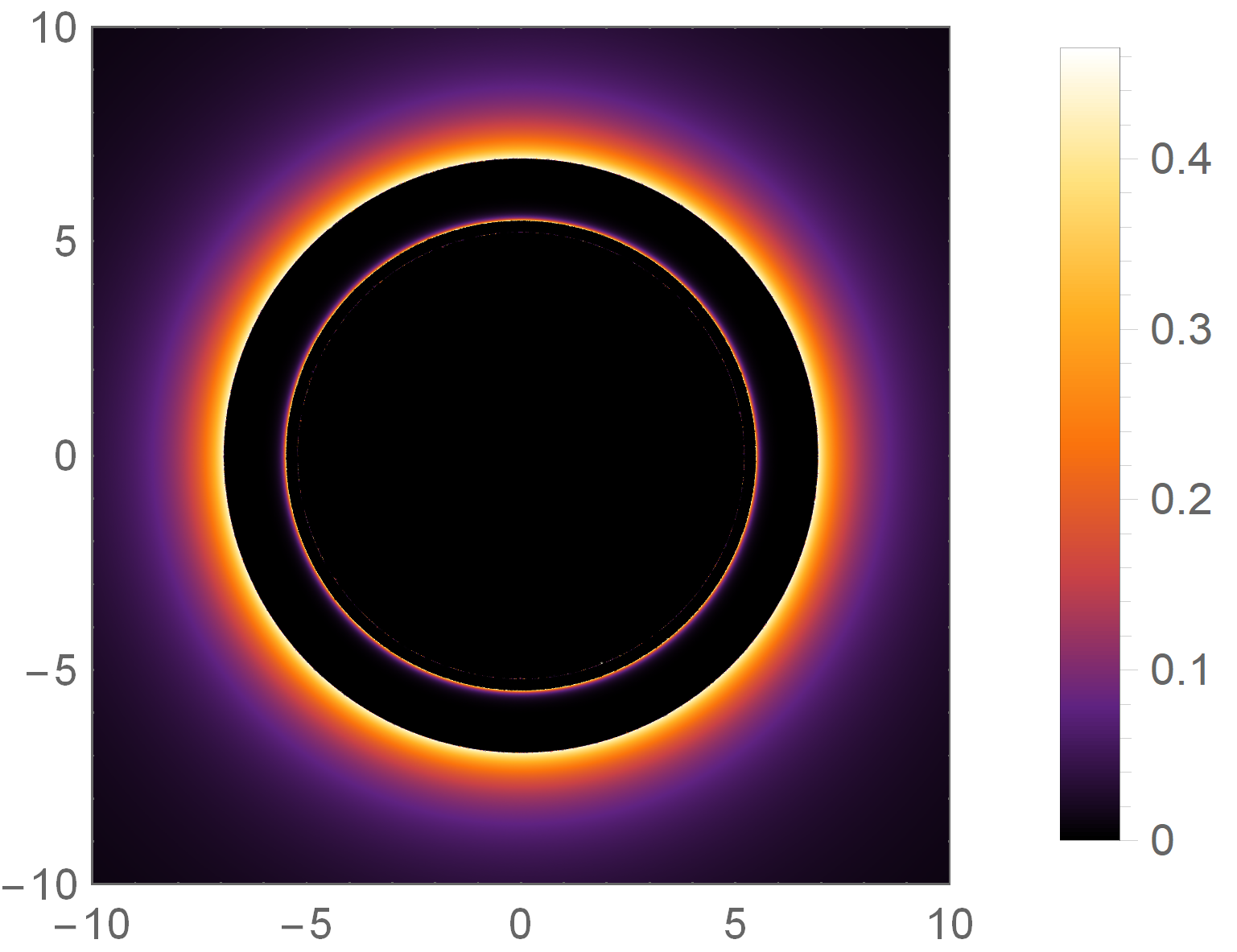}
\includegraphics[width=3.65cm]{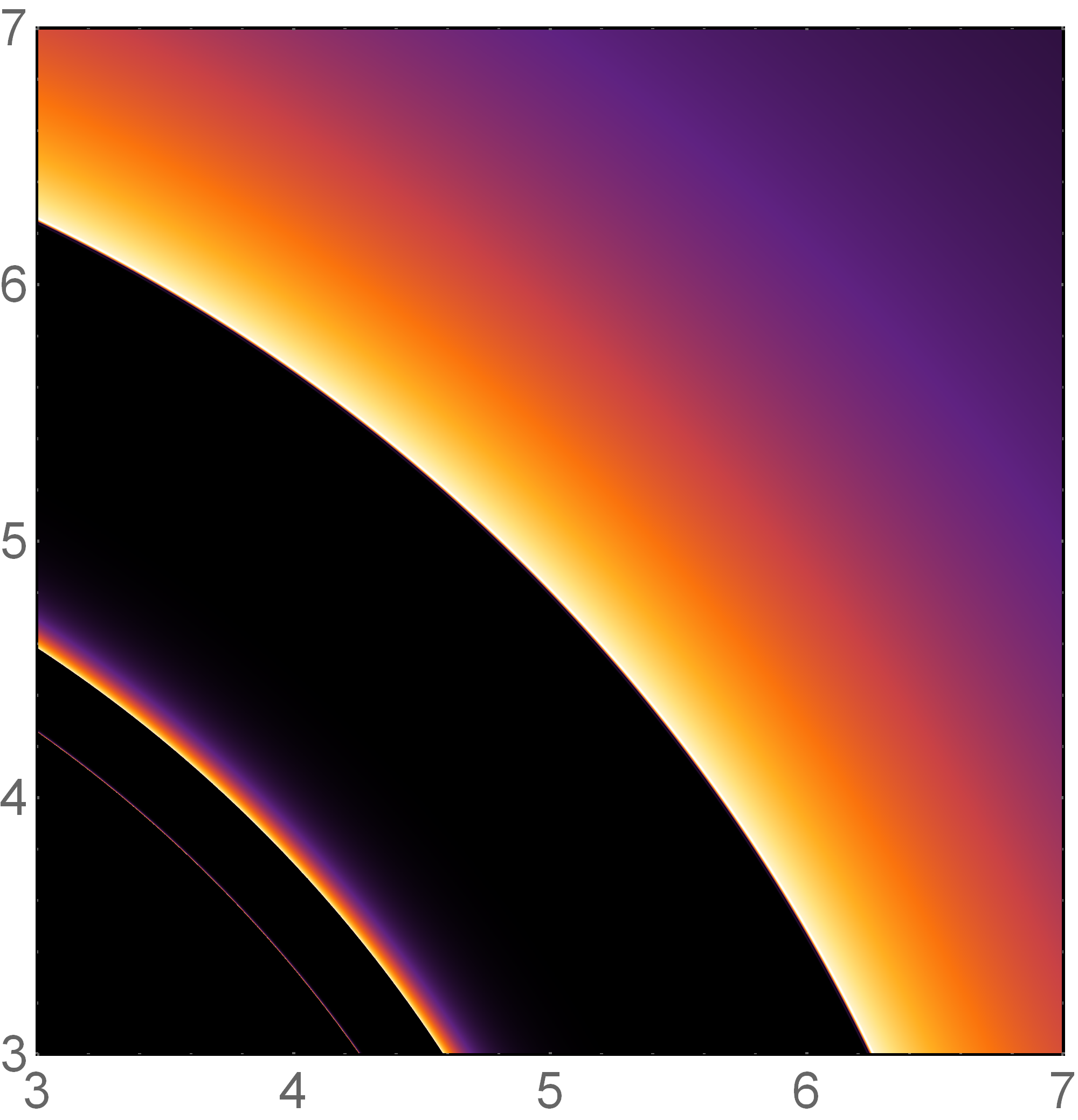}}\\
\subfigure[\, $h=-1$]
{\includegraphics[width=4.cm]{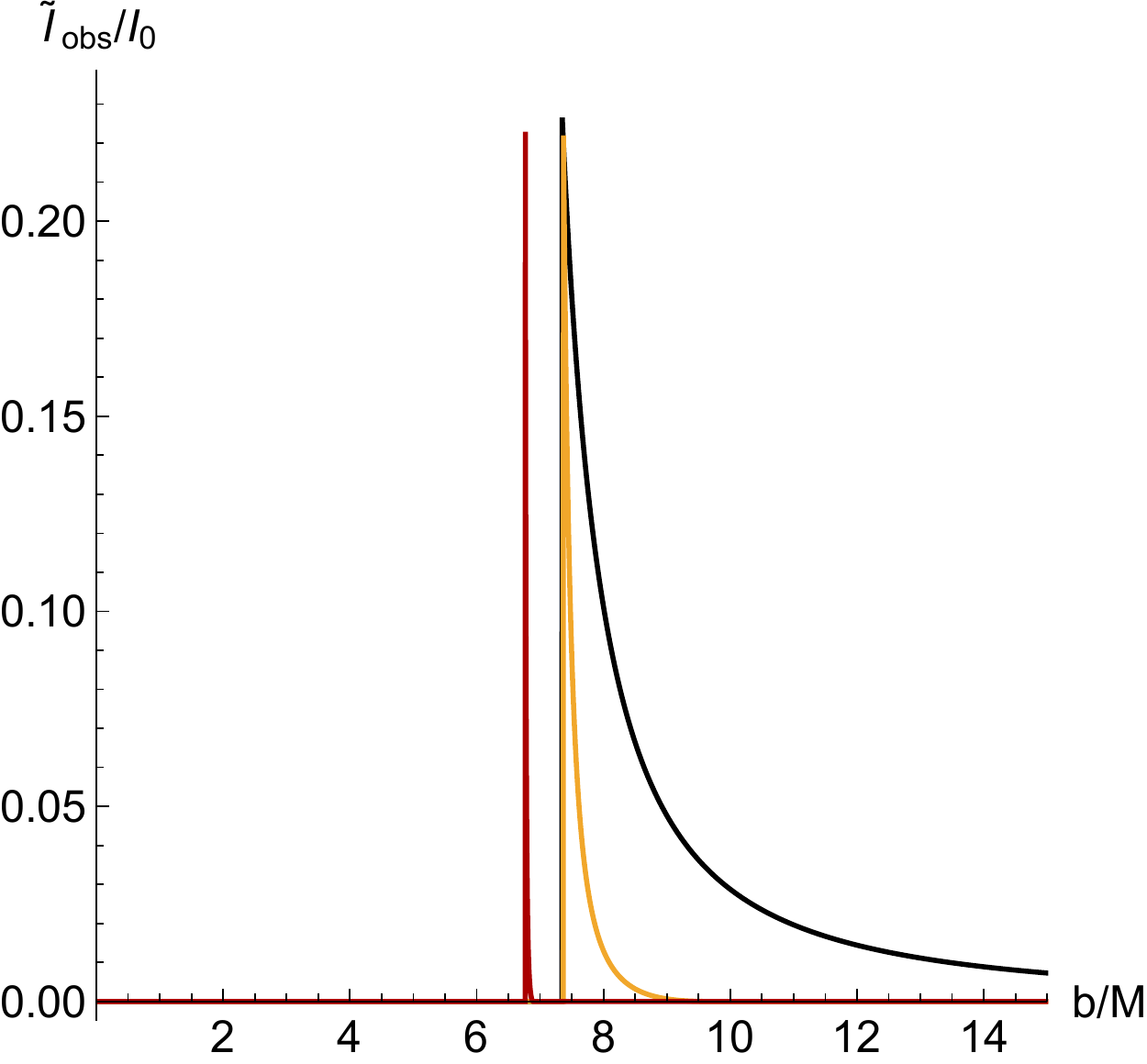} \label{}\hspace{2mm} \includegraphics[width=4.cm]{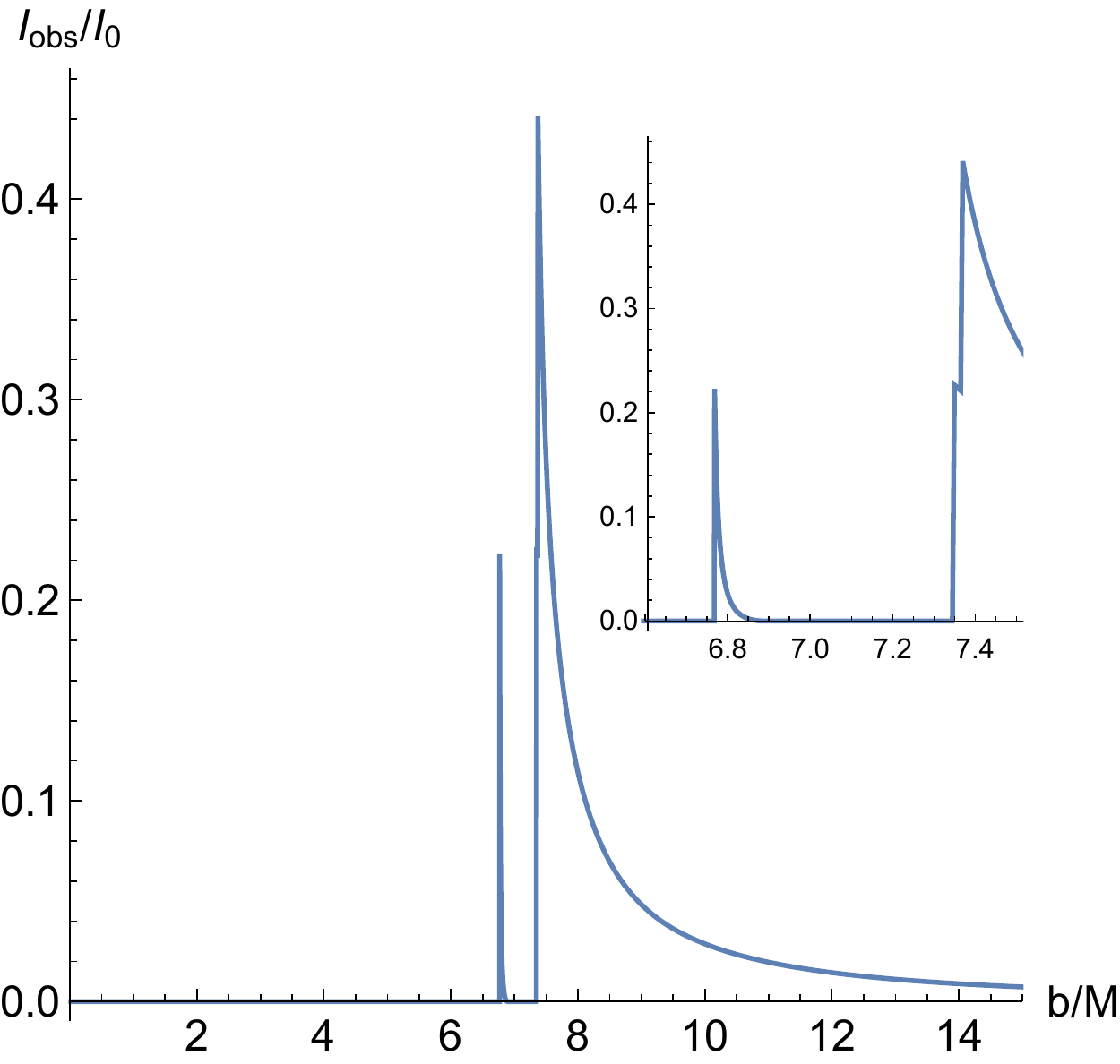}\hspace{2mm} \includegraphics[width=5cm]{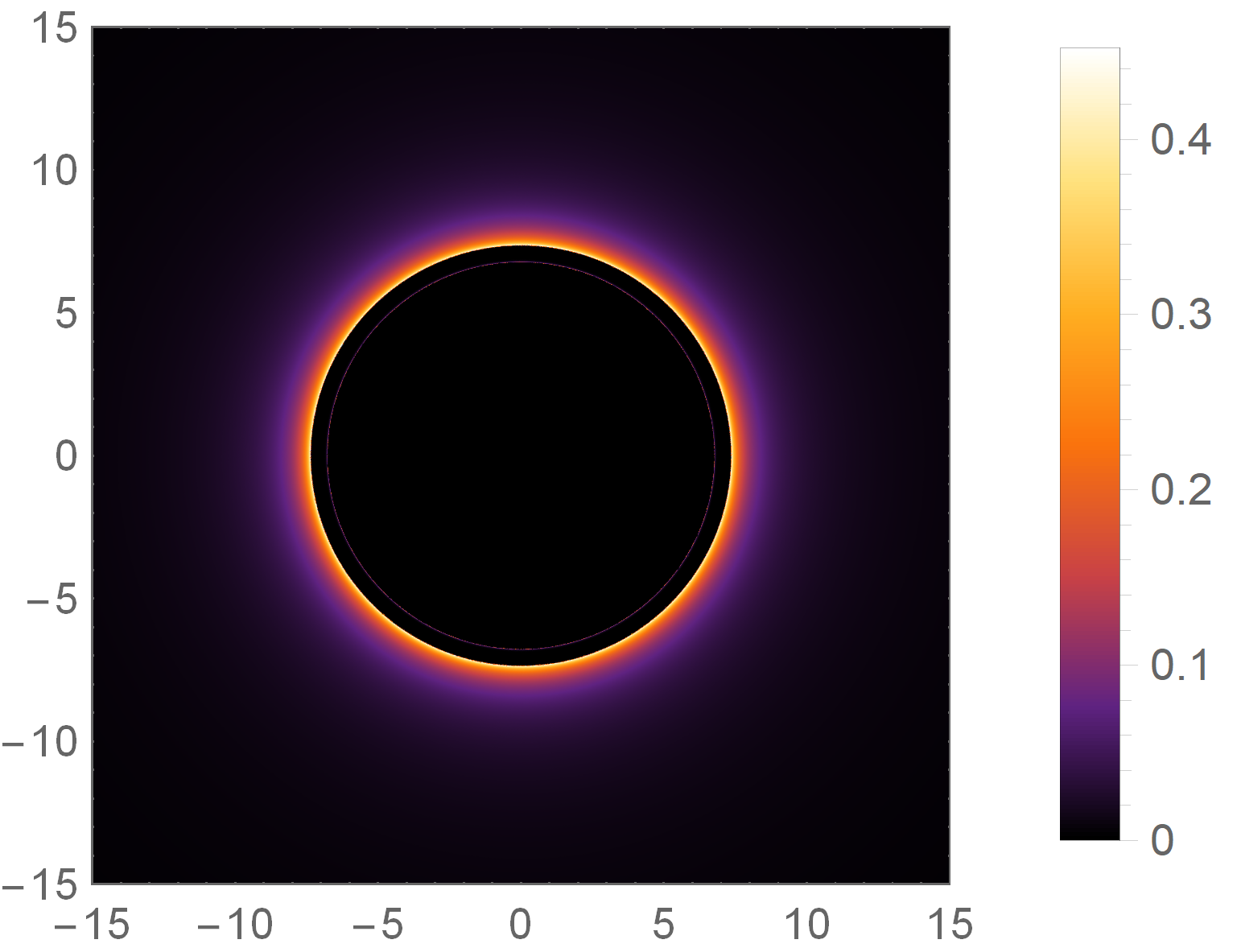}
\includegraphics[width=3.85cm]{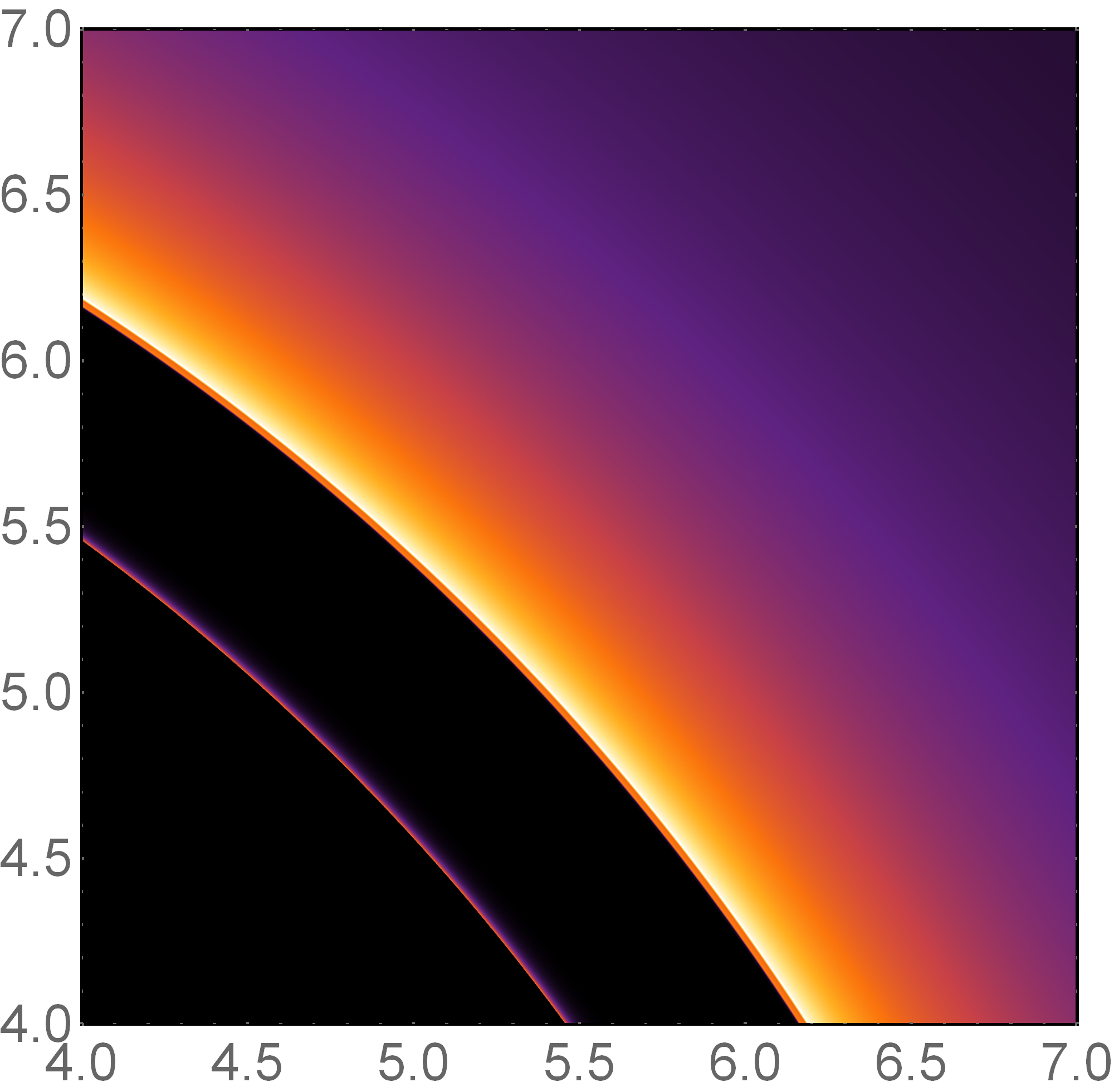}}\\
\subfigure[\, $h=-2$]
{\includegraphics[width=4.cm]{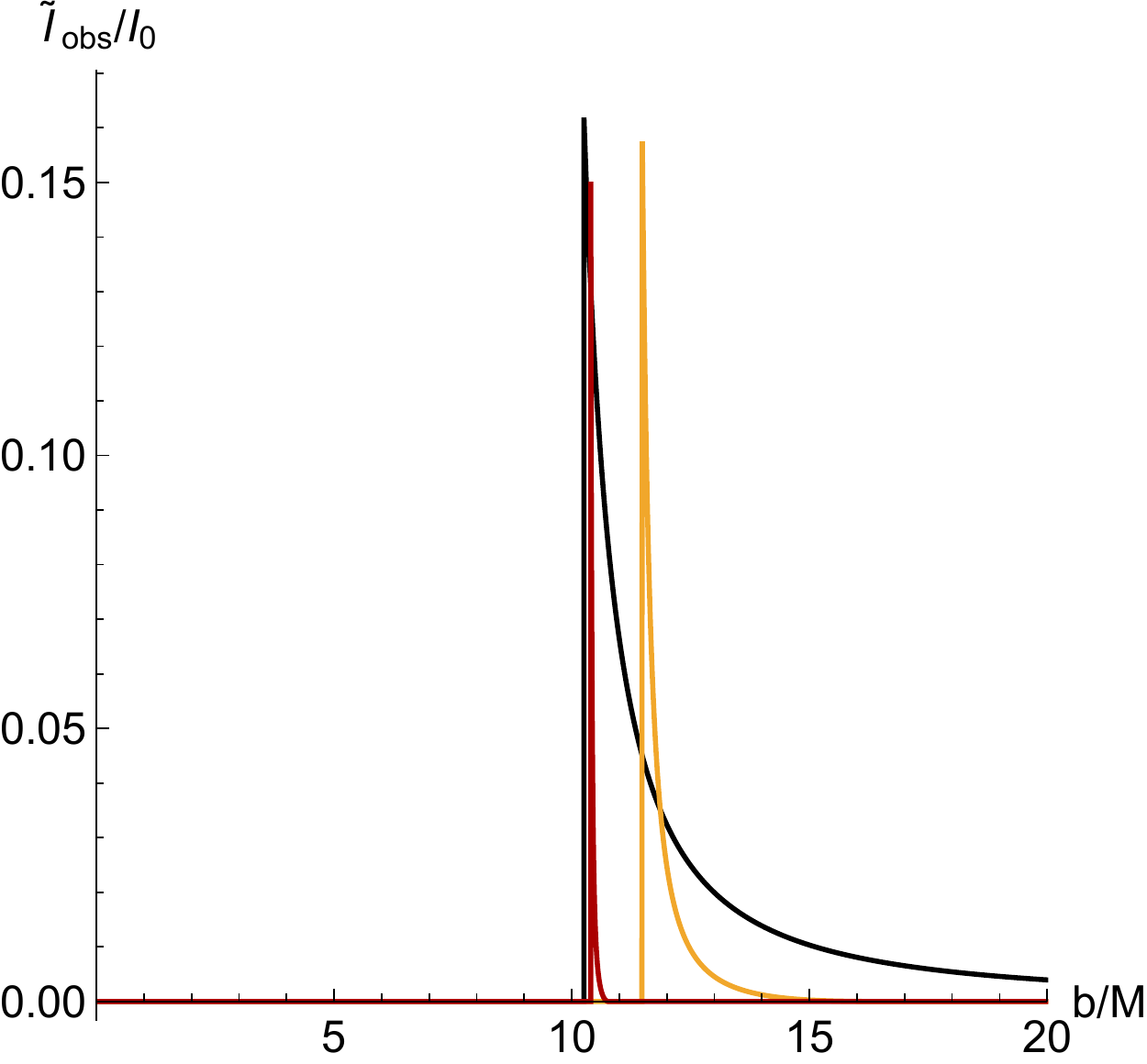} \label{}\hspace{2mm} \includegraphics[width=4.cm]{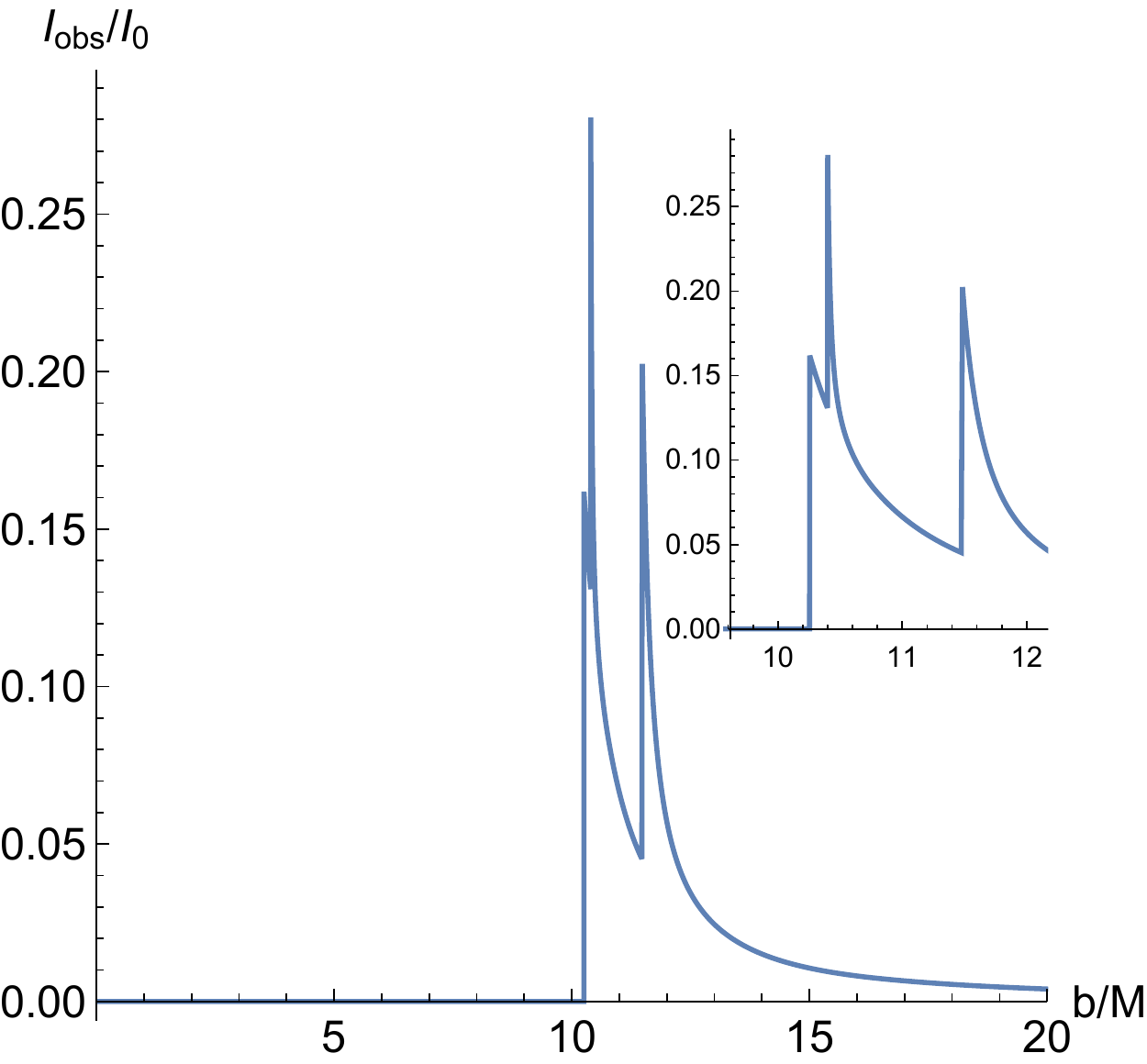}\hspace{2mm} \includegraphics[width=5cm]{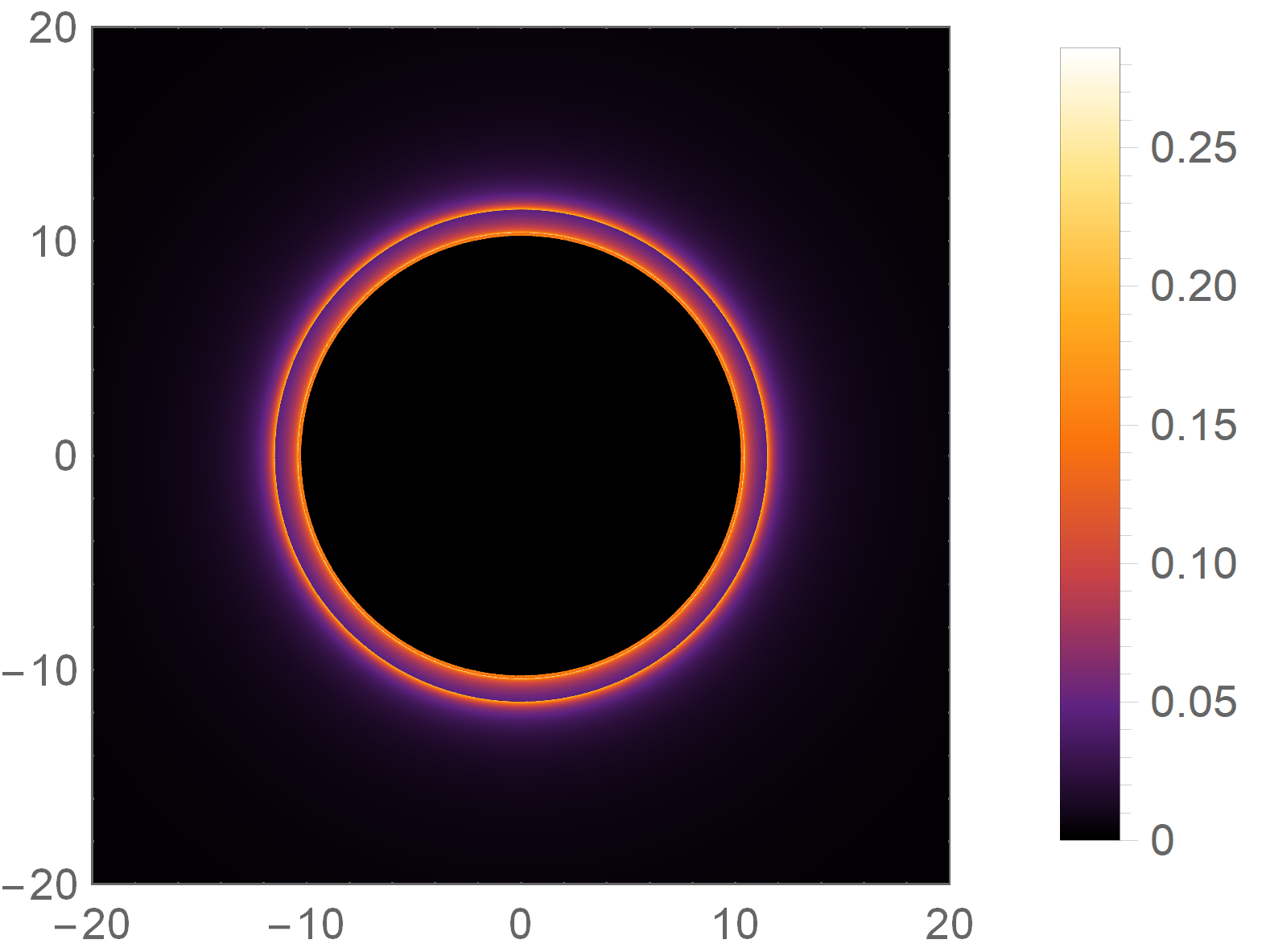}
\includegraphics[width=3.7cm]{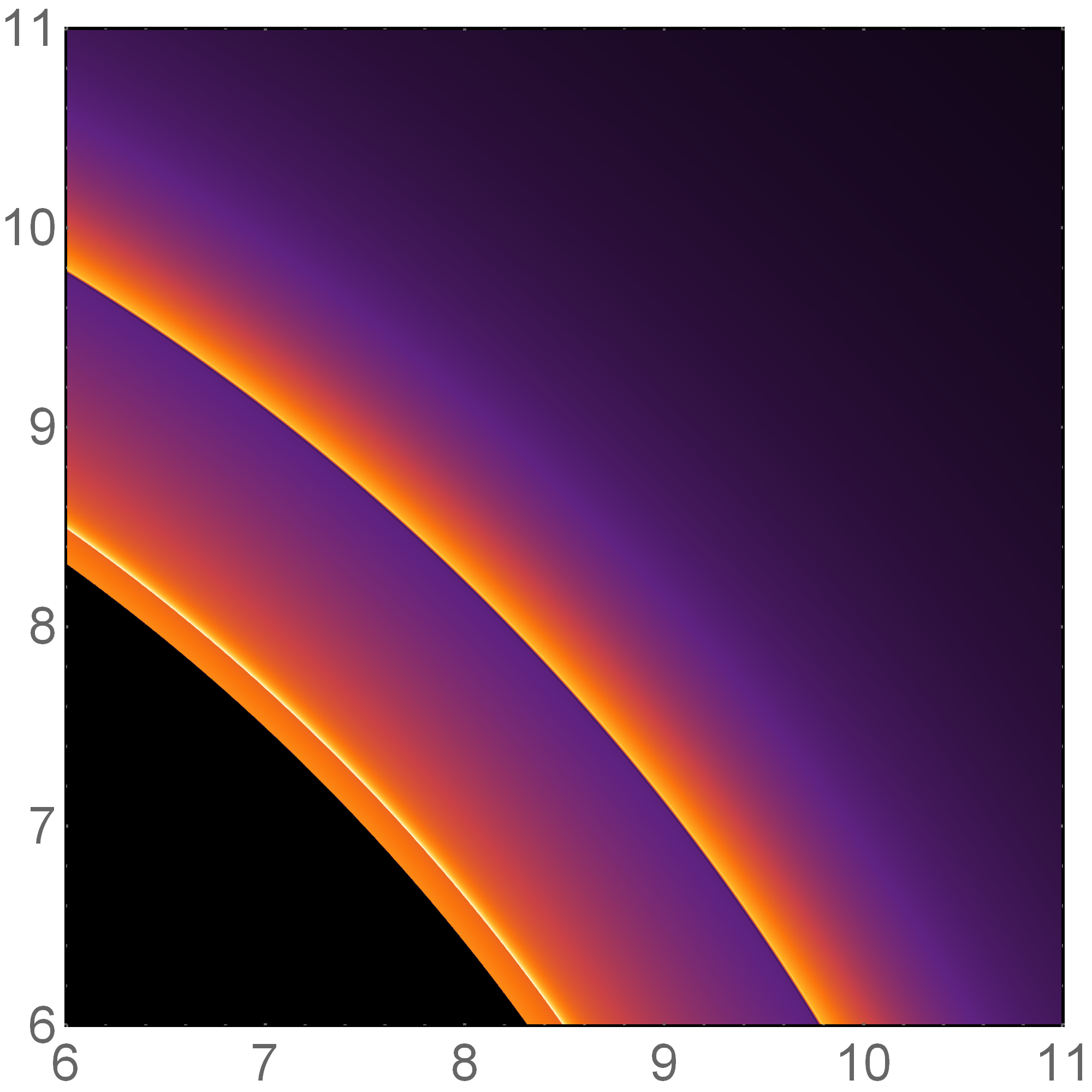}}\\
\caption{Observational appearances of the profile 1 \eqref{diskprofile1} of a thin disk for different $h$ with $M=1$. \textbf{First column}: the different observed intensities originated from the first (black), second (gold) and third (red) transfer function in Eq.\eqref{eqtransfer} respectively. \textbf{Second column}: the total observed intensities $I_{obs}/I_0$ as a function of impact parameter $b$. \textbf{Third column}: optical appearance: the distribution of observed intensities into two-dimensional disks. \textbf{Fourth column}: the zoomed in sectors.}
\label{figprofile1}
\end{figure}
\begin{figure}[htbp]
\centering
\subfigure[\, $h=0$]
{\includegraphics[width=4.cm]{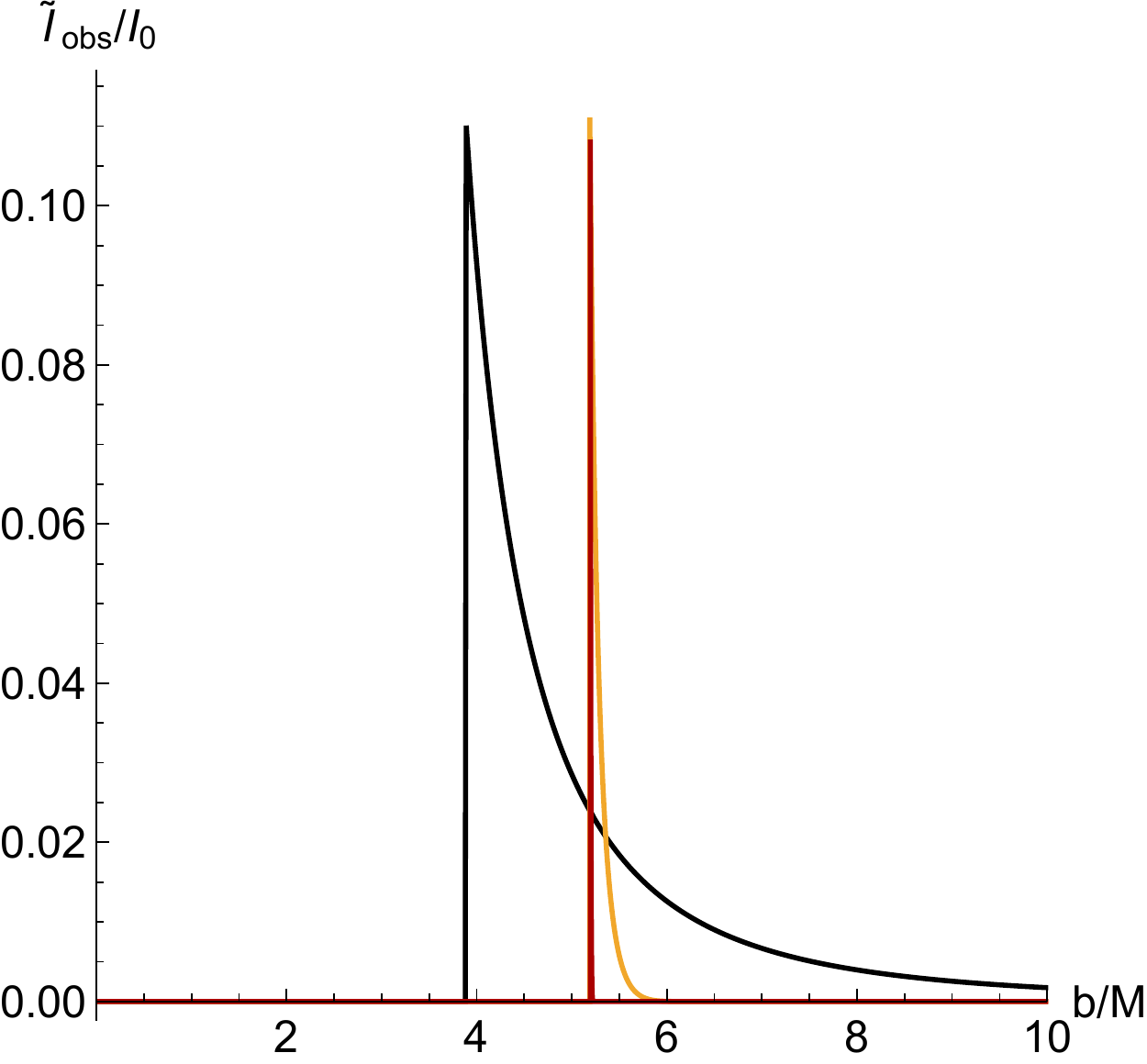} \label{}\hspace{2mm} \includegraphics[width=4.cm]{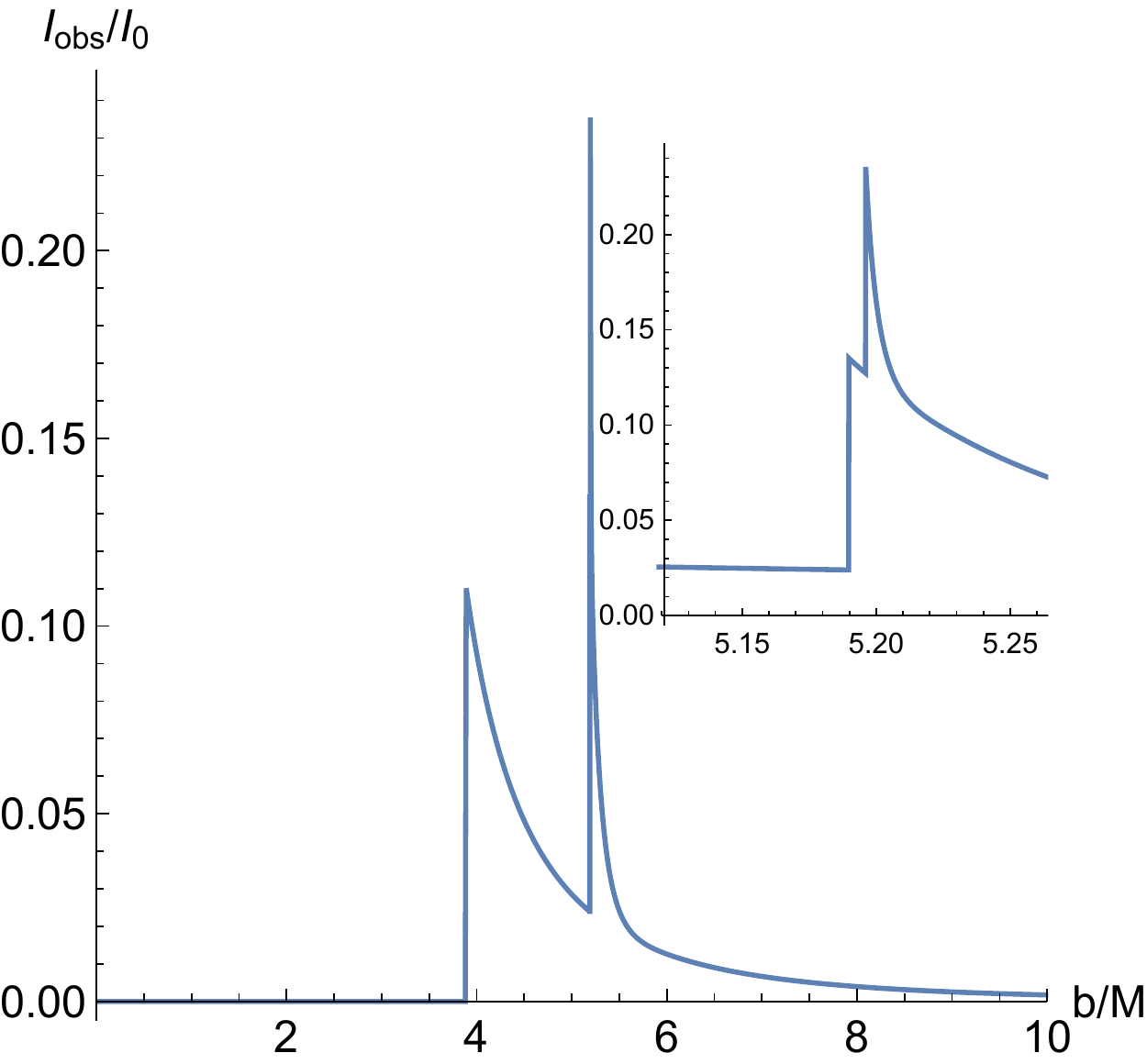}\hspace{2mm} \includegraphics[width=5cm]{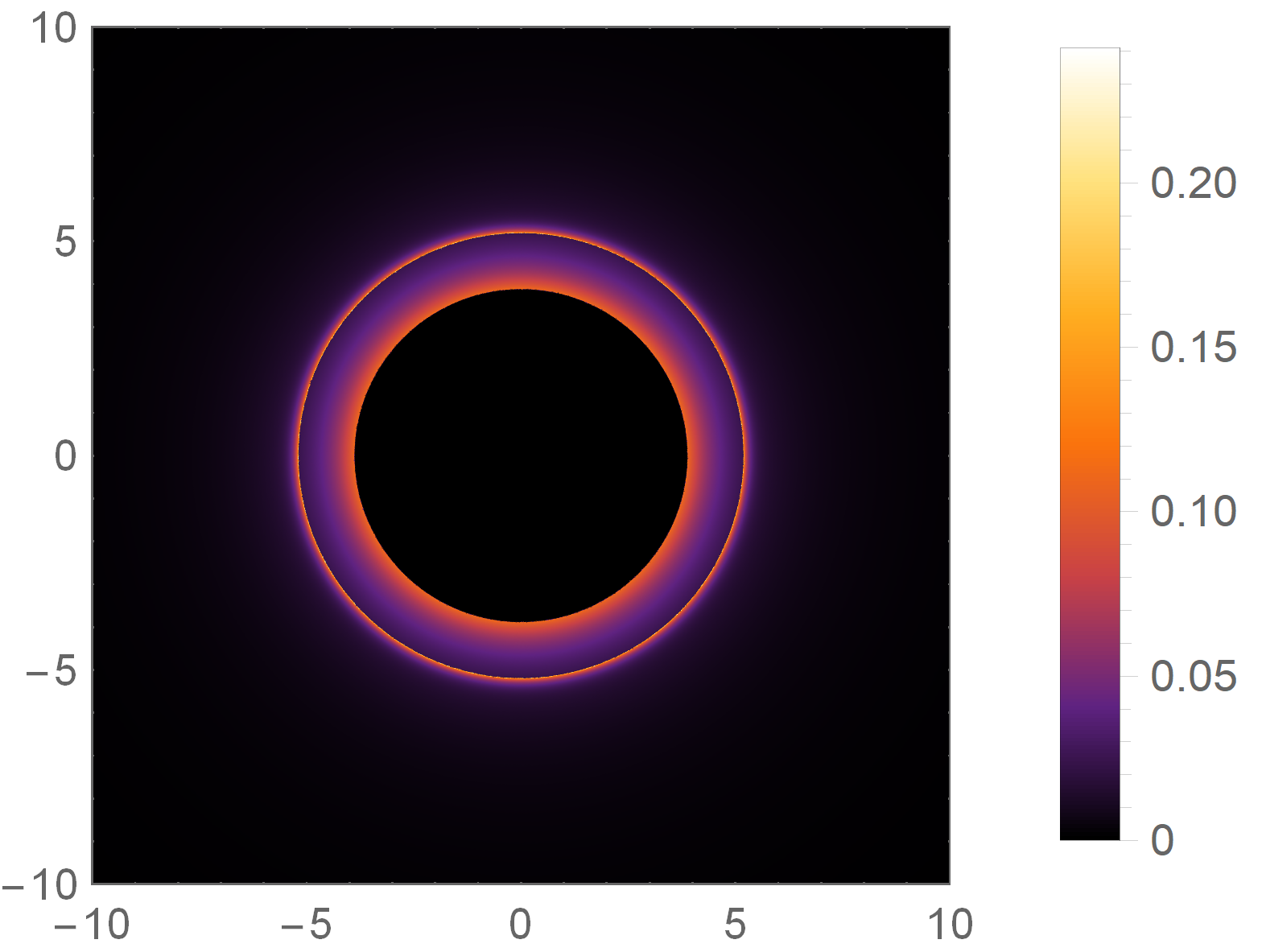}
\includegraphics[width=3.8cm]{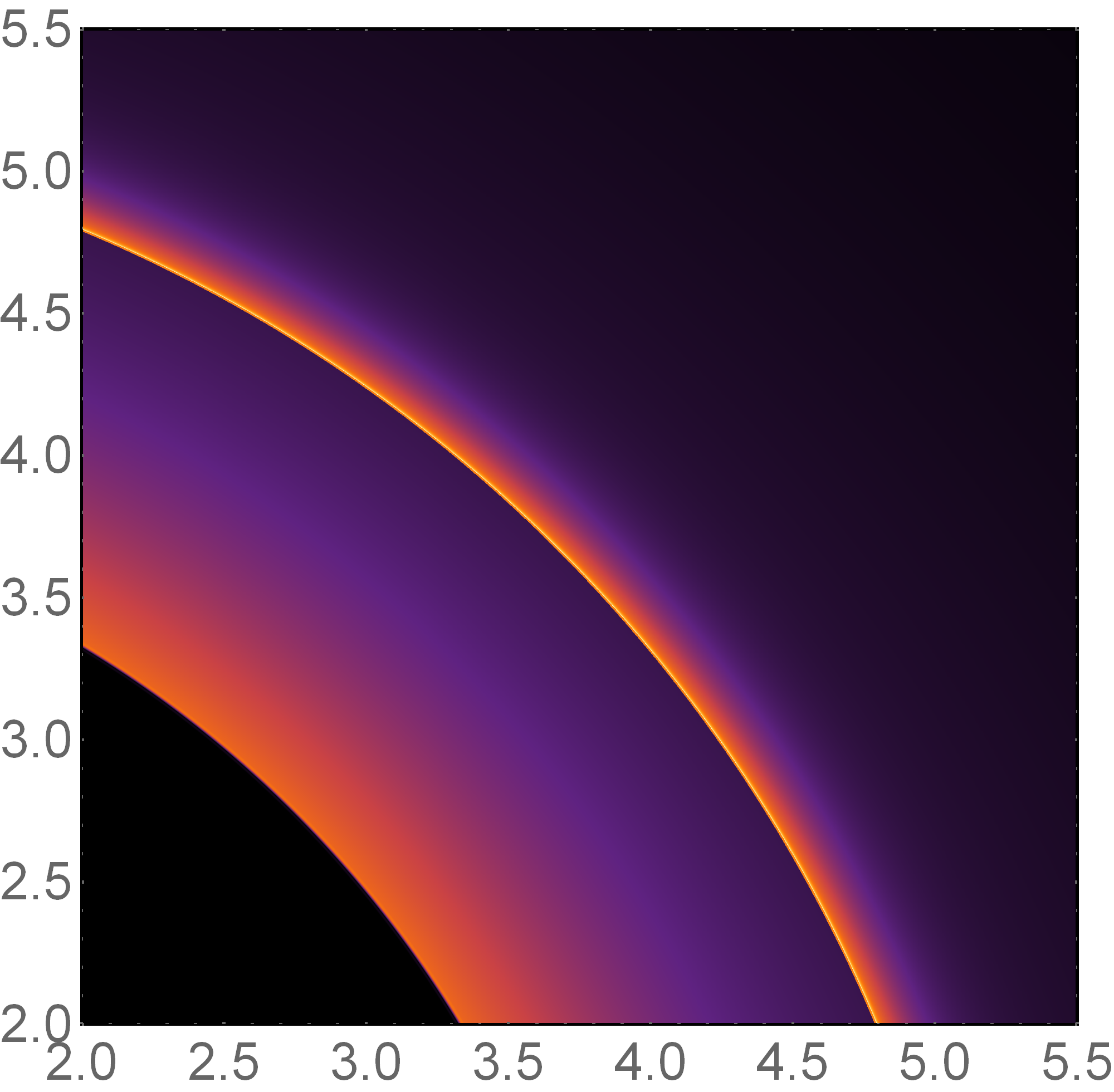}}\\
\subfigure[\, $h=-1$]
{\includegraphics[width=4.cm]{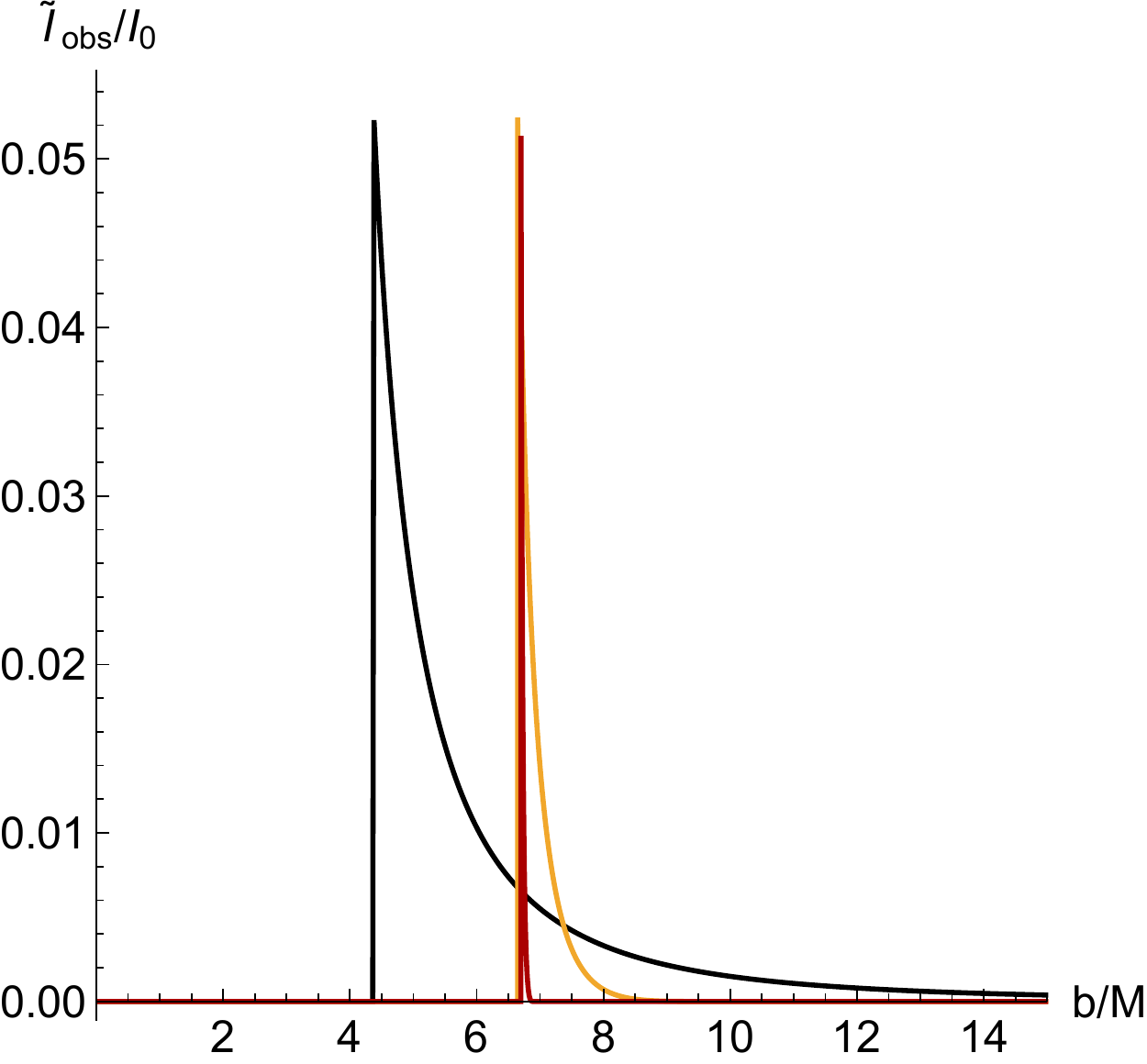} \label{}\hspace{2mm} \includegraphics[width=4.cm]{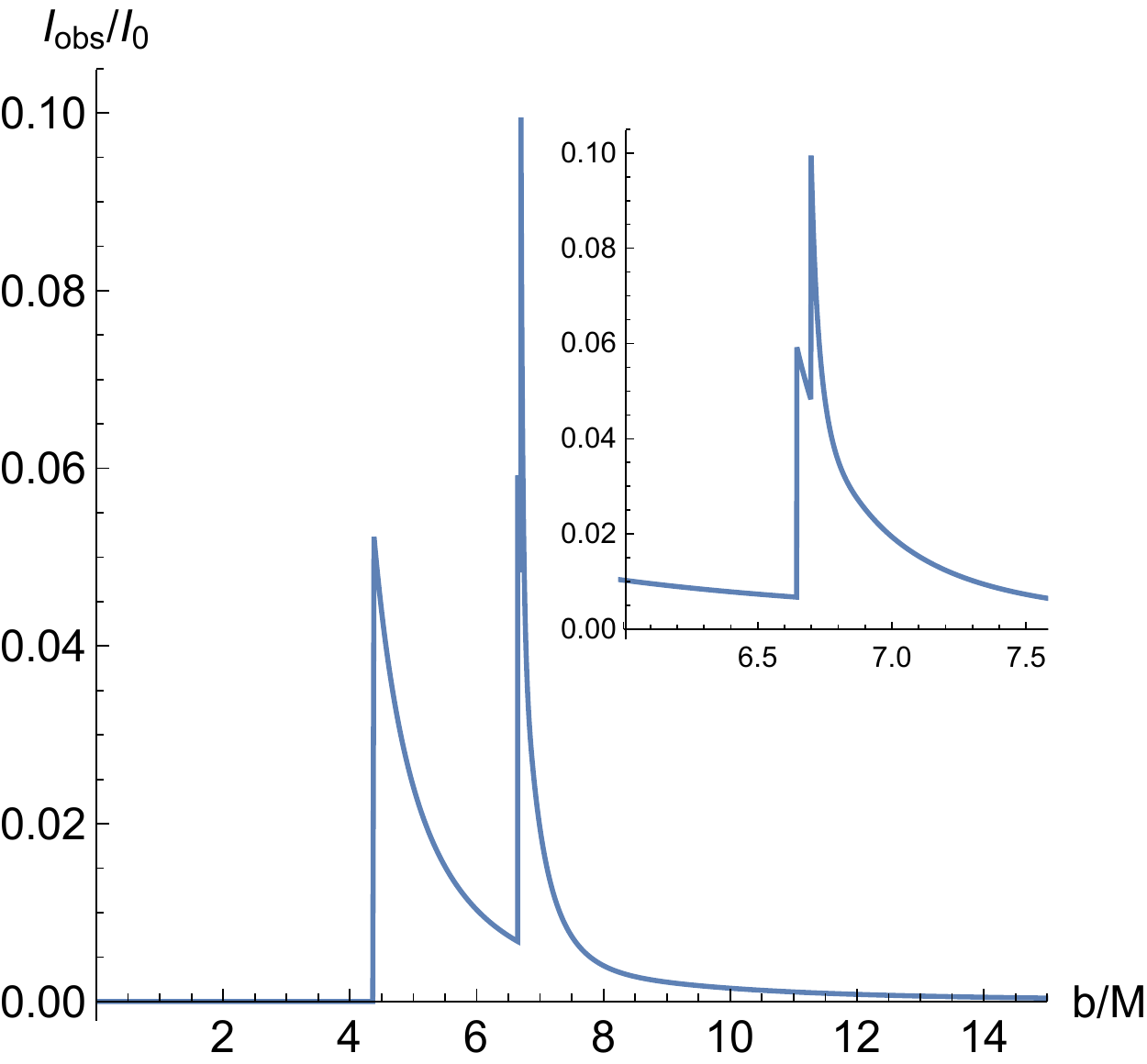}\hspace{2mm} \includegraphics[width=5cm]{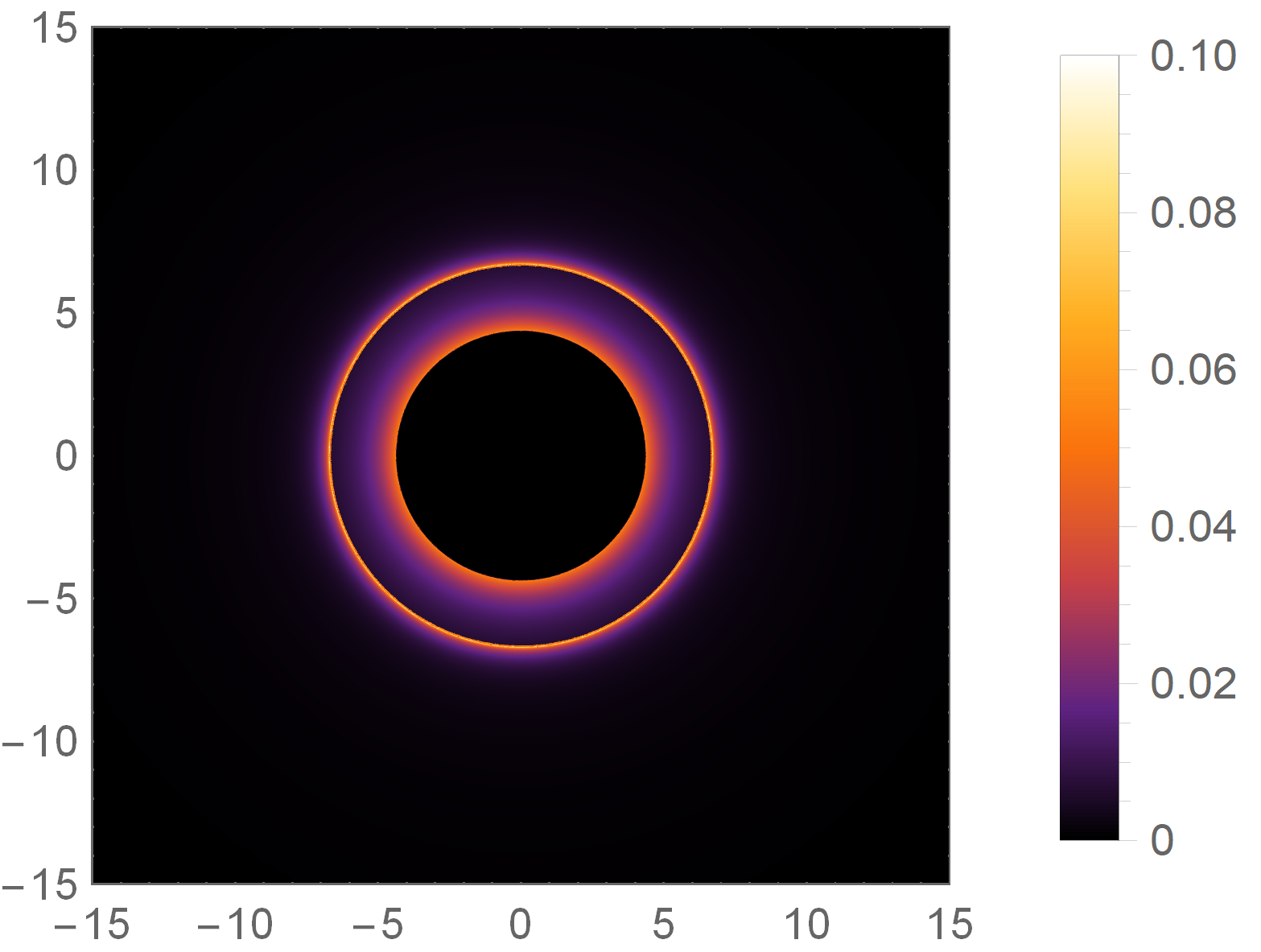}
\includegraphics[width=3.6cm]{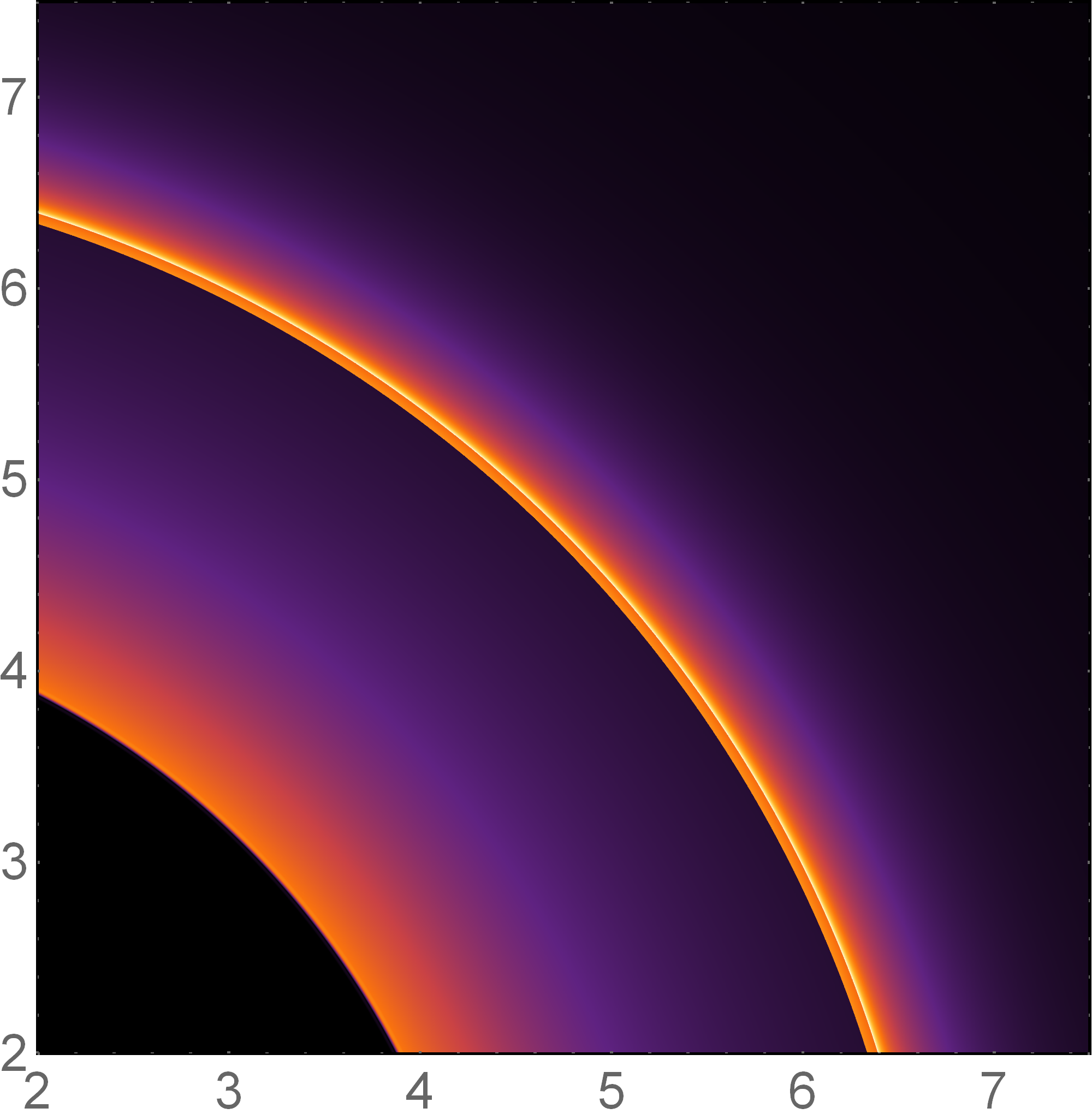}}\\
\subfigure[\, $h=-2$]
{\includegraphics[width=4.cm]{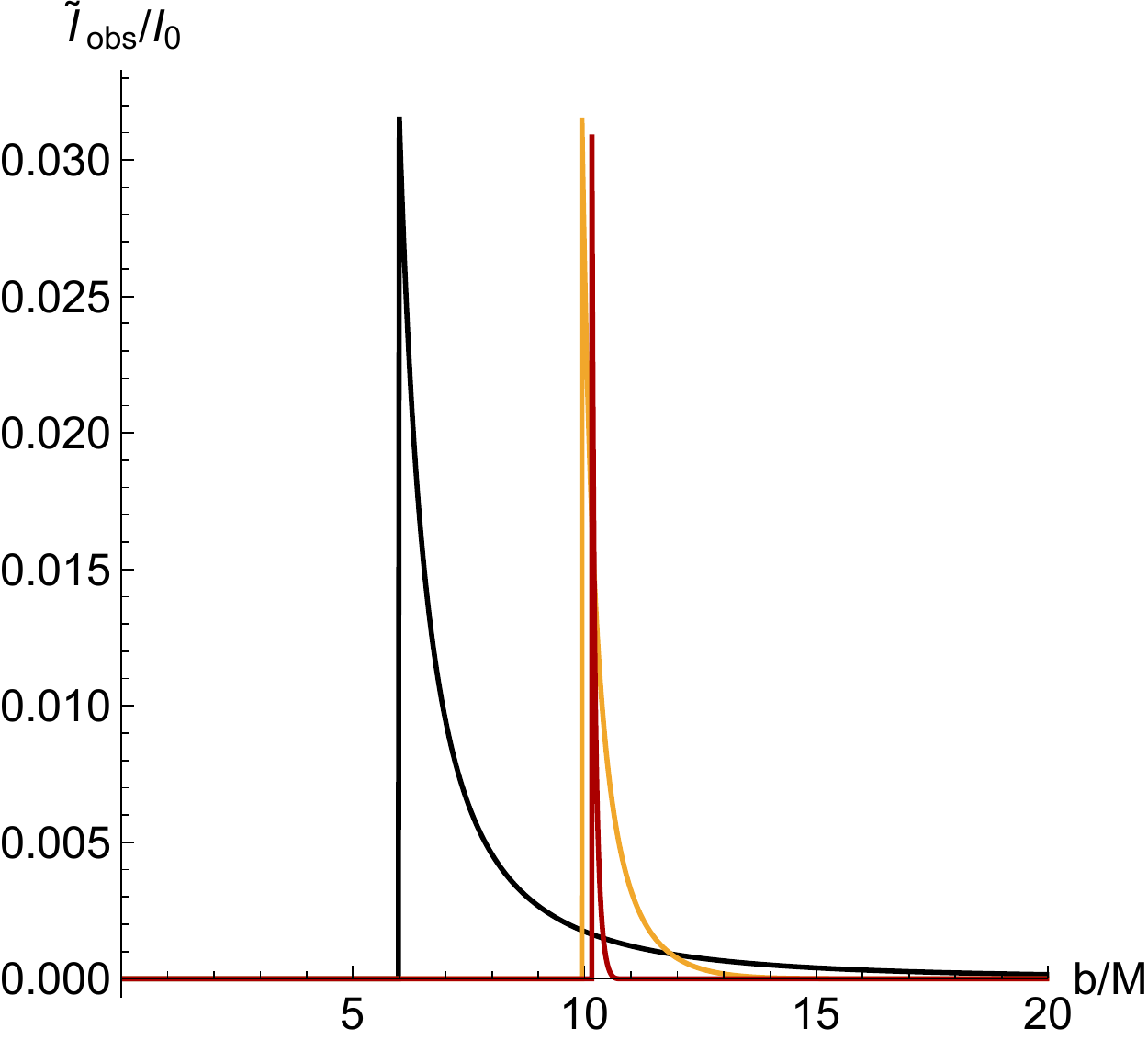} \label{}\hspace{2mm} \includegraphics[width=4.cm]{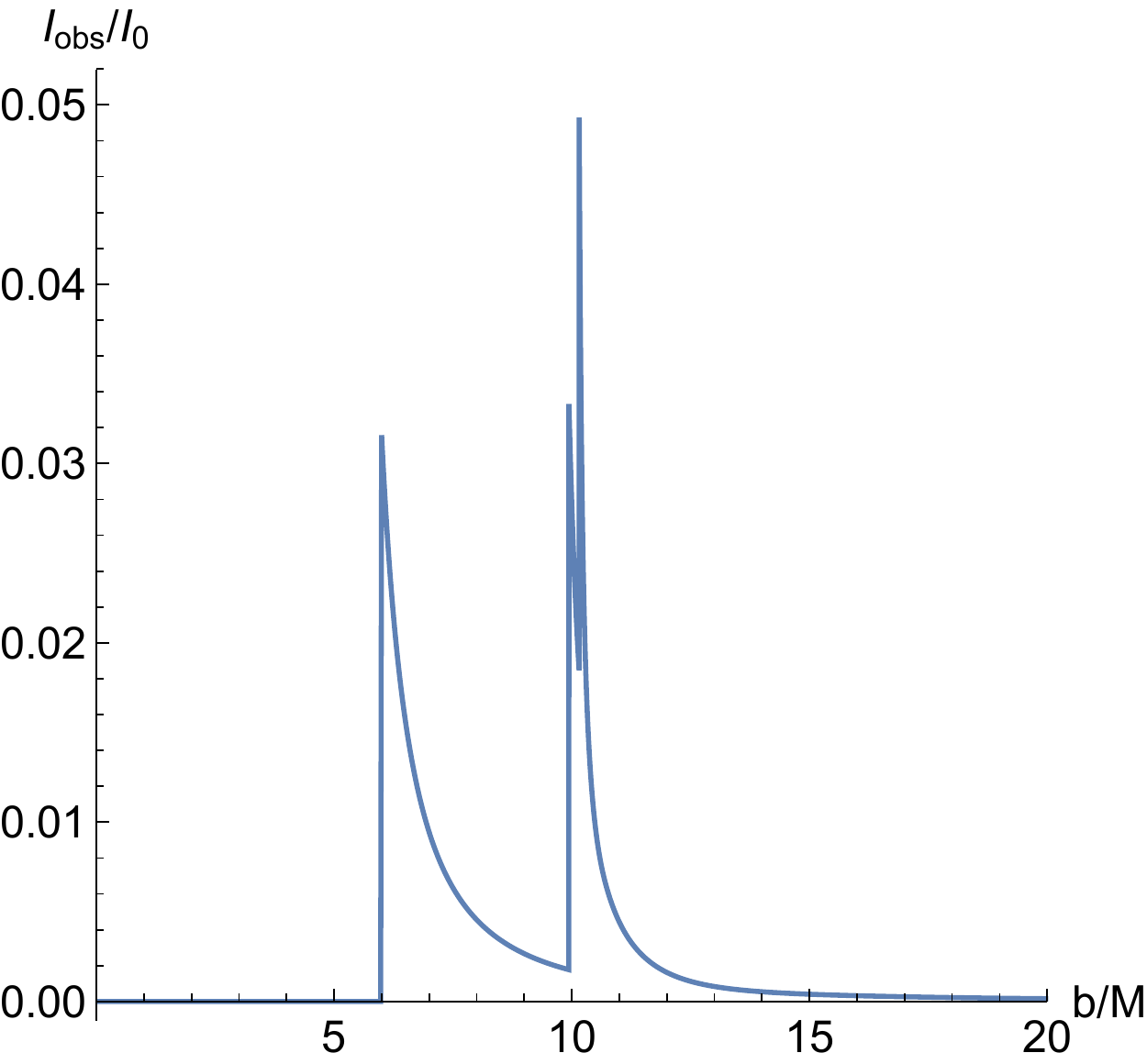}\hspace{2mm} \includegraphics[width=5cm]{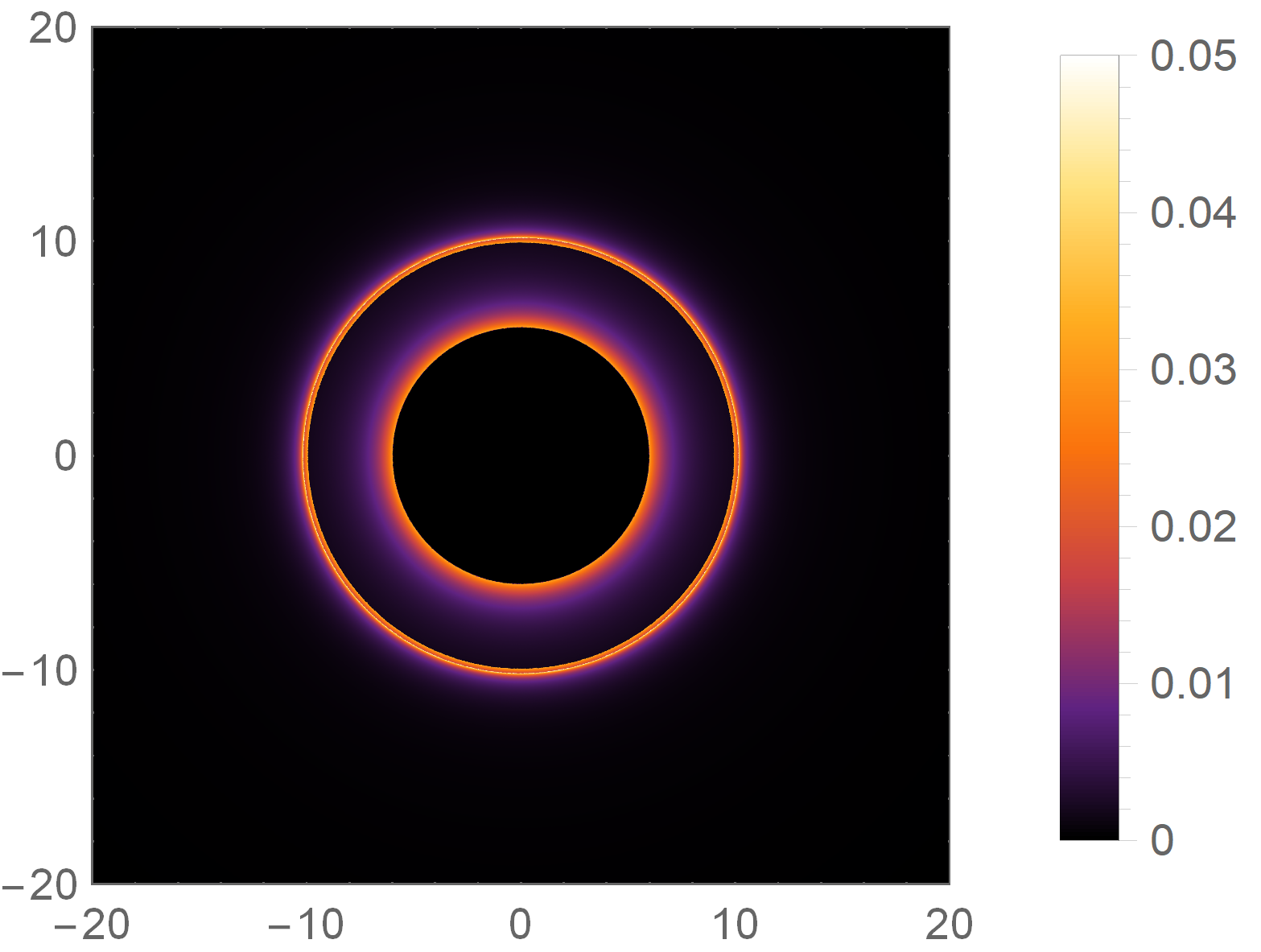}
\includegraphics[width=3.7cm]{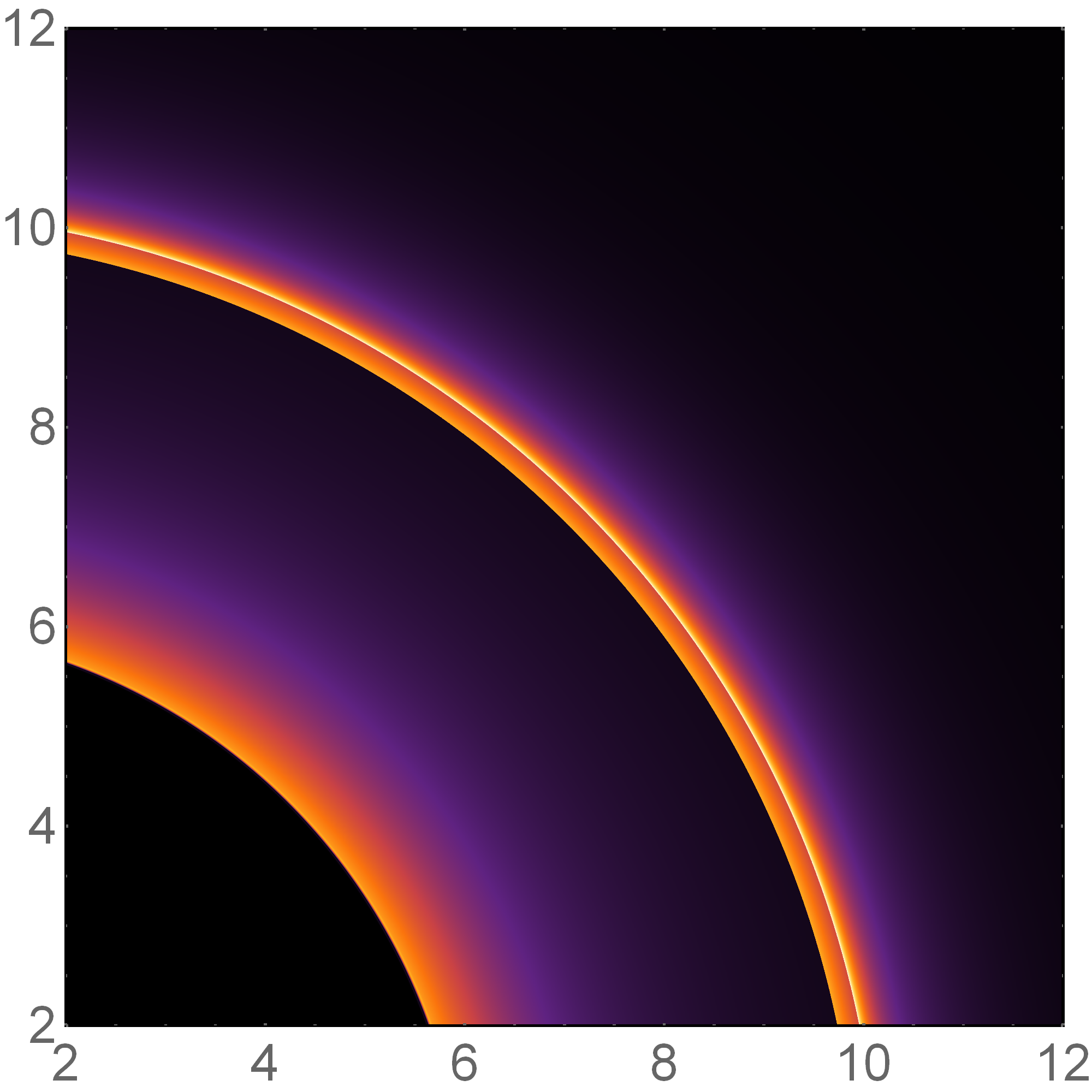}}\\
\caption{Observational appearances of the profile 2 \eqref{diskprofile2} of a thin disk for different $h$ with $M=1$.  \textbf{First column}: the different observed intensities originated from the first (black), second (gold) and third (red) transfer function in Eq.\eqref{eqtransfer} respectively. \textbf{Second column}: the total observed intensities $I_{obs}/I_0$ as a function of impact parameter $b$. \textbf{Third column}: optical appearance: the distribution of observed intensities into two-dimensional disks. \textbf{Fourth column}: the zoomed in sectors.}
\label{figprofile2}
\end{figure}
\begin{figure}[htbp]
\centering
\subfigure[\, $h=0$]
{\includegraphics[width=4.cm]{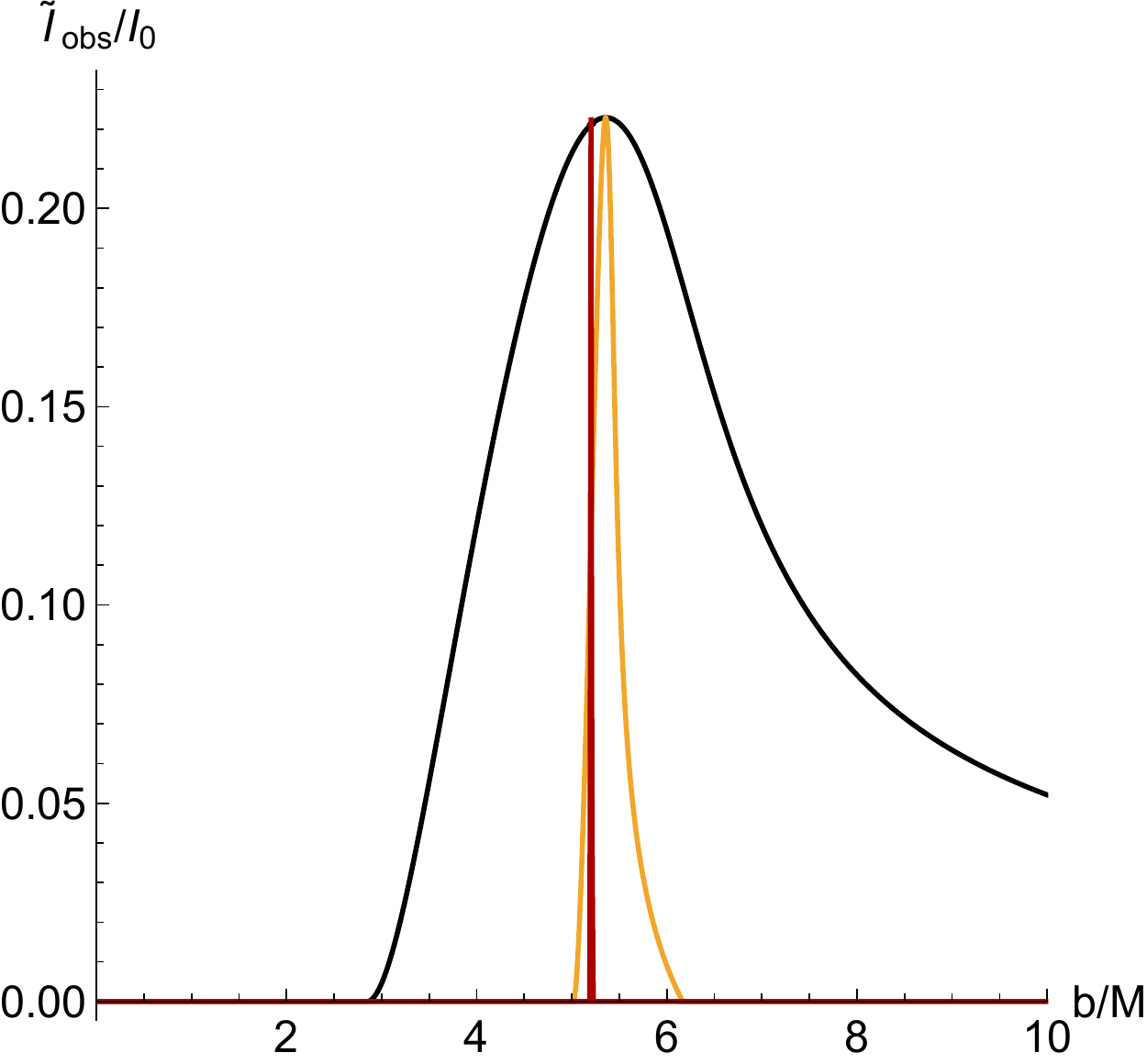} \label{}\hspace{2mm} \includegraphics[width=4.cm]{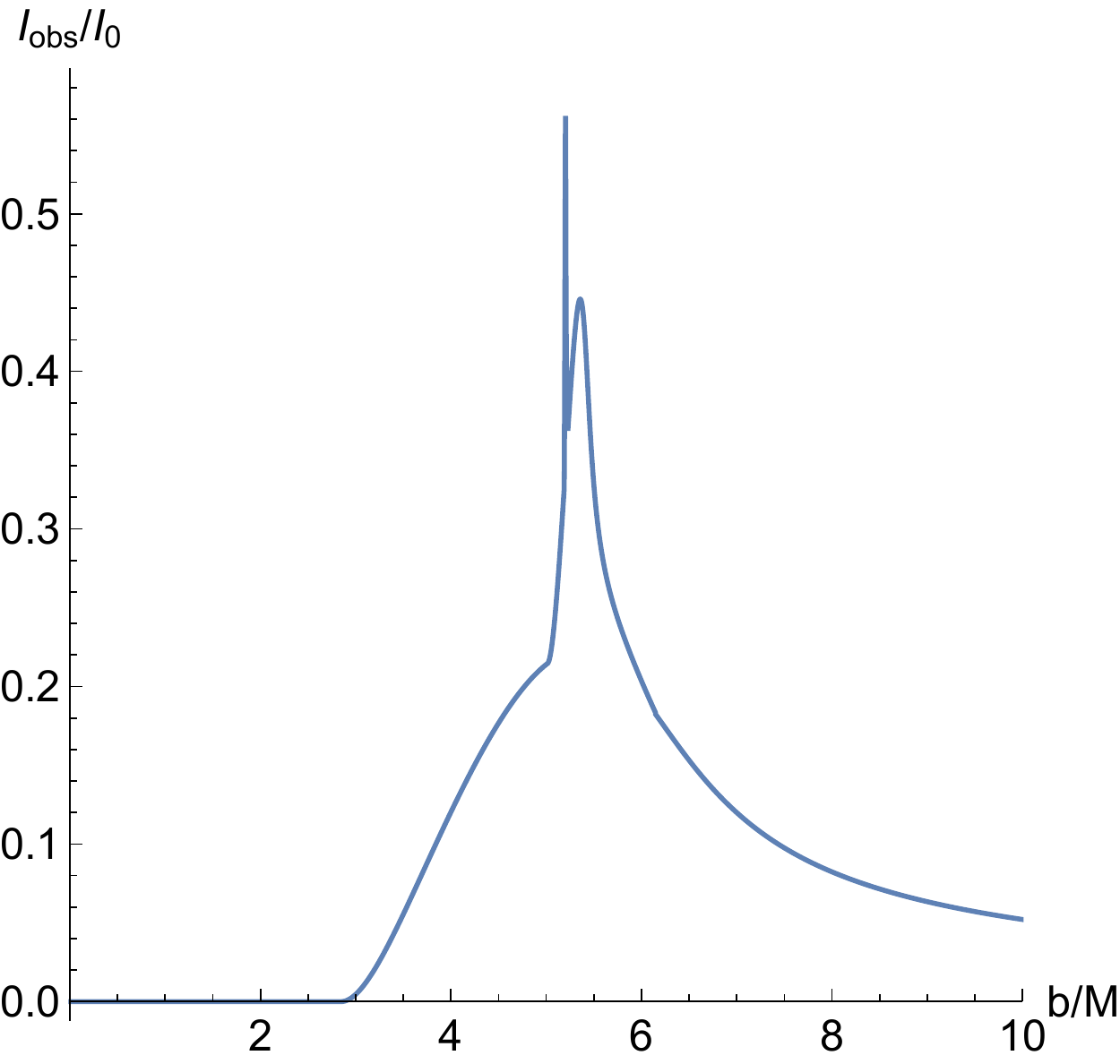}\hspace{2mm} \includegraphics[width=5cm]{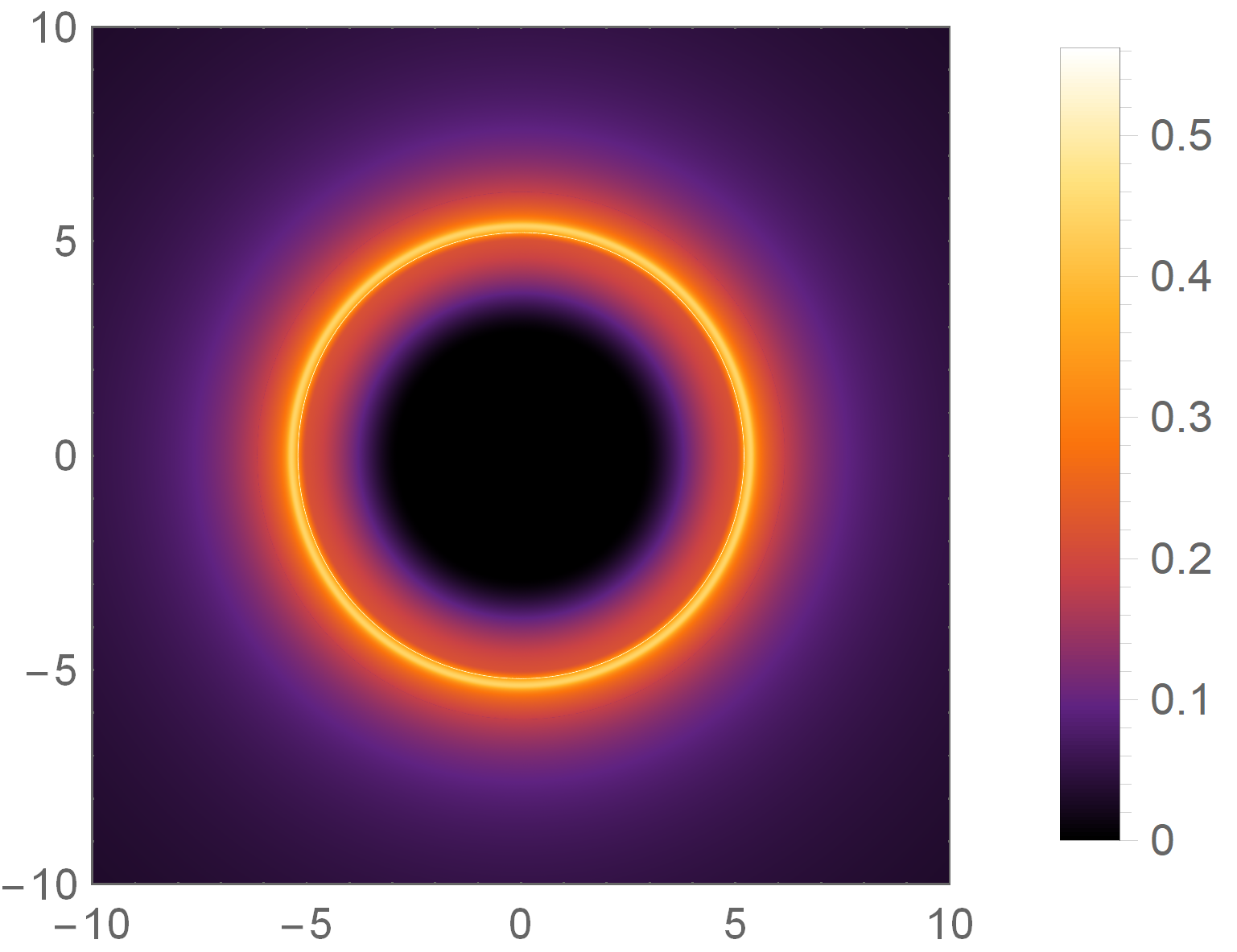}
\includegraphics[width=3.65cm]{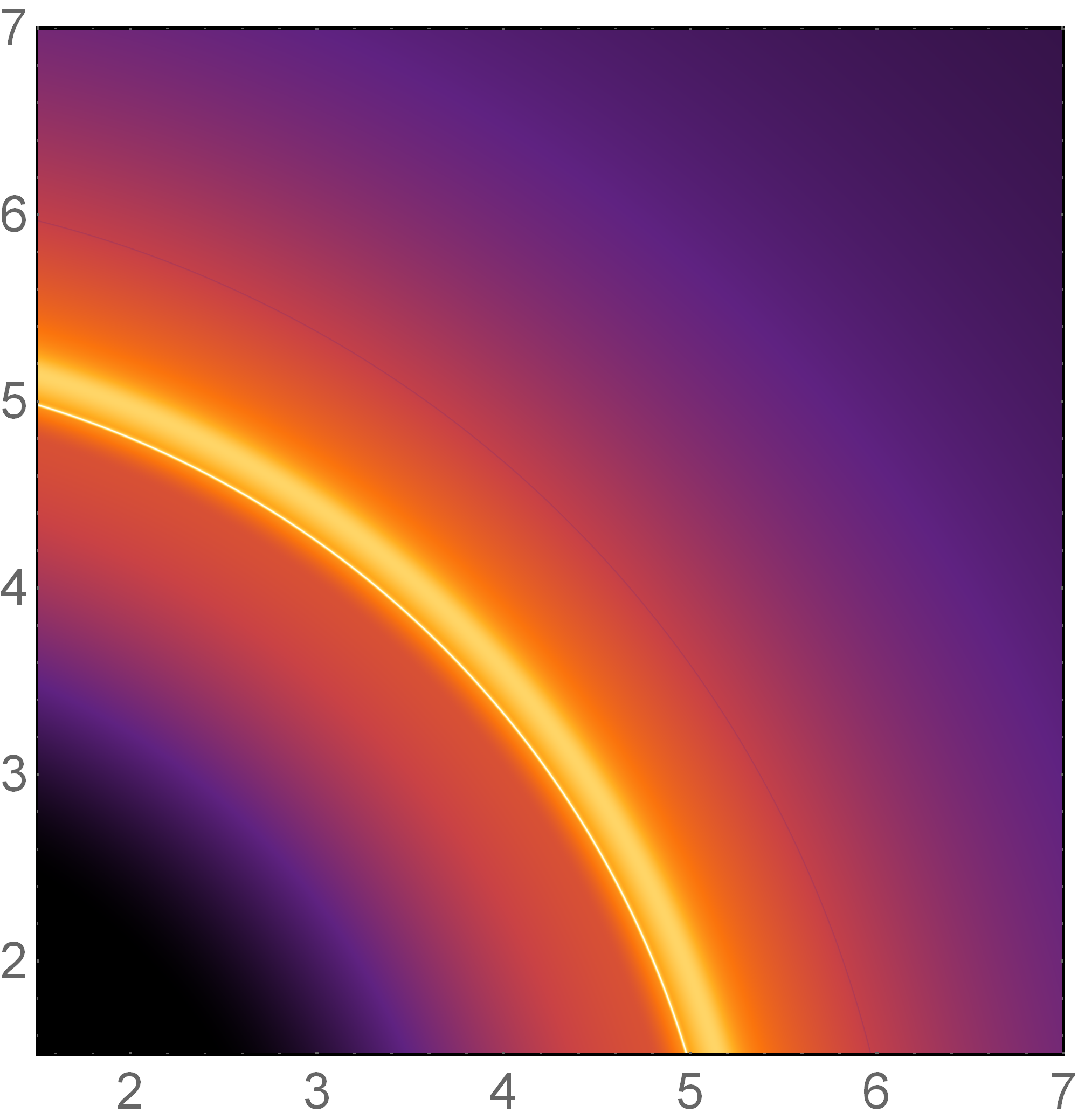}}\\
\subfigure[\, $h=-1$]
{\includegraphics[width=4.cm]{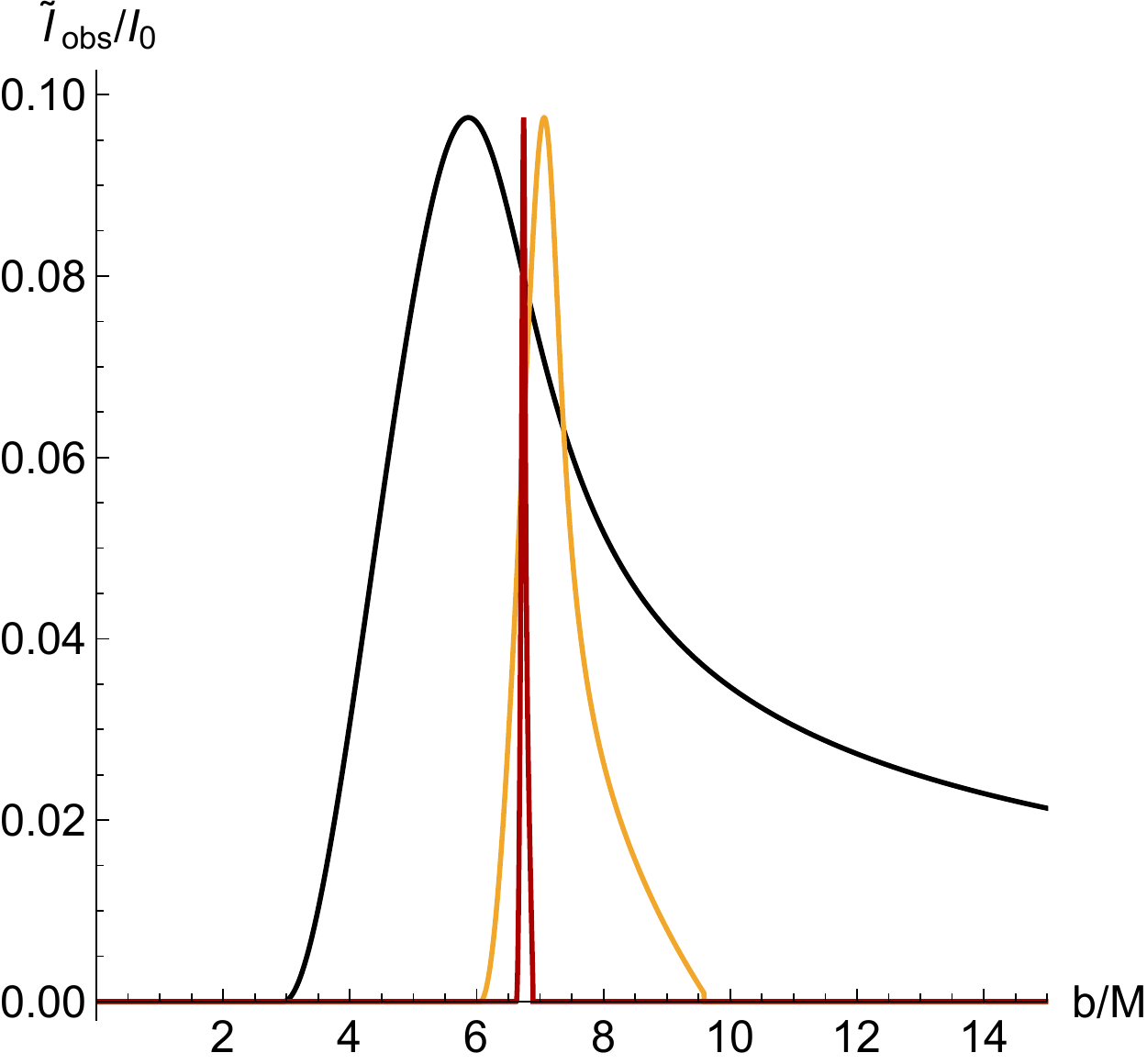} \label{}\hspace{5mm} \includegraphics[width=4.cm]{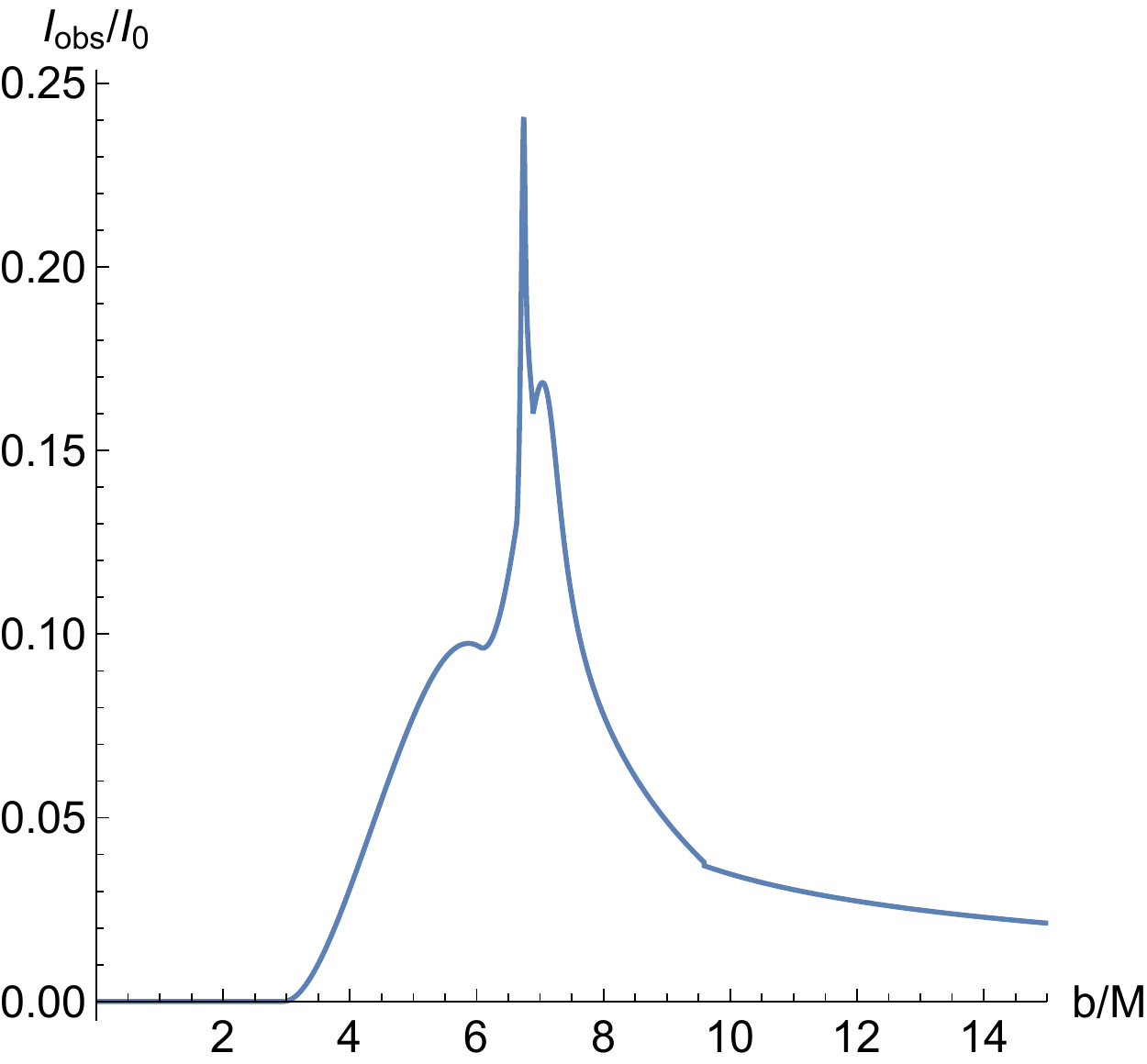}\hspace{2mm} \includegraphics[width=5cm]{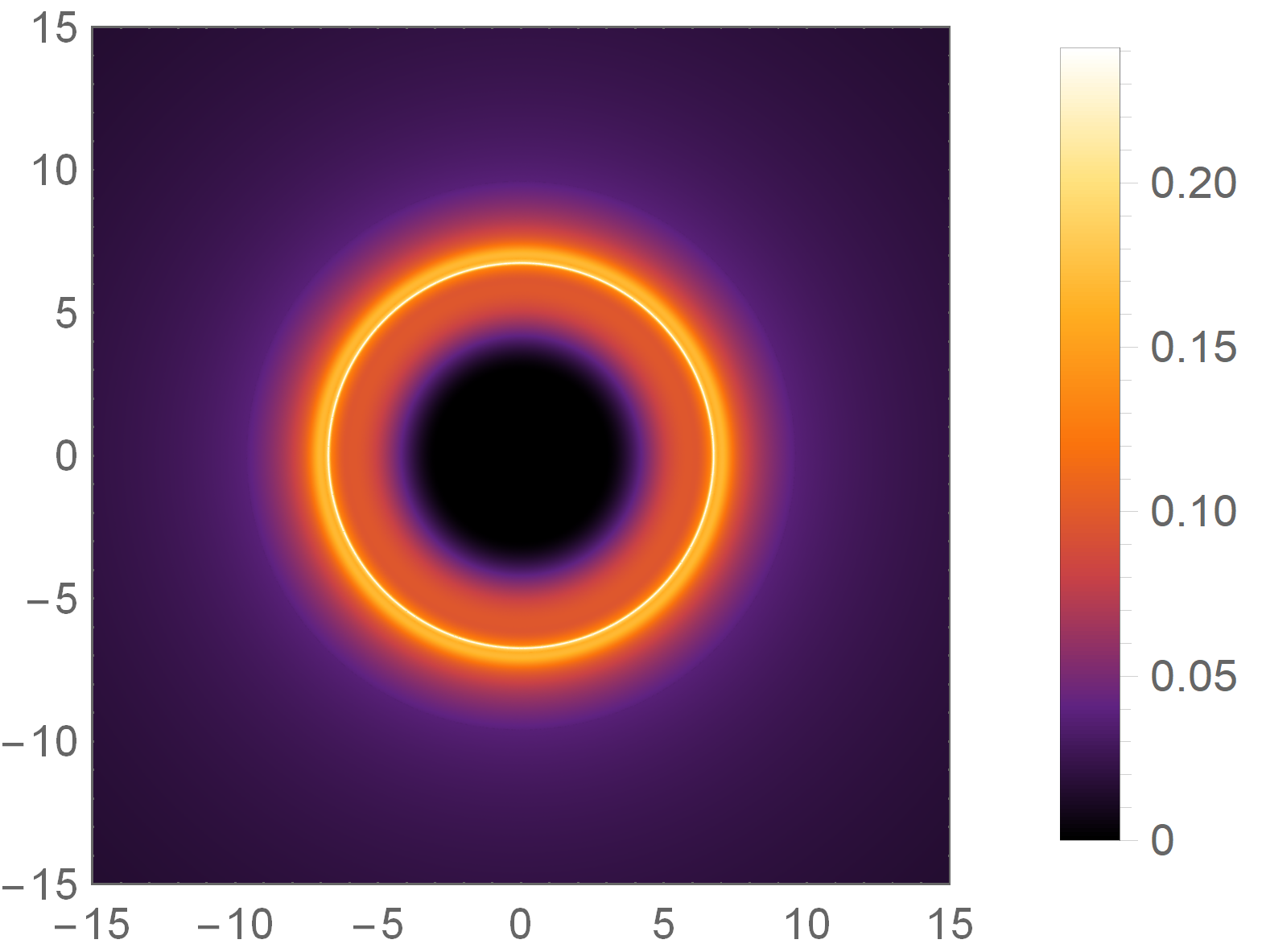}
\includegraphics[width=3.6cm]{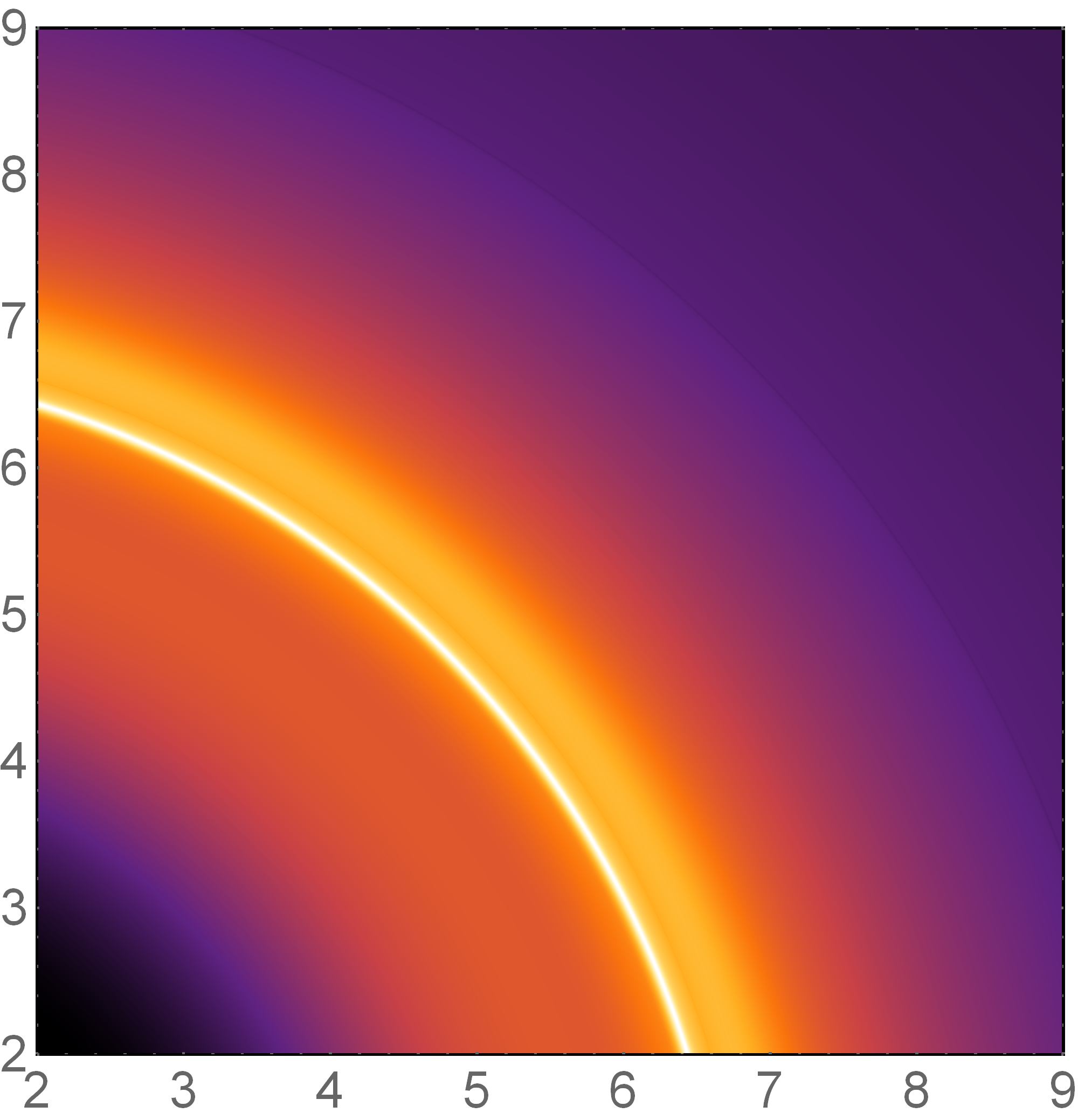}}\\
\subfigure[\, $h=-2$]
{\includegraphics[width=4.cm]{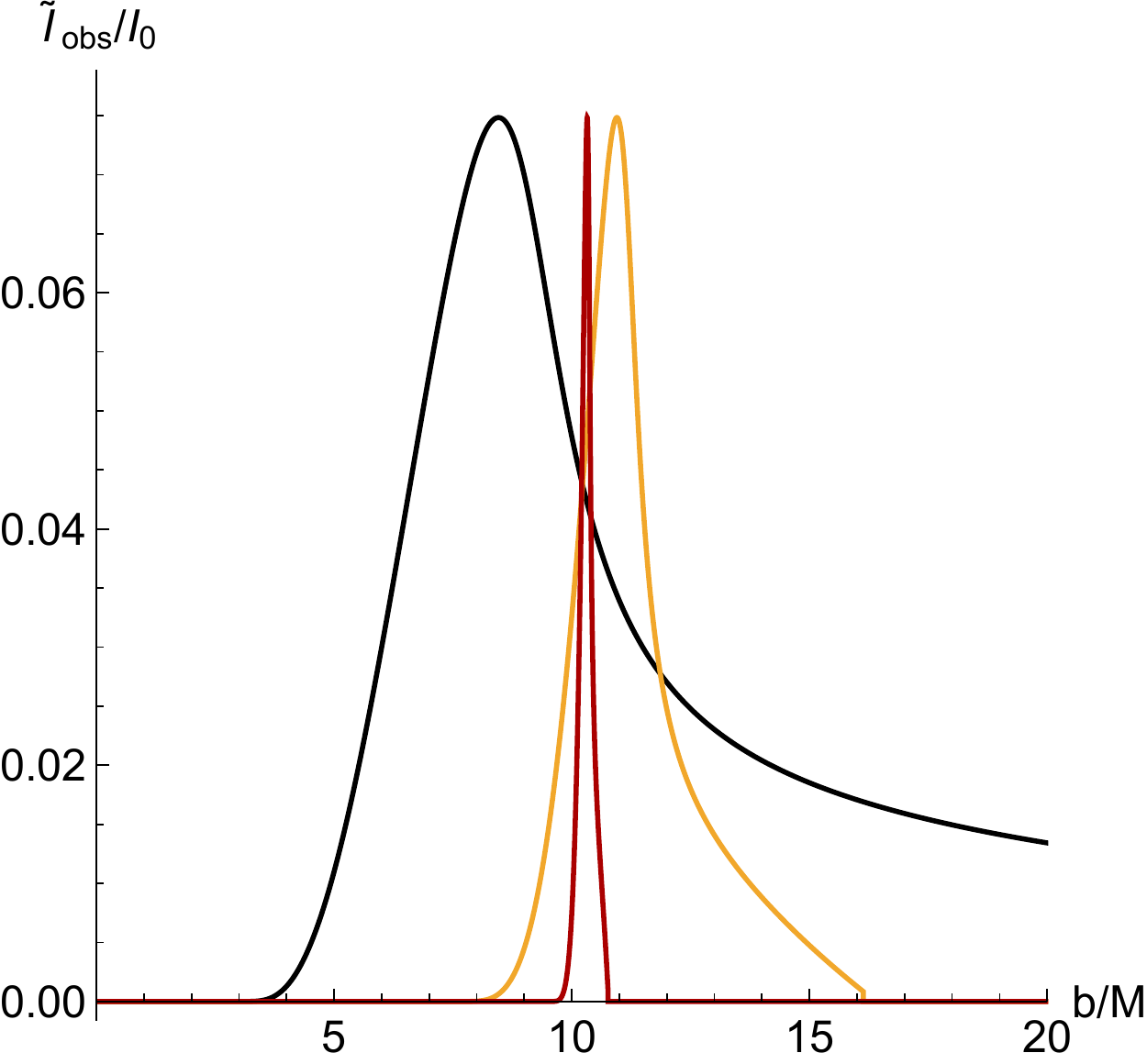} \label{}\hspace{2mm} \includegraphics[width=4.cm]{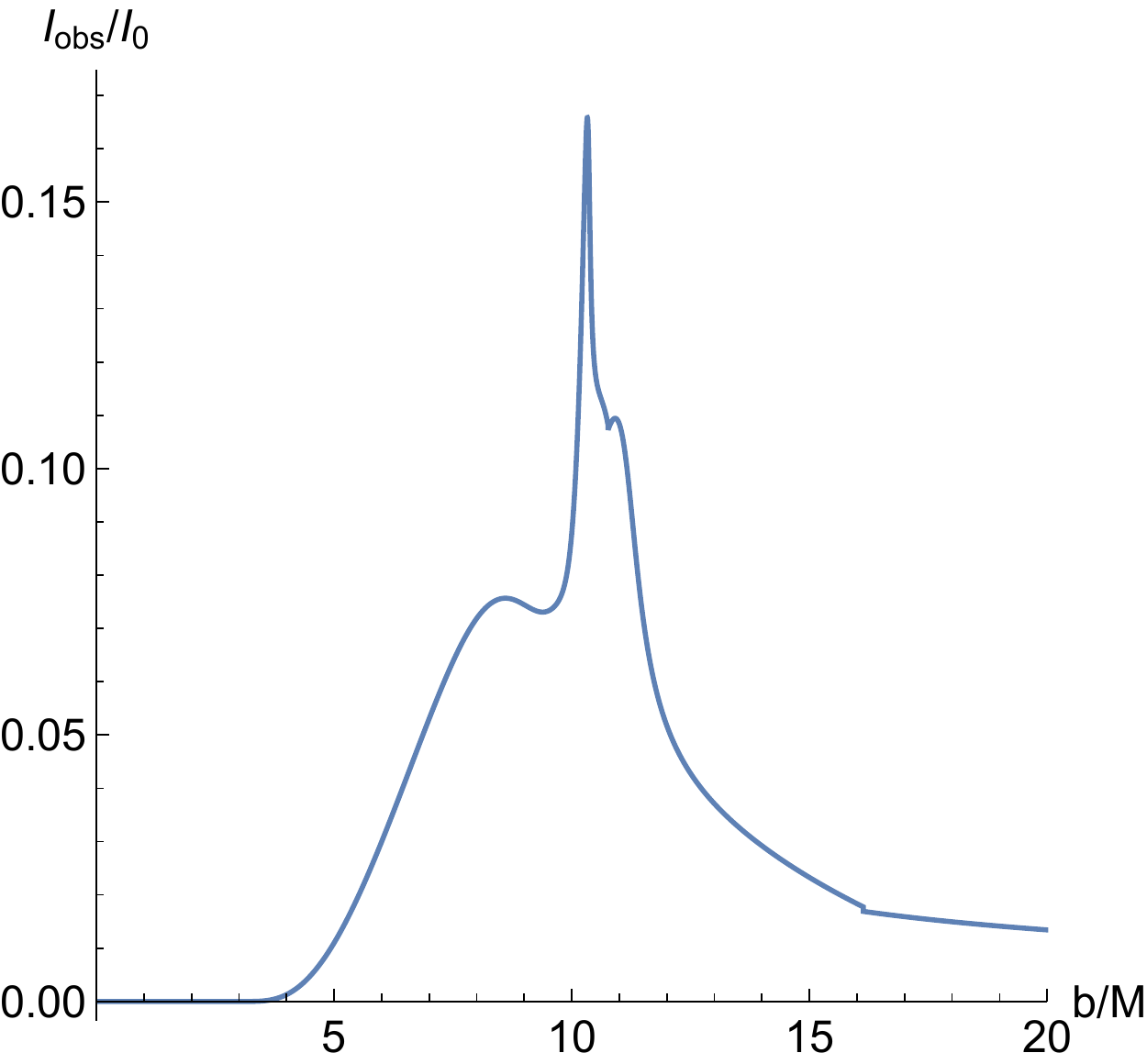}\hspace{2mm} \includegraphics[width=5cm]{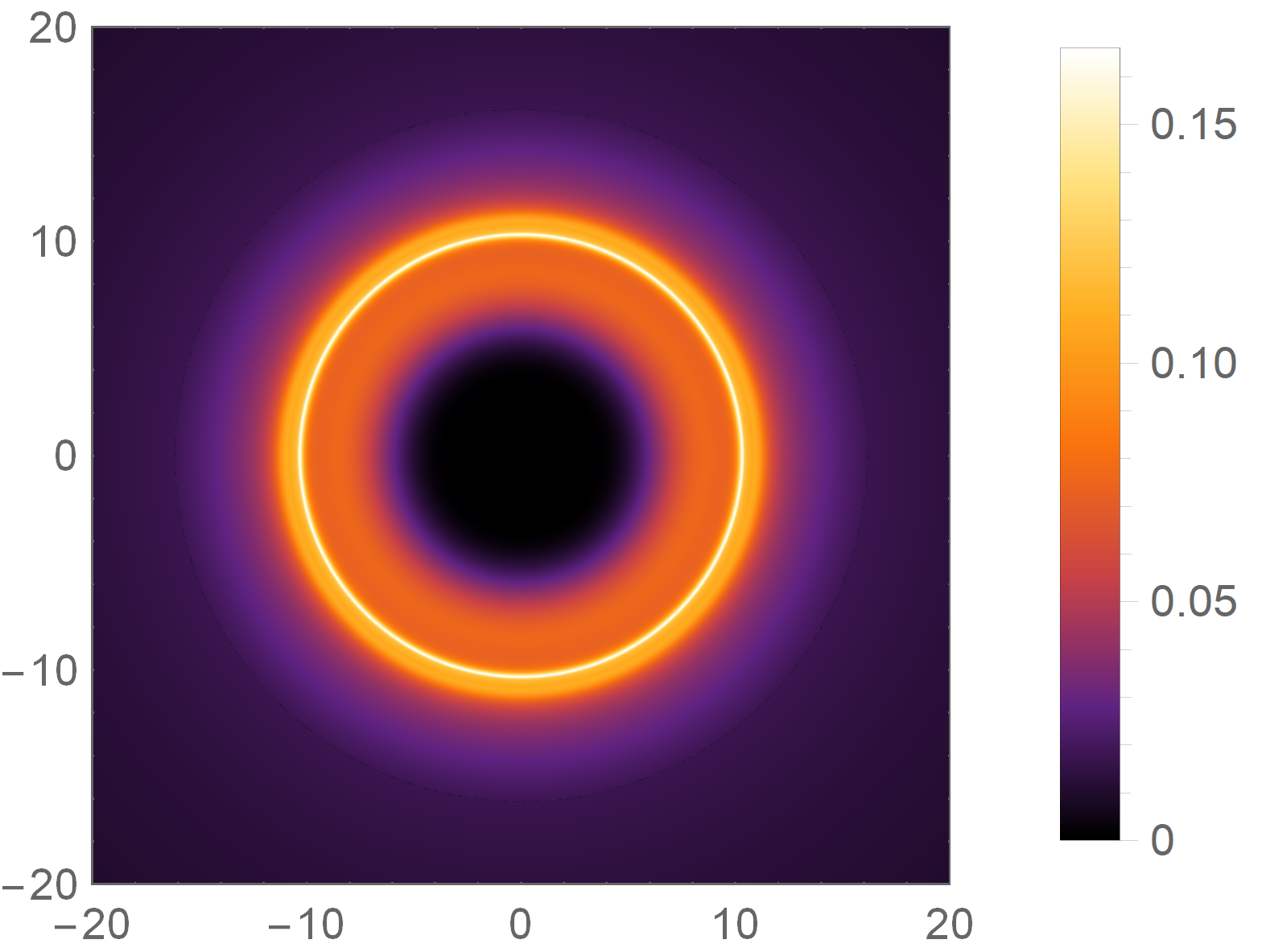}
\includegraphics[width=3.7cm]{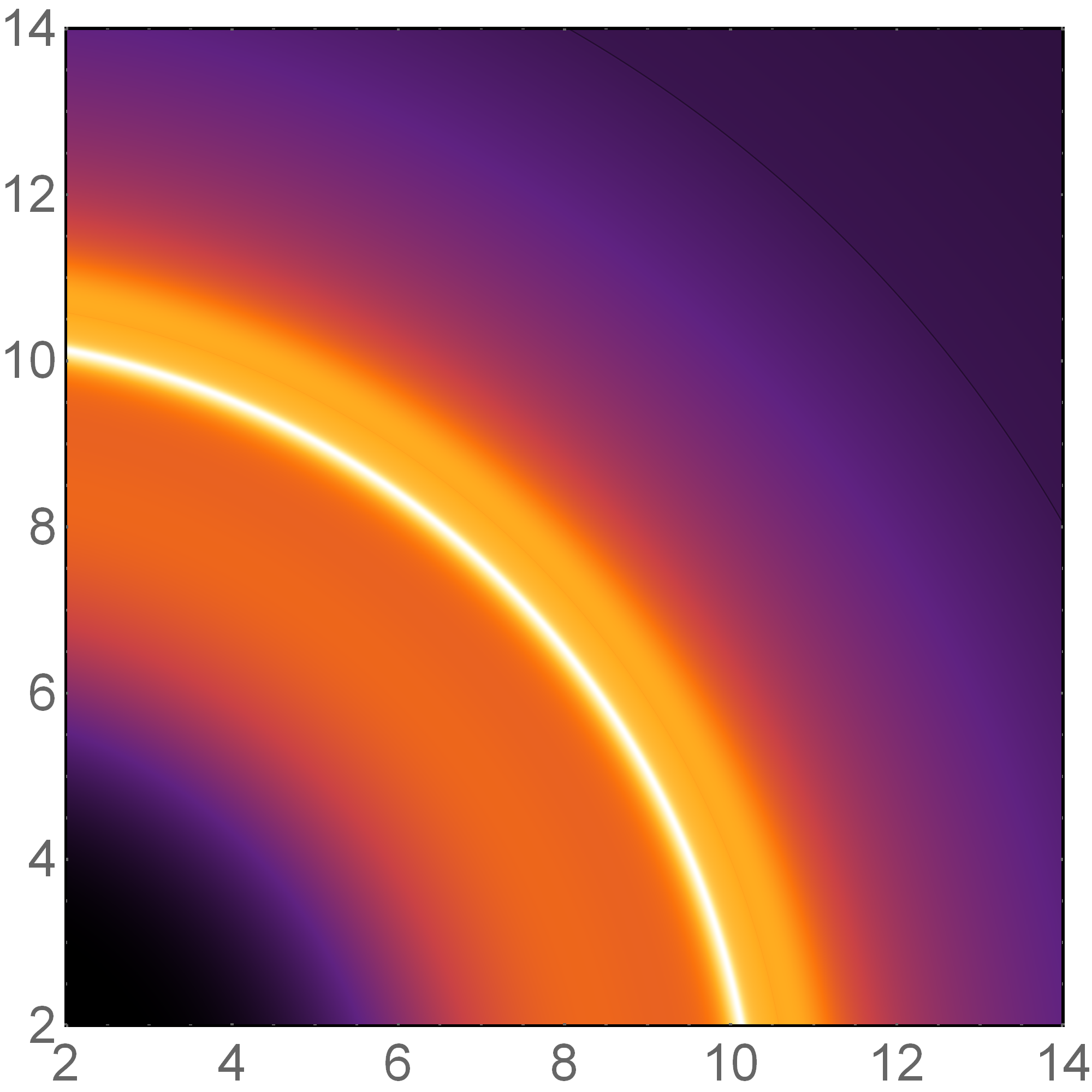}}\\
\caption{Observational appearances of the profile 3 \eqref{diskprofile3} of a thin disk for different $h$ with $M=1$.  \textbf{First column}: the different observed intensities originated from the first (black), second (gold) and third (red) transfer function in Eq.\eqref{eqtransfer} respectively. \textbf{Second column}: the total observed intensities $I_{obs}/I_0$ as a function of impact parameter $b$. \textbf{Third column}: optical appearance: the distribution of observed intensities into two-dimensional disks. \textbf{Fourth column}: the zoomed in sectors.}
\label{figprofile3}
\end{figure}

Firstly we analyze the observed intensities and black hole images caused by the first emission profile of accretion disk \eqref{diskprofile1} which are shown in Fig.\ref{figprofile1}. In leftmost figures, we find that the direct intensity moves towards the photon ring intensity relatively. However, the relative positions of lensed and photon ring intensities remain unchanged. In the movement of direct intensity, it combines with different ring intensities which construct different optical features. For example when $h=0$ (the Schwarzschild cases \cite{Gralla:2019xty}), the total observed intensities show the three peaks without almost superposition effect. But the photon ring intensity is the spike with a narrow width at $b\sim 5.21M$, which contributes a totally negligible flux to the observed feature and is only zoomed in in the rightmost plot. On the contrary, the direct and lensed ring intensities make a prominent and small contributions to the total brightness respectively that are clearly shown in the image. For $h=-1$, we find that the maximum peak of the observed intensities originated from the superposition of direct and lensed ring intensities. Besides, these two intensities also contribute an extra tiny peak at $b\sim 7.35 M$ which is very close to the maximum peak and hard to see. Similarly to the Schwarzschild case, the photon ring intensity also makes the negligible contribution to the black hole image. However, for $h=-2$, the direct and photon ring intensities make together a maximum peak with an extra peak, where the latter can be seen clearly in the zoomed in sector. For the lensed ring intensity combined with the direct intensity, it also can be seen in the image due to its larger width.

Then we discuss the optical features of the Horndeski hairy black hole with the second accretion disk profile \eqref{diskprofile2} shown in Fig.\ref{figprofile2}. One can see that unlike the results of the first accretion profile, the positions of direct intensities are always located inside the lensed and photon ring intensities. However, the lensed ring intensity moves towards the photon ring intensity. For $h=0$, the lensed ring intensity combined with photon ring intensity construct together a maximum peak with an extra peak. But the extra peak is very close to the maximum peak and difficult to be distinguished in the black hole image. For the Horndeski hairy black hole, the extra peak gradually moves away from the maximum peak and becomes visible even in the wider zoomed in sector.

Finally, we analyze the results of Horndeski hairy black hole with the third accretion profile \eqref{diskprofile3} shown in Fig.\ref{figprofile3}. We find that the three intensities moving away from each other as the hairy parameter $h$ decreases, but the photon ring intensity always makes the dominant contribution to the maximum intensity. Moreover, the contribution of direct intensity to the maximum intensity is larger than that of lensed ring intensity for $h=0$ and $h=-1$, but the situation changes for $h=-2$ due to the intensities' movement. Meanwhile, comparing to the Schwarzschild case, the bright ring appeared around the hairy black hole is more clear to see due to its larger width of photon ring intensity.

In short, comparing to the Schwarzschild black hole under each thin disk accretion, the Horndeski hair has significant influence on the brightness distribution which can be arbitrary combinations among direct, lensed ring and photon ring intensities, depending on each impact parameter. So, in Horndeski hairy black hole's optical appearance, the origination of the observed intensity for each impact parameter, especially its maximum values which correspond to rings, could be dependent of the Horndeski hair. Moreover, in  all three emission profiles, we find that the total observed intensity of the Horndeski hairy black hole is always smaller than that of Schwarzschild black hole.

\section{Rings and images of hairy black hole illuminated by thin spherical accretions}\label{spherical}
{The disk-shaped accretion flow surrounding the black hole usually forms when the materials in the Universe are trapped by a black hole and rotate with a large angular momentum. However, if the angular momentum is extremely small, the matters will flow radially to the black hole and form spherically symmetric accretion \cite{Yuan:2014gma}. Thus, in this section, we will investigate the optical features of black hole surrounded by an optically and geometrically thin accretion flow with spherically symmetric. In this case,} the specific intensity $I(\nu_o)$ observed by a distant observer at $r=\infty$ (measured in erg s$^{-1}$ cm$^{-2}$ str$^{-1}$ Hz$^{-1}$ ) radiated by the accretion flow can be obtained by integrating the specific emissivity along the photon path $\gamma$ \cite{Jaroszynski:1997bw,Bambi:2013nla}
\begin{equation}
I(\nu_o)=\int_{\gamma} g^3 j_e(\nu_e)dl_{prop},\label{eq-Intensity}
\end{equation}
where $g=\nu_o/\nu_e$ is again the redshift factor,  $\nu_o$ and $\nu_e$ are observed photon frequency and emitted photon frequency respectively. $j_e(\nu_e)$ is the emissivity per unit
volume in the rest frame and we will set {$j_e(\nu_e)\propto \delta(\nu_r-\nu_e)/r^2$ as usual \cite{Bambi:2013nla}, where $\nu_r$ is the emitter’s rest-frame frequency}. $dl_{prop}$ is the infinitesimal proper length. {Integrating
Eq.(\ref{eq-Intensity}) over all the observed frequencies, we get the total observed intensity given by}
\begin{equation}
I_{obs}=\int_{\nu_o} I(\nu_o)d\nu_o=\int_{\nu_e}\int_\gamma g^4 j_e(\nu_e)dl_{prop} d\nu_e.\label{eq-Intensity2}
\end{equation}
Then, we will explore the influence of Horndeski hair on the black hole image when the spherical accretion is static and radially free-infalling, respectively.

\subsection{Static spherical accretions}
We firstly consider the static spherical accretions which are distributed outside the event horizon of the hairy black hole \eqref{eq:metric}. In this case, the redshift factor is $g=f(r)^{1/2}$, and the proper length is
\begin{equation}
dl_{prop}=\sqrt{\frac{1}{f(r)}dr^2+r^2d\phi^2}=\sqrt{\frac{1}{f(r)}+r^2\left(\frac{d\phi}{dr}\right)^2}dr.
\end{equation}
Therefore the total observed intensity observed by a distant observer is given by
\begin{equation}
I_{obs}=\int_{\gamma}\frac{f(r)^{2}}{r^2}\sqrt{\frac{1}{f(r)}+r^2\left(\frac{d\phi}{dr}\right)^2}dr.\label{eqintensity}
\end{equation}
It is obvious that the observed intensity depends on the radial distance $r$ and its integral interval, which depends on the impact parameter $b$ as we analyzed in the previous section.
\begin{figure}[htbp]
\centering
\includegraphics[width=6.5cm]{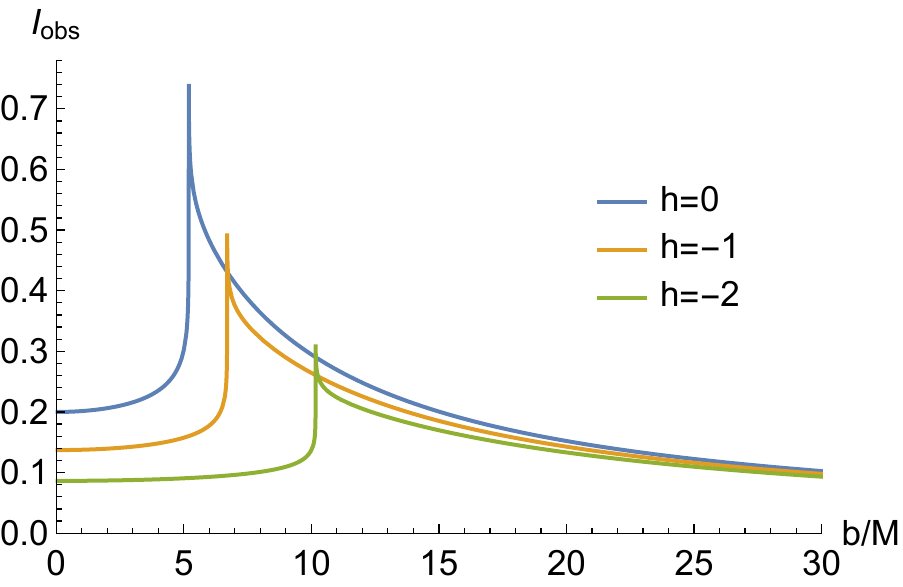}
\caption{The total observed intensity $I_{obs}$ with respect to impact parameters $b$ for the hairy black hole surrounded by the static spherical accretion flow.}
\label{fig-intensity-static}
\end{figure}

Then we plot the observed intensity with respect to $b$ in Fig.\ref{fig-intensity-static}, which shows that $I_{obs}$ for hairy black hole is always smaller than that for Schwarzschild black hole. It is noted that with the decrease of parameter $h$, the peak value of observed intensity at the $b=b_{ph}$ decreases which means that the maximum brightness of Horndeski hairy black hole image is also lower than that of the Schwarzschild case. As illustrated in \cite{Zeng:2020dco}, the intensity at $b=b_{ph}$ should be infinite because the photon travels around the black hole infinite times which should contribute an arbitrarily large intensity. However, the obtained intensity never goes to infinity due to the numerical limitations.

\begin{figure}[htbp]
\centering
\subfigure[\, $h=0$]
{\includegraphics[width=5.5cm]{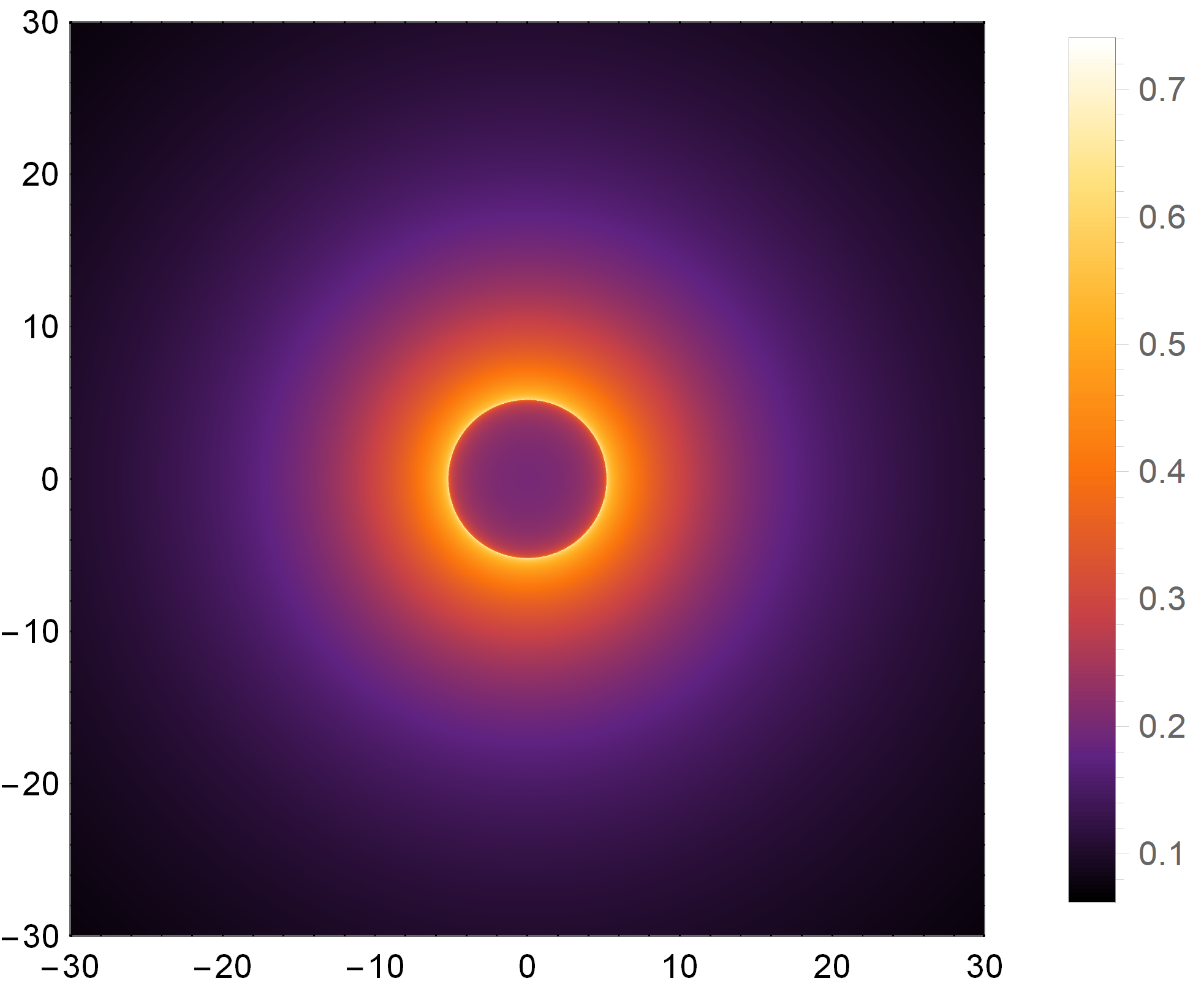}}
\subfigure[\, $h=-1$]
{\includegraphics[width=5.5cm]{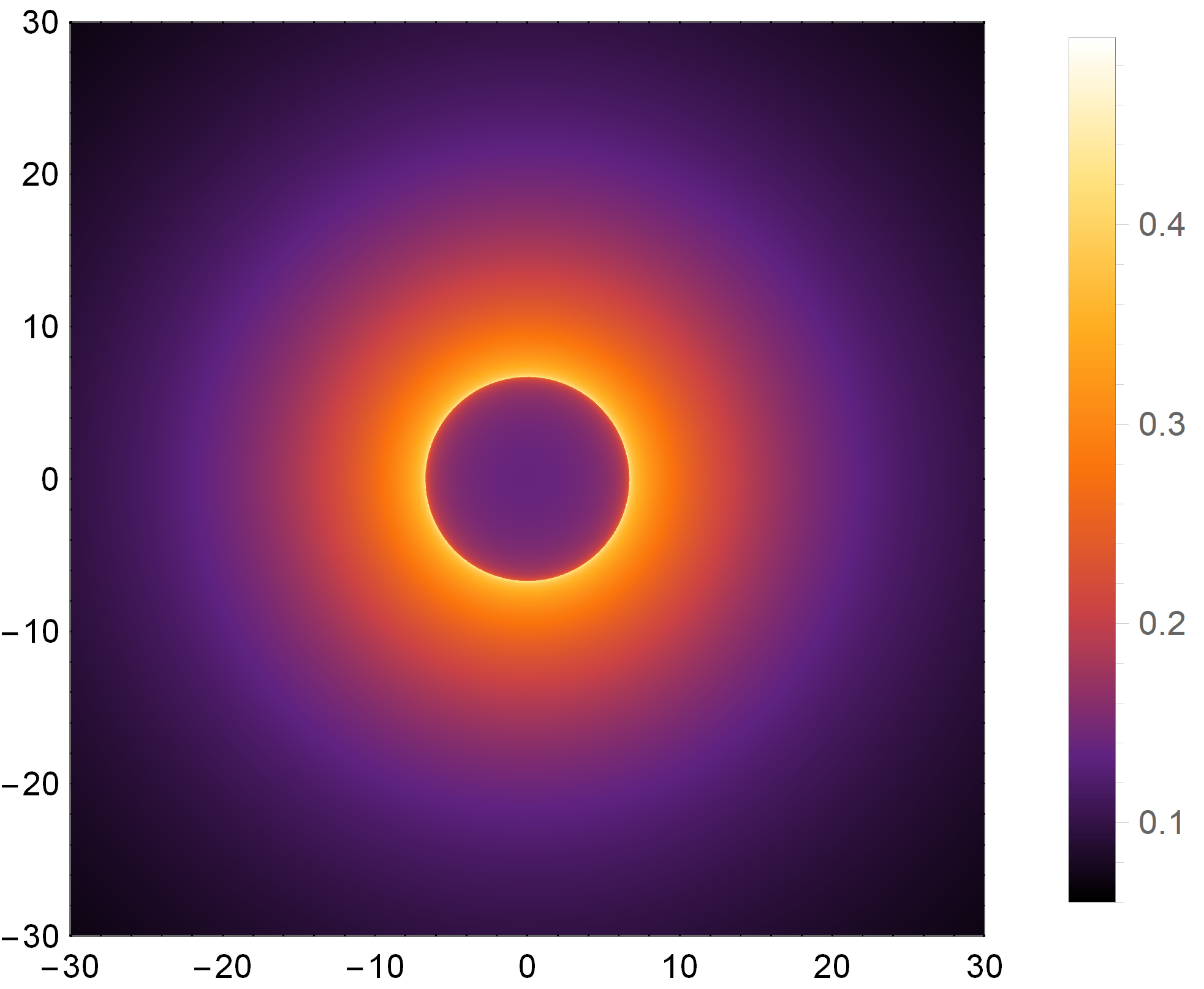}}
\subfigure[\, $h=-2$]
{\includegraphics[width=5.5cm]{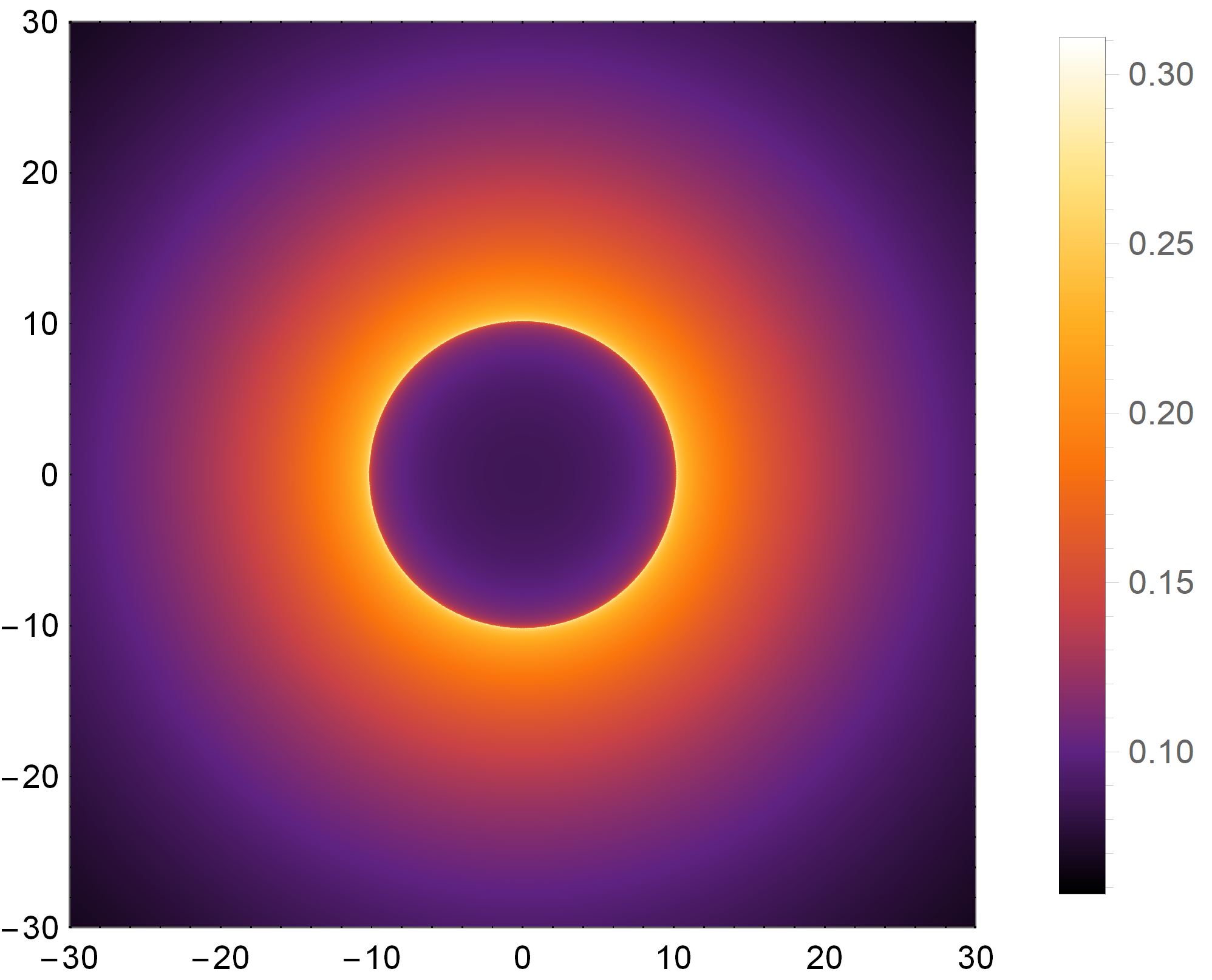}}
\caption{The black hole images with static spherical accretions for different $h$ with $M=1$.}
\label{fig-image-static}
\end{figure}

The distribution of observed intensity \eqref{eqintensity} in two-dimensional plane should depict the image of hairy black hole illuminated by the static spherical accretion for equatorial observers, as shown in Fig.\ref{fig-image-static}. The faint illuminating region in the center is the black hole shadow and the surrounding represents the spherical accretion flow. It is obvious that a bright photon ring can be observed very close to the shadow. Moreover, comparing to the Schwarzschild black hole,  the Horndeski hairy black hole corresponds to fainter photon ring but larger shadow which are consistent with the results reflected in Fig.\ref{fig-intensity-static}.

\subsection{Radially infalling spherical accretions}
We then consider more realistic description that the black hole is surrounded by the radially infalling freely spherical accretion because the most of matters are dynamical in the Universe. Therefore, the redshift factor $g$ should be replaced by \cite{Bambi:2013nla}
\begin{equation}
g=\frac{k_\mu u_o^\mu}{k_\nu u_e^\nu}.
\end{equation}
Here, $k_\mu$ is the four-momentum of photon emitted from accretion matter determined by $k_\mu=\partial\mathcal{L}/\partial\dot{x}^\mu$, which give us
\begin{equation}
k_t=-\frac{1}{b}, \qquad k_r=\pm \frac{1}{f(r)}\sqrt{\frac{1}{b^2}-\frac{f(r)}{r^2}}, \qquad k_\theta=0, \qquad k_\phi=\pm 1, \label{eq-k}
\end{equation}
where we have inserted the Eqs.(\ref{eq1}-\ref{eq3}). It is noted that $\pm$ in $k_r$ describe the photons moving radially inward and outward, while in $k_\phi$ they describe the photons moving counterclockwise and clockwise direction respectively.
$u_o^\mu$ and $u_e^\nu$ are four-velocity of distant observer and accretion matters respectively. Considering an stationary distant observer, we have $u_o^\mu=(1,0,0,0)$.  $u_e^\nu$ is evaluated as
\begin{equation}
u_e^t=\frac{1}{f(r)}, \qquad u_e^r=-\sqrt{1-f(r)}, \qquad u_e^\theta=u_e^\phi=0.
\end{equation}
Subsequently, the redshift factor $g$ can be expressed as
\begin{equation}
g=\frac{k_t}{k_t u_e^t+k_r u_e^r}=\frac{1}{\frac{1}{f(r)}\pm\sqrt{1-f(r)} \sqrt{\frac{1}{f(r)}\left(\frac{1}{f(r)}-\frac{b^2}{r^2}\right)}}.
\end{equation}
Moreover, the proper length can be defined as \cite{Bambi:2013nla}
\begin{equation}
dl_{prop}=k_\mu u_e^\mu d\lambda=\frac{k_t}{g|k_r|}dr.
\end{equation}
Thus, the total observed intensity \eqref{eq-Intensity2} for the radially infalling accretion flow is
\begin{equation}
I_{obs}=\int_{\gamma}  \frac{g^3 k_t}{r^2 |k_r|} dr,
\end{equation}
which is shown in the Fig.\ref{infallingIntensity1}. Similar to the static case, the observed intensity increases firstly, then reaches a maximum at photon sphere $b=b_{ph}$ and finally declines with the increase of $b$ for a fixed parameter $h$. As $h$ decreases, the intensity will become weak for all the $b$. To compare the difference of observed intensities between the static and infalling spherical accretions, we collect them in Fig.\ref{comparisonh0}-Fig.\ref{comparisonh3}. We can see that for the same parameter $h$, the intensity of the black hole image in the infalling spherical accretion is weaker than that in the static case, which is due to the Doppler effect. Additionally, whether the black hole is surrounded by the static or infalling spherical accretion, the shadow radius is always the same as photon sphere, which implies that the shadow size is independent of spherical accretion flow.
Then, the image of hairy black hole surrounded by the infalling accretion flow in the two-dimensional plane are presented in Fig.\ref{fig-image-infalling}.   It is obvious that comparing to that in the static case, the central black hole shadow surrounded by a bright photon ring has a lower brightness.
\begin{figure}[htbp]
\centering
\subfigure[\, total intensity $I_{obs}$]
{\includegraphics[width=4cm]{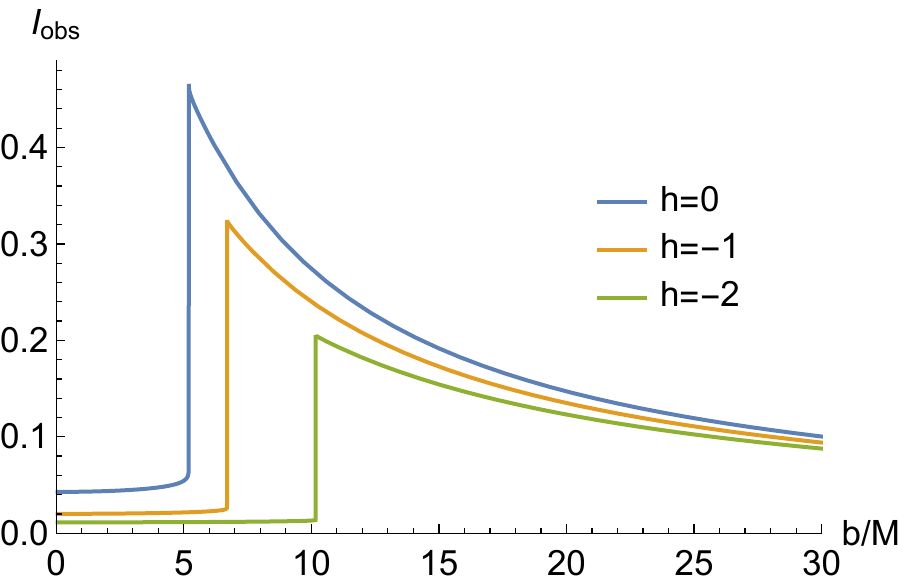}\label{infallingIntensity1}}\hspace{5mm}
\subfigure[\, $h=0$]
{\includegraphics[width=4cm]{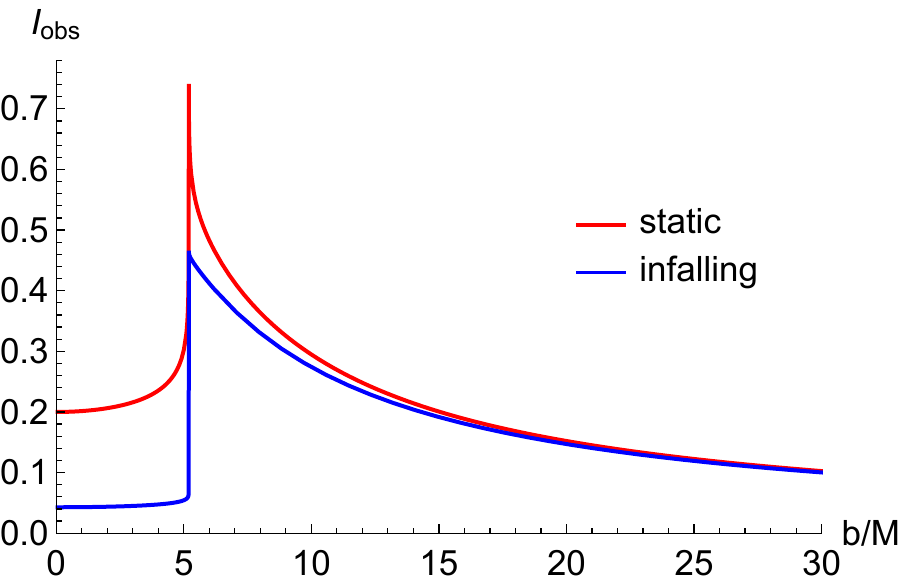}\label{comparisonh0}}\hspace{5mm}
\subfigure[\, $h=-1$]
{\includegraphics[width=4cm]{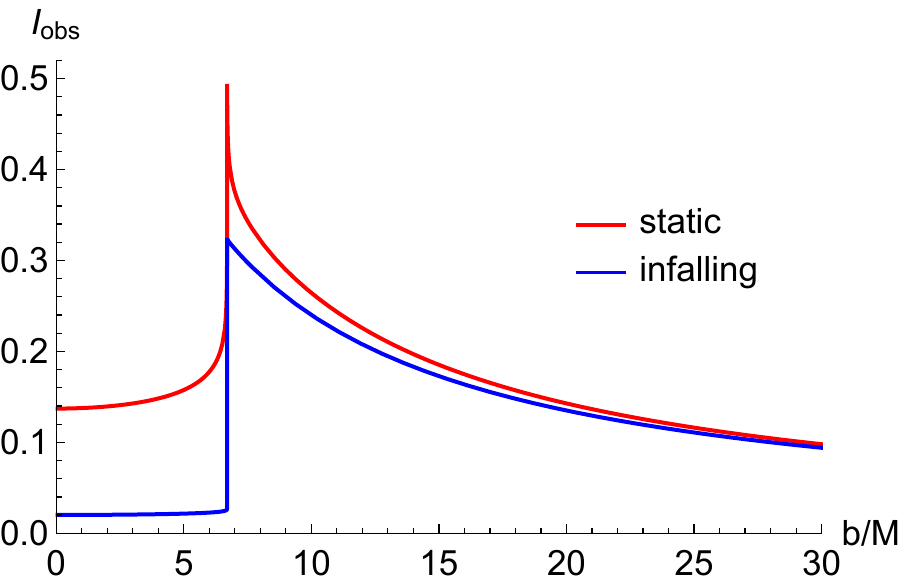}\label{comparisonh1}}\hspace{5mm}
\subfigure[\, $h=-2$]
{\includegraphics[width=4cm]{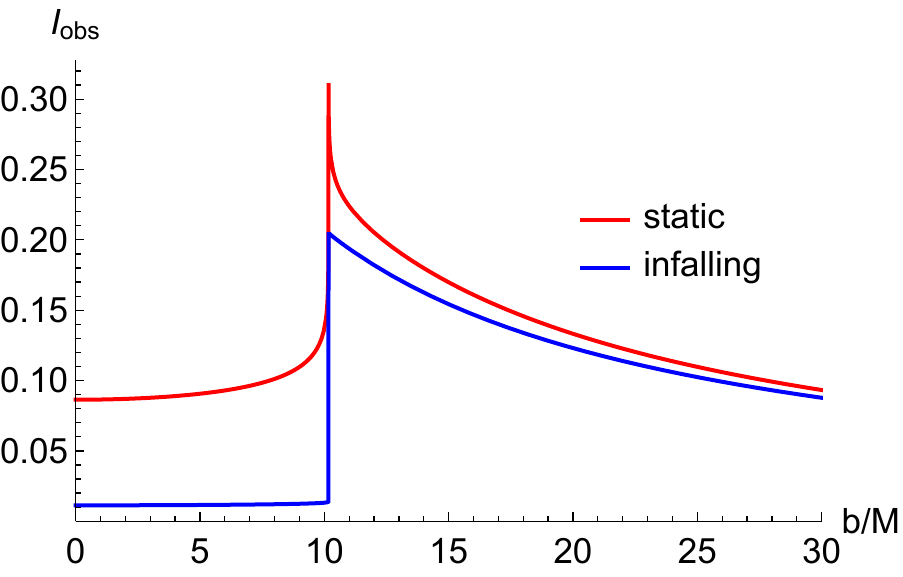}\label{comparisonh3}}\hspace{5mm}
\caption{(a) The observed intensities for the infalling spherical accretion. (b)-(d) The comparisons of the observed intensities between the static (red curves) and radially infalling (blue curves) spherical accretions for different $h$.}
\label{comparison}
\end{figure}
\begin{figure}[htbp]
\centering
\subfigure[\, $h=0$]
{\includegraphics[width=5.5cm]{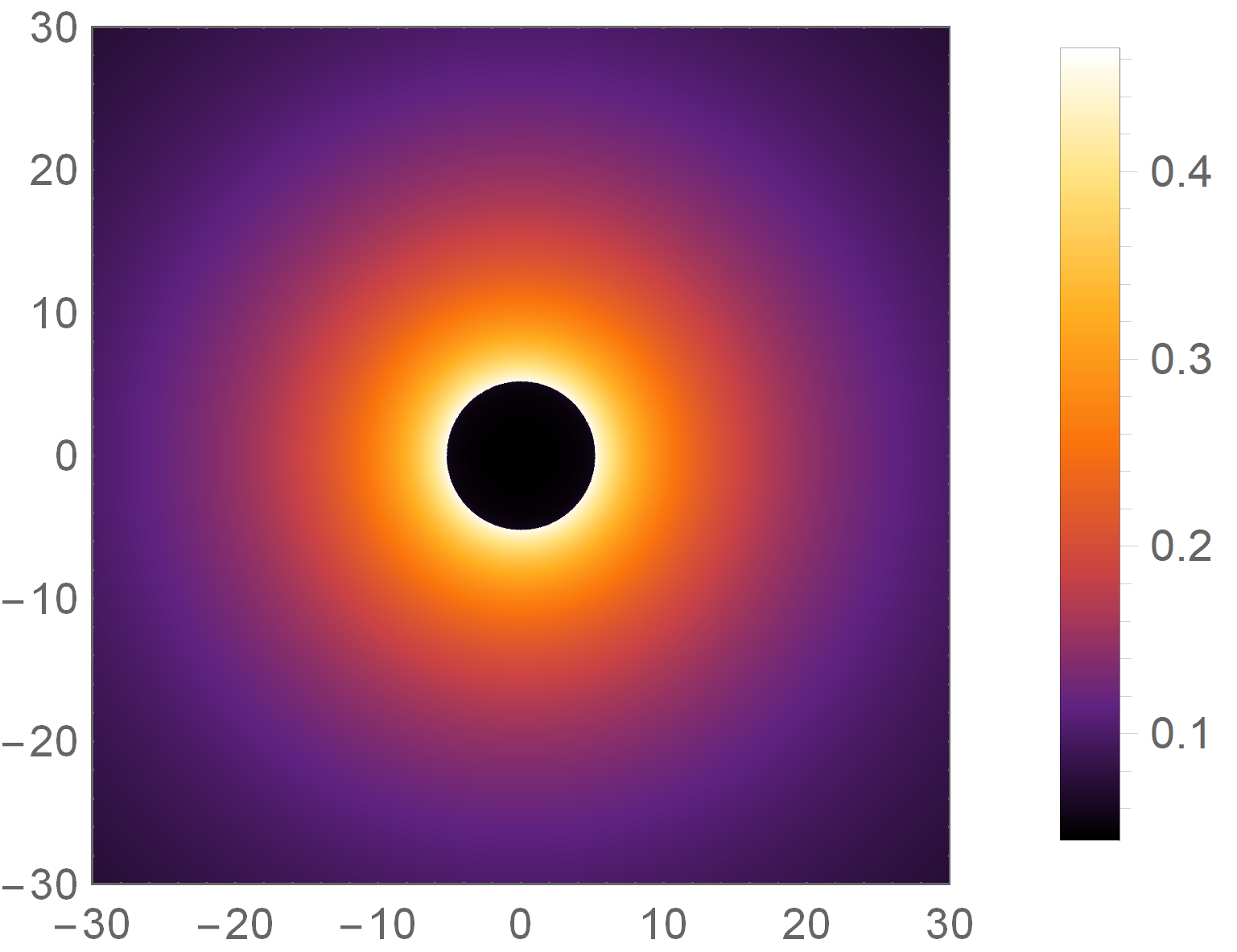}\label{infallingIntensity2}}
\subfigure[\, $h=-1$]
{\includegraphics[width=5.5cm]{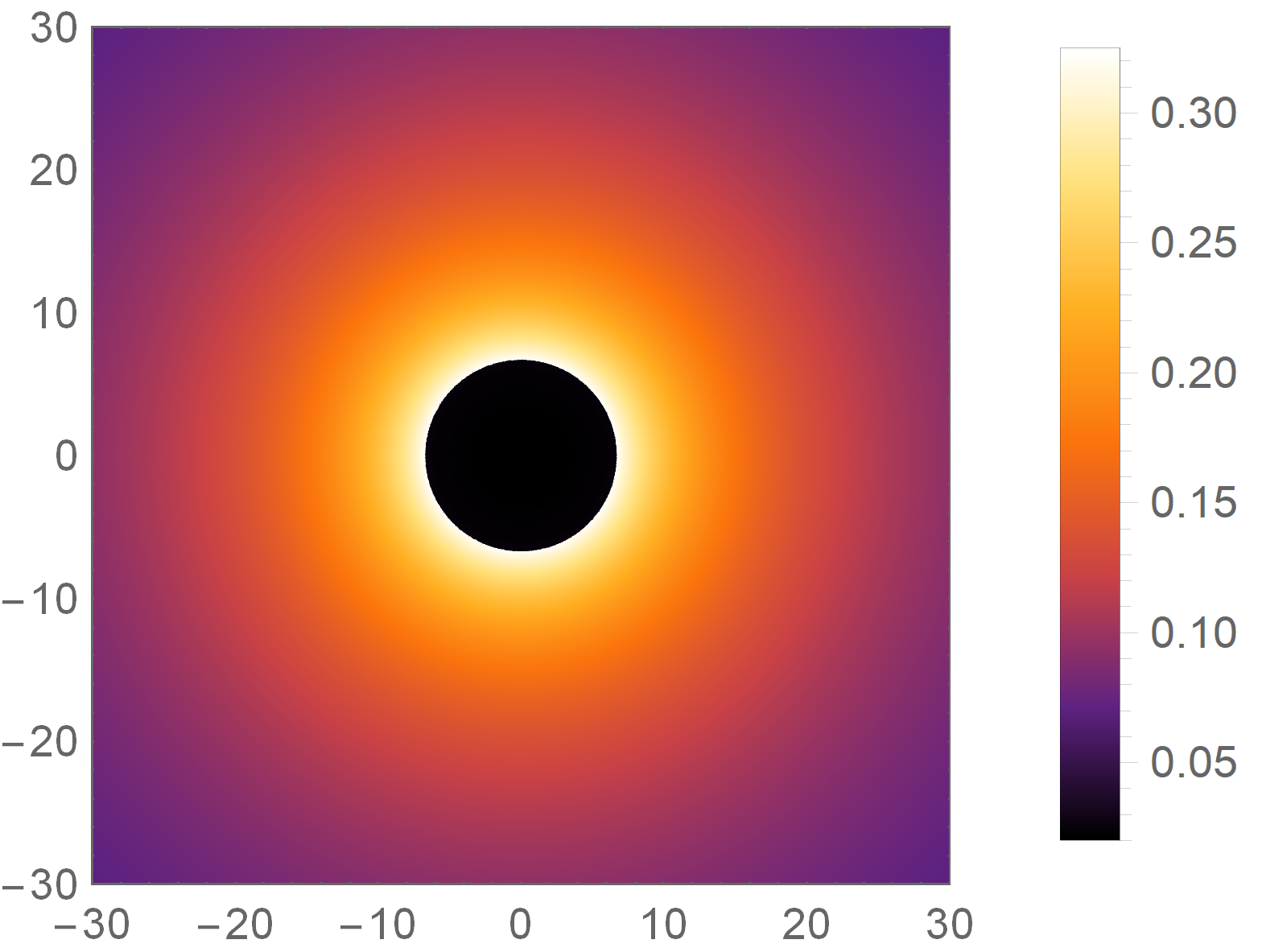}}
\subfigure[\, $h=-2$]
{\includegraphics[width=5.5cm]{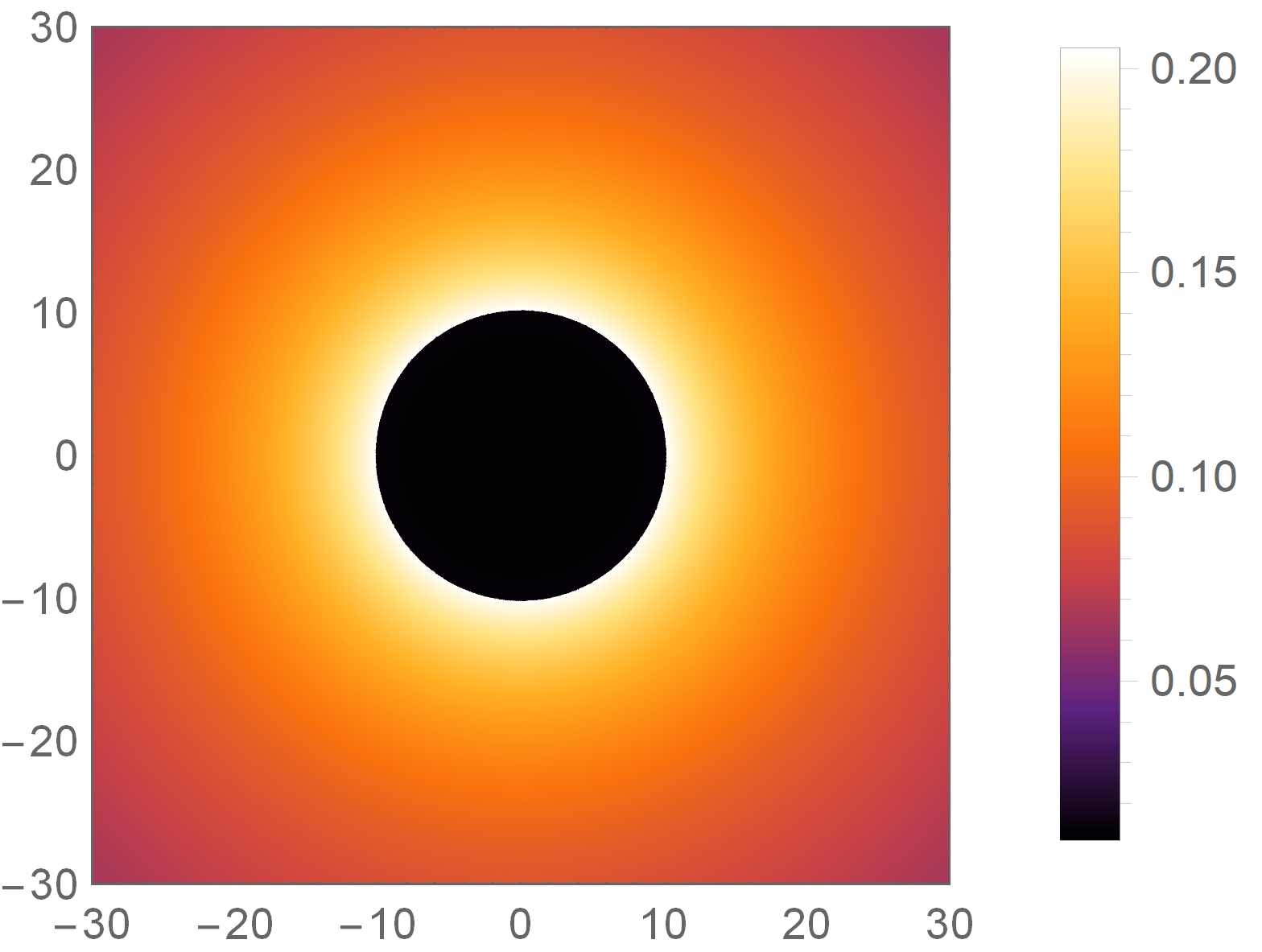}}
\caption{The hairy black hole images with radially infalling spherical accretions for different $h$ with $M=1$.}
\label{fig-image-infalling}
\end{figure}

Thus, comparing to the Schwarzschild black hole, the existence of Horndeski hair will enlarge the photon ring, which then corresponds to a larger black hole shadow for a distant observer. Under the illumination of spherical accretion flow, the deviations of shadow region, bright photon ring and the brightness distribution due to the effect of Horndeski hair from the Schwarzschild black hole are significant.

\section{Conclusion and discussion}\label{conclusion}
Horndeski gravity has significant consequences in describing the accelerated expansion and other interesting observations, so it attracts considerable attentions in cosmological and astrophysical communities. Meanwhile, the black hole solutions in Horndeski gravity may be a good platform to  test the no-hair theorem of classical black hole which also triggers great interest among physicists, so their theoretical and observational properties have also been widely studied. In this paper, we explored  the optical appearance of  a static hairy black hole in Horndeski gravity, and analyzed the influence of Horndeski hair on the photon rings and black hole image illuminated by various accretions.

Firstly, we investigated the Horndeski hairy black hole image illuminated by an optical and geometrically thin accretion disk, which  located at rest on the equatorial plane of hairy black hole. {Comparing to Schwarzschild black hole, both the photon sphere and the critical impact parameter are larger, while the lensed ring and photon ring emissions from the thin accretion disk always correspond to wider range of impact parameter for the hairy black hole. All those quantities depending on the Horndeski parameter can be found from Fig.\ref{figveff} and Fig.\ref{orbitNo} (see also Table \ref{BHquantity} and Table \ref{tableb}), which we have argued to be able to easily understand from the shape of the  potential function of the photon's radial motion.} In addition, the Horndeski hair tends to suppress the demagnification factor in the second and third transfer function, so we predicted that due to the Horndeski hair,  the lensed ring and photon
ring could be more easily to be observed. Then by taking three toy models of emission profile as examples, we carefully analyzed how the brightness from the direct, lensed ring and photon ring intensity contribute to the total observed intensities. Our novel analysis clearly show why the rings in the images form and how is effected by the Horndeski hair. Our results show that the Horndeski hair could significantly affect the brightness distribution which can be arbitrary combinations among direct, lensed ring and photon ring intensity, depending on each impact parameter, so the Horndeski hairy black hole has a completely different optical appearance in contract to the Schwarzschild black hole under illumination of all three emission profiles (Figs.\ref{figprofile1}-\ref{figprofile3}). {In addition, in  all three emission profiles, we found that the total observed intensity of the Horndeski hairy black hole is always smaller than that of Schwarzschild black hole. It means that if we mimic the EHT resolution  of the black hole images by blurring them, the hairy black hole will corresponds to a dimmer realistic observation.}
It is noted that here we focused on the distance static observer viewing face-on in the north pole direction, it is straightforward to extend our study into any observational angles.

Then, we considered that the black hole is illuminated by a static and radially free-falling spherical accretions flow, respectively. In both cases, we can see that a bright photon ring closely surrounds a dark region indicated the black hole shadow (Fig.\ref{fig-image-static} and Fig.\ref{fig-image-infalling}); and comparing to the Schwarzschild black hole, the shadow size for the hairy black hole is larger but the photon ring is fainter. For fixed hairy parameter, the size of shadow keeps the same in the two accretions cases, but its brightness for the infalling accretion is fainter than the static accretion because of the Doppler effect.

In conclusion, according to the light rays distributions, we figured out the optical appearances of Horndeski hairy black hole surrounded by various thin disk or spherical accretion flow, which differentiate the Schwarzschild black hole in GR. {It is worthwhile to point out that the rich features of rings and images caused by the Horndeski hairy parameter cannot be observed for Schwarzschild black hole, i.e, no  degeneracy exists between the Horndeski hairy parameter and the Schwarzschild mass parameter in our scenario. This is because we are working with all dimensionless quantities rescaled by the mass parameter $M$, and so all the results will not depend on the values of mass in the calculation.} Our theoretical studies are ideal, especially, the accretion profiles we consider are far away from the realistic astrophysical environments, but this study provides potential way to distinguish Horndeski hairy black hole from static black hole in GR.  We expect that these preliminary results could shed light on the future test of no-hair theorem using black hole shadow.

\begin{acknowledgments}
This work is partly supported by  Natural Science Foundation of Jiangsu Province under Grant No.BK20211601, Fok Ying Tung Education Foundation under Grant No.171006, the Postgraduate Research \& Practice Innovation Program of Jiangsu Province under Grant No. KYCX22\_3452 and KYCX21\_3192, and Top Talent Support Program from Yangzhou University.
\end{acknowledgments}


\begin{appendices}
\section{Inner-most stable circular orbit (ISCO)}
For the motion of massive particle, it requires the Lagrangian $\mathcal{L}=-1/2$. Similar to the calculations of null geodesic in section \ref{sec-Trajectories}, one can also obtain the orbit equation $(dr/d\phi)^2+\tilde{V}_{\text{eff}}(r)=0$  and introduce another effective potential $\tilde{V}_{\text{eff}}$ given by
\begin{equation}
\tilde{V}_{\text{eff}}=r^4\left(\frac{\tilde{E}^2}{\tilde{L}_z^2}-\frac{f(r)}{\tilde{L}_z^2}-\frac{f(r)}{r^2}\right),
\end{equation}
which is different from the photon's. $\tilde{E}$ and $\tilde{L}_z$ are the energy and angular momentum for the massive particle respectively. The ISCO is determined by $\tilde{V}_{\text{eff}}=0$, $\tilde{V}_{\text{eff}}'=0$ and $\tilde{V}_{\text{eff}}''=0$, where the prime represents the derivative with respect to $r$. Finally the radius $r_{isco}$ of ISCO satisfies the following relation
\begin{equation}
r_{isco}=\frac{3f(r_{isco})f'(r_{isco})}{2f'(r_{isco})^2-f(r_{isco})f''(r_{isco})},\label{formularisco}
\end{equation}
which is numerically solved and its function of the hairy parameter is shown in Fig.\ref{fig-ISCO}.

\begin{figure}[htbp]
\centering
\includegraphics[width=6cm]{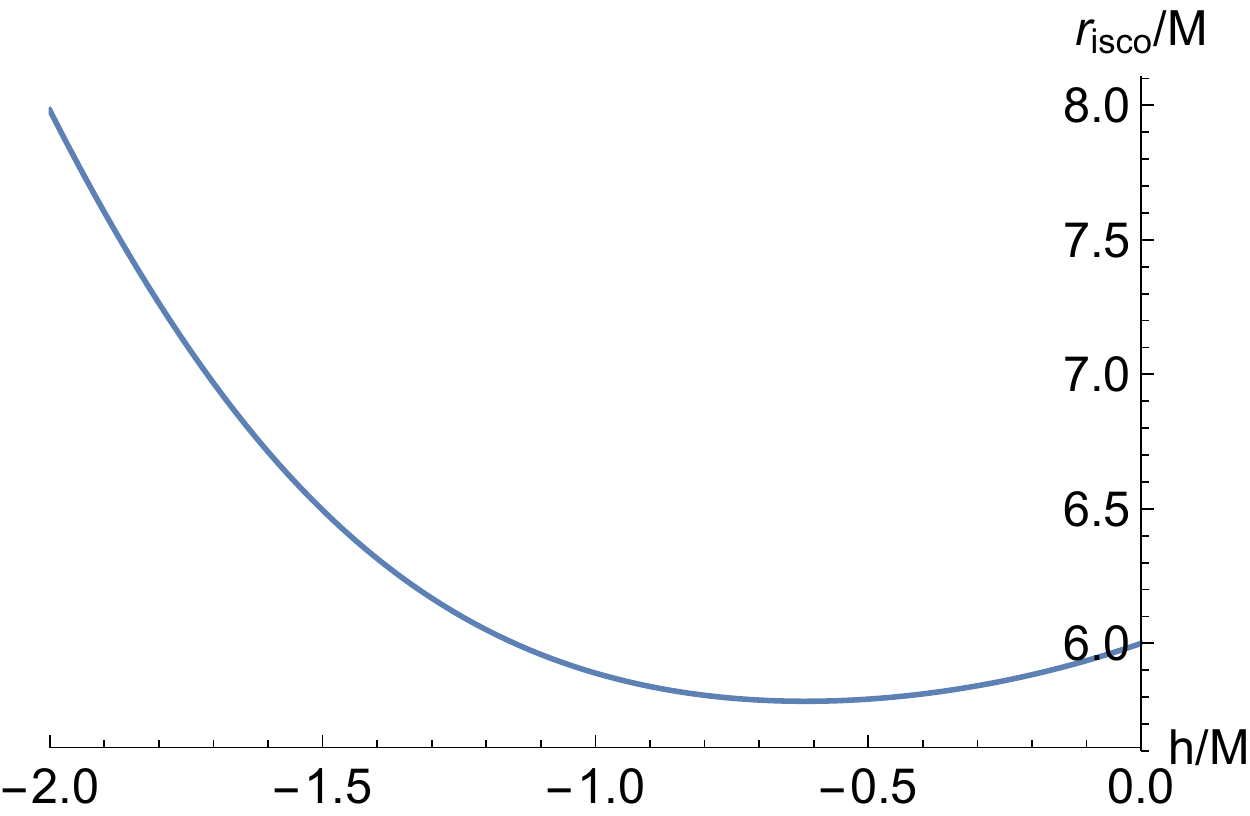}
\caption{The radius of innermost stable circular orbit $r_{isco}$ as the function of the hairy parameter $h$.}
\label{fig-ISCO}
\end{figure}

\end{appendices}

\bibliographystyle{utphys}
\bibliography{ref}

\providecommand{\href}[2]{#2}\begingroup\raggedright\begin{thebibliography}{100}

\bibitem{Damour:1992we}
T.~Damour and G.~Esposito-Farese, ``{Tensor multiscalar theories of
  gravitation},'' \href{http://dx.doi.org/10.1088/0264-9381/9/9/015}{{\em
  Class. Quant. Grav.} {\bfseries 9} (1992) 2093--2176}.

\bibitem{Horndeski:1974wa}
G.~W. Horndeski, ``{Second-order scalar-tensor field equations in a
  four-dimensional space},'' \href{http://dx.doi.org/10.1007/BF01807638}{{\em
  Int. J. Theor. Phys.} {\bfseries 10} (1974) 363--384}.

\bibitem{Bellini:2015xja}
E.~Bellini, A.~J. Cuesta, R.~Jimenez, and L.~Verde, ``{Constraints on
  deviations from \ensuremath{\Lambda}CDM within Horndeski gravity},''
  \href{http://dx.doi.org/10.1088/1475-7516/2016/06/E01}{{\em JCAP} {\bfseries
  02} (2016) 053}, \href{http://arxiv.org/abs/1509.07816}{{\ttfamily
  arXiv:1509.07816 [astro-ph.CO]}}. [Erratum: JCAP 06, E01 (2016)].

\bibitem{Bhattacharya:2016naa}
S.~Bhattacharya and S.~Chakraborty, ``{Constraining some Horndeski gravity
  theories},'' \href{http://dx.doi.org/10.1103/PhysRevD.95.044037}{{\em Phys.
  Rev. D} {\bfseries 95} no.~4, (2017) 044037},
  \href{http://arxiv.org/abs/1607.03693}{{\ttfamily arXiv:1607.03693 [gr-qc]}}.

\bibitem{Kreisch:2017uet}
C.~D. Kreisch and E.~Komatsu, ``{Cosmological Constraints on Horndeski Gravity
  in Light of GW170817},''
  \href{http://dx.doi.org/10.1088/1475-7516/2018/12/030}{{\em JCAP} {\bfseries
  12} (2018) 030}, \href{http://arxiv.org/abs/1712.02710}{{\ttfamily
  arXiv:1712.02710 [astro-ph.CO]}}.

\bibitem{Hou:2017cjy}
S.~Hou and Y.~Gong, ``{Constraints on Horndeski Theory Using the Observations
  of Nordtvedt Effect, Shapiro Time Delay and Binary Pulsars},''
  \href{http://dx.doi.org/10.1140/epjc/s10052-018-5738-8}{{\em Eur. Phys. J. C}
  {\bfseries 78} no.~3, (2018) 247},
  \href{http://arxiv.org/abs/1711.05034}{{\ttfamily arXiv:1711.05034 [gr-qc]}}.

\bibitem{SpurioMancini:2019rxy}
A.~Spurio~Mancini, F.~K\"ohlinger, B.~Joachimi, V.~Pettorino, B.~M. Sch\"afer,
  R.~Reischke, E.~van Uitert, S.~Brieden, M.~Archidiacono, and J.~Lesgourgues,
  ``{KiDS + GAMA: constraints on horndeski gravity from combined large-scale
  structure probes},'' \href{http://dx.doi.org/10.1093/mnras/stz2581}{{\em Mon.
  Not. Roy. Astron. Soc.} {\bfseries 490} no.~2, (2019) 2155--2177},
  \href{http://arxiv.org/abs/1901.03686}{{\ttfamily arXiv:1901.03686
  [astro-ph.CO]}}.

\bibitem{Allahyari:2020jkn}
A.~Allahyari, M.~A. Gorji, and S.~Mukohyama, ``{Bounds on the Horndeski
  Gauge-Gravity Coupling},''
  \href{http://dx.doi.org/10.1088/1475-7516/2020/05/013}{{\em JCAP} {\bfseries
  05} (2020) 013}, \href{http://arxiv.org/abs/2002.11932}{{\ttfamily
  arXiv:2002.11932 [astro-ph.CO]}}. [Erratum: JCAP 05, E02 (2021)].

\bibitem{Kobayashi:2019hrl}
T.~Kobayashi, ``{Horndeski theory and beyond: a review},''
  \href{http://dx.doi.org/10.1088/1361-6633/ab2429}{{\em Rept. Prog. Phys.}
  {\bfseries 82} no.~8, (2019) 086901},
  \href{http://arxiv.org/abs/1901.07183}{{\ttfamily arXiv:1901.07183 [gr-qc]}}.

\bibitem{Feng:2015oea}
X.-H. Feng, H.-S. Liu, H.~L\"u, and C.~N. Pope, ``{Black Hole Entropy and
  Viscosity Bound in Horndeski Gravity},''
  \href{http://dx.doi.org/10.1007/JHEP11(2015)176}{{\em JHEP} {\bfseries 11}
  (2015) 176}, \href{http://arxiv.org/abs/1509.07142}{{\ttfamily
  arXiv:1509.07142 [hep-th]}}.

\bibitem{Kuang:2016edj}
X.-M. Kuang and E.~Papantonopoulos, ``{Building a Holographic Superconductor
  with a Scalar Field Coupled Kinematically to Einstein Tensor},''
  \href{http://dx.doi.org/10.1007/JHEP08(2016)161}{{\em JHEP} {\bfseries 08}
  (2016) 161}, \href{http://arxiv.org/abs/1607.04928}{{\ttfamily
  arXiv:1607.04928 [hep-th]}}.

\bibitem{Jiang:2017imk}
W.-J. Jiang, H.-S. Liu, H.~Lu, and C.~N. Pope, ``{DC Conductivities with
  Momentum Dissipation in Horndeski Theories},''
  \href{http://dx.doi.org/10.1007/JHEP07(2017)084}{{\em JHEP} {\bfseries 07}
  (2017) 084}, \href{http://arxiv.org/abs/1703.00922}{{\ttfamily
  arXiv:1703.00922 [hep-th]}}.

\bibitem{Baggioli:2017ojd}
M.~Baggioli and W.-J. Li, ``{Diffusivities bounds and chaos in holographic
  Horndeski theories},'' \href{http://dx.doi.org/10.1007/JHEP07(2017)055}{{\em
  JHEP} {\bfseries 07} (2017) 055},
  \href{http://arxiv.org/abs/1705.01766}{{\ttfamily arXiv:1705.01766
  [hep-th]}}.

\bibitem{Feng:2018sqm}
X.-H. Feng and H.-S. Liu, ``{Holographic Complexity Growth Rate in Horndeski
  Theory},'' \href{http://dx.doi.org/10.1140/epjc/s10052-019-6547-4}{{\em Eur.
  Phys. J. C} {\bfseries 79} no.~1, (2019) 40},
  \href{http://arxiv.org/abs/1811.03303}{{\ttfamily arXiv:1811.03303
  [hep-th]}}.

\bibitem{Wang:2019jyw}
X.-J. Wang, H.-S. Liu, and W.-J. Li, ``{AC charge transport in holographic
  Horndeski gravity},''
  \href{http://dx.doi.org/10.1140/epjc/s10052-019-7460-6}{{\em Eur. Phys. J. C}
  {\bfseries 79} no.~11, (2019) 932},
  \href{http://arxiv.org/abs/1909.00224}{{\ttfamily arXiv:1909.00224
  [hep-th]}}.

\bibitem{Zhang:2022hxl}
D.~Zhang, G.~Fu, X.-J. Wang, Q.~Pan, and J.-P. Wu, ``{Transport properties in
  the Horndeski holographic two-currents model},''
  \href{http://dx.doi.org/10.1140/epjc/s10052-023-11444-8}{{\em Eur. Phys. J.
  C} {\bfseries 83} no.~4, (2023) 316},
  \href{http://arxiv.org/abs/2211.07074}{{\ttfamily arXiv:2211.07074
  [hep-th]}}.

\bibitem{Bravo-Gaete:2020lzs}
M.~Bravo-Gaete and F.~F. Santos, ``{Complexity of four-dimensional hairy
  anti-de-Sitter black holes with a rotating string and shear viscosity in
  generalized scalar\textendash{}tensor theories},''
  \href{http://dx.doi.org/10.1140/epjc/s10052-022-10064-y}{{\em Eur. Phys. J.
  C} {\bfseries 82} no.~2, (2022) 101},
  \href{http://arxiv.org/abs/2010.10942}{{\ttfamily arXiv:2010.10942
  [hep-th]}}.

\bibitem{Bravo-Gaete:2022lno}
M.~Bravo-Gaete, F.~F. Santos, and H.~Boschi-Filho, ``{Shear viscosity from
  black holes in generalized scalar-tensor theories in arbitrary dimensions},''
  \href{http://dx.doi.org/10.1103/PhysRevD.106.066010}{{\em Phys. Rev. D}
  {\bfseries 106} no.~6, (2022) 066010},
  \href{http://arxiv.org/abs/2201.07961}{{\ttfamily arXiv:2201.07961
  [hep-th]}}.

\bibitem{Rinaldi:2012vy}
M.~Rinaldi, ``{Black holes with non-minimal derivative coupling},''
  \href{http://dx.doi.org/10.1103/PhysRevD.86.084048}{{\em Phys. Rev. D}
  {\bfseries 86} (2012) 084048},
  \href{http://arxiv.org/abs/1208.0103}{{\ttfamily arXiv:1208.0103 [gr-qc]}}.

\bibitem{Cisterna:2014nua}
A.~Cisterna and C.~Erices, ``{Asymptotically locally AdS and flat black holes
  in the presence of an electric field in the Horndeski scenario},''
  \href{http://dx.doi.org/10.1103/PhysRevD.89.084038}{{\em Phys. Rev. D}
  {\bfseries 89} (2014) 084038},
  \href{http://arxiv.org/abs/1401.4479}{{\ttfamily arXiv:1401.4479 [gr-qc]}}.

\bibitem{Sotiriou:2013qea}
T.~P. Sotiriou and S.-Y. Zhou, ``{Black hole hair in generalized scalar-tensor
  gravity},'' \href{http://dx.doi.org/10.1103/PhysRevLett.112.251102}{{\em
  Phys. Rev. Lett.} {\bfseries 112} (2014) 251102},
  \href{http://arxiv.org/abs/1312.3622}{{\ttfamily arXiv:1312.3622 [gr-qc]}}.

\bibitem{Miao:2016aol}
Y.-G. Miao and Z.-M. Xu, ``{Thermodynamics of Horndeski black holes with
  non-minimal derivative coupling},''
  \href{http://dx.doi.org/10.1140/epjc/s10052-016-4482-1}{{\em Eur. Phys. J. C}
  {\bfseries 76} no.~11, (2016) 638},
  \href{http://arxiv.org/abs/1607.06629}{{\ttfamily arXiv:1607.06629
  [hep-th]}}.

\bibitem{Babichev:2016rlq}
E.~Babichev, C.~Charmousis, and A.~Leh\'ebel, ``{Black holes and stars in
  Horndeski theory},''
  \href{http://dx.doi.org/10.1088/0264-9381/33/15/154002}{{\em Class. Quant.
  Grav.} {\bfseries 33} no.~15, (2016) 154002},
  \href{http://arxiv.org/abs/1604.06402}{{\ttfamily arXiv:1604.06402 [gr-qc]}}.

\bibitem{Benkel:2016rlz}
R.~Benkel, T.~P. Sotiriou, and H.~Witek, ``{Black hole hair formation in
  shift-symmetric generalised scalar-tensor gravity},''
  \href{http://dx.doi.org/10.1088/1361-6382/aa5ce7}{{\em Class. Quant. Grav.}
  {\bfseries 34} no.~6, (2017) 064001},
  \href{http://arxiv.org/abs/1610.09168}{{\ttfamily arXiv:1610.09168 [gr-qc]}}.

\bibitem{Filios:2018xvy}
G.~Filios, P.~A. Gonz\'alez, X.-M. Kuang, E.~Papantonopoulos, and Y.~V\'asquez,
  ``{Spontaneous Momentum Dissipation and Coexistence of Phases in Holographic
  Horndeski Theory},'' \href{http://dx.doi.org/10.1103/PhysRevD.99.046017}{{\em
  Phys. Rev. D} {\bfseries 99} no.~4, (2019) 046017},
  \href{http://arxiv.org/abs/1808.07766}{{\ttfamily arXiv:1808.07766
  [hep-th]}}.

\bibitem{Cisterna:2018hzf}
A.~Cisterna, C.~Erices, X.-M. Kuang, and M.~Rinaldi, ``{Axionic black branes
  with conformal coupling},''
  \href{http://dx.doi.org/10.1103/PhysRevD.97.124052}{{\em Phys. Rev. D}
  {\bfseries 97} no.~12, (2018) 124052},
  \href{http://arxiv.org/abs/1803.07600}{{\ttfamily arXiv:1803.07600
  [hep-th]}}.

\bibitem{Giusti:2021sku}
A.~Giusti, S.~Zentarra, L.~Heisenberg, and V.~Faraoni, ``{First-order
  thermodynamics of Horndeski gravity},''
  \href{http://dx.doi.org/10.1103/PhysRevD.105.124011}{{\em Phys. Rev. D}
  {\bfseries 105} no.~12, (2022) 124011},
  \href{http://arxiv.org/abs/2108.10706}{{\ttfamily arXiv:2108.10706 [gr-qc]}}.

\bibitem{Babichev:2013cya}
E.~Babichev and C.~Charmousis, ``{Dressing a black hole with a time-dependent
  Galileon},'' \href{http://dx.doi.org/10.1007/JHEP08(2014)106}{{\em JHEP}
  {\bfseries 08} (2014) 106}, \href{http://arxiv.org/abs/1312.3204}{{\ttfamily
  arXiv:1312.3204 [gr-qc]}}.

\bibitem{Babichev:2017lmw}
E.~Babichev, C.~Charmousis, G.~Esposito-Far\`ese, and A.~Leh\'ebel,
  ``{Stability of Black Holes and the Speed of Gravitational Waves within
  Self-Tuning Cosmological Models},''
  \href{http://dx.doi.org/10.1103/PhysRevLett.120.241101}{{\em Phys. Rev.
  Lett.} {\bfseries 120} no.~24, (2018) 241101},
  \href{http://arxiv.org/abs/1712.04398}{{\ttfamily arXiv:1712.04398 [gr-qc]}}.

\bibitem{BenAchour:2018dap}
J.~Ben~Achour and H.~Liu, ``{Hairy Schwarzschild-(A)dS black hole solutions in
  degenerate higher order scalar-tensor theories beyond shift symmetry},''
  \href{http://dx.doi.org/10.1103/PhysRevD.99.064042}{{\em Phys. Rev. D}
  {\bfseries 99} no.~6, (2019) 064042},
  \href{http://arxiv.org/abs/1811.05369}{{\ttfamily arXiv:1811.05369 [gr-qc]}}.

\bibitem{Takahashi:2019oxz}
K.~Takahashi, H.~Motohashi, and M.~Minamitsuji, ``{Linear stability analysis of
  hairy black holes in quadratic degenerate higher-order scalar-tensor
  theories: Odd-parity perturbations},''
  \href{http://dx.doi.org/10.1103/PhysRevD.100.024041}{{\em Phys. Rev. D}
  {\bfseries 100} no.~2, (2019) 024041},
  \href{http://arxiv.org/abs/1904.03554}{{\ttfamily arXiv:1904.03554 [gr-qc]}}.

\bibitem{Minamitsuji:2019shy}
M.~Minamitsuji and J.~Edholm, ``{Black hole solutions in shift-symmetric
  degenerate higher-order scalar-tensor theories},''
  \href{http://dx.doi.org/10.1103/PhysRevD.100.044053}{{\em Phys. Rev. D}
  {\bfseries 100} no.~4, (2019) 044053},
  \href{http://arxiv.org/abs/1907.02072}{{\ttfamily arXiv:1907.02072 [gr-qc]}}.

\bibitem{Arkani-Hamed:2003juy}
N.~Arkani-Hamed, P.~Creminelli, S.~Mukohyama, and M.~Zaldarriaga, ``{Ghost
  inflation},'' \href{http://dx.doi.org/10.1088/1475-7516/2004/04/001}{{\em
  JCAP} {\bfseries 04} (2004) 001},
  \href{http://arxiv.org/abs/hep-th/0312100}{{\ttfamily arXiv:hep-th/0312100}}.

\bibitem{Khoury:2020aya}
J.~Khoury, M.~Trodden, and S.~S.~C. Wong, ``{Existence and instability of hairy
  black holes in shift-symmetric Horndeski theories},''
  \href{http://dx.doi.org/10.1088/1475-7516/2020/11/044}{{\em JCAP} {\bfseries
  11} (2020) 044}, \href{http://arxiv.org/abs/2007.01320}{{\ttfamily
  arXiv:2007.01320 [astro-ph.CO]}}.

\bibitem{Hui:2012qt}
L.~Hui and A.~Nicolis, ``{No-Hair Theorem for the Galileon},''
  \href{http://dx.doi.org/10.1103/PhysRevLett.110.241104}{{\em Phys. Rev.
  Lett.} {\bfseries 110} (2013) 241104},
  \href{http://arxiv.org/abs/1202.1296}{{\ttfamily arXiv:1202.1296 [hep-th]}}.

\bibitem{Babichev:2017guv}
E.~Babichev, C.~Charmousis, and A.~Leh\'ebel, ``{Asymptotically flat black
  holes in Horndeski theory and beyond},''
  \href{http://dx.doi.org/10.1088/1475-7516/2017/04/027}{{\em JCAP} {\bfseries
  04} (2017) 027}, \href{http://arxiv.org/abs/1702.01938}{{\ttfamily
  arXiv:1702.01938 [gr-qc]}}.

\bibitem{Bergliaffa:2021diw}
S.~E.~P. Bergliaffa, R.~Maier, and N.~d.~O. Silvano, ``{Hairy Black Holes from
  Horndeski Theory},'' \href{http://arxiv.org/abs/2107.07839}{{\ttfamily
  arXiv:2107.07839 [gr-qc]}}.

\bibitem{Walia:2021emv}
R.~K. Walia, S.~D. Maharaj, and S.~G. Ghosh, ``{Rotating Black Holes in
  Horndeski Gravity: Thermodynamic and Gravitational Lensing},''
  \href{http://dx.doi.org/10.1140/epjc/s10052-022-10451-5}{{\em Eur. Phys. J.
  C} {\bfseries 82} (2022) 547},
  \href{http://arxiv.org/abs/2109.08055}{{\ttfamily arXiv:2109.08055 [gr-qc]}}.

\bibitem{Jha:2022tdl}
S.~K. Jha, M.~Khodadi, A.~Rahaman, and A.~Sheykhi, ``{Superradiant energy
  extraction from rotating hairy Horndeski black holes},''
  \href{http://dx.doi.org/10.1103/PhysRevD.107.084052}{{\em Phys. Rev. D}
  {\bfseries 107} no.~8, (2023) 084052},
  \href{http://arxiv.org/abs/2212.13051}{{\ttfamily arXiv:2212.13051 [gr-qc]}}.

\bibitem{EventHorizonTelescope:2019dse}
{\bfseries Event Horizon Telescope} Collaboration, K.~Akiyama {\em et~al.},
  ``{First M87 Event Horizon Telescope Results. I. The Shadow of the
  Supermassive Black Hole},''
  \href{http://dx.doi.org/10.3847/2041-8213/ab0ec7}{{\em Astrophys. J. Lett.}
  {\bfseries 875} (2019) L1}, \href{http://arxiv.org/abs/1906.11238}{{\ttfamily
  arXiv:1906.11238 [astro-ph.GA]}}.

\bibitem{EventHorizonTelescope:2019uob}
{\bfseries Event Horizon Telescope} Collaboration, K.~Akiyama {\em et~al.},
  ``{First M87 Event Horizon Telescope Results. II. Array and
  Instrumentation},'' \href{http://dx.doi.org/10.3847/2041-8213/ab0c96}{{\em
  Astrophys. J. Lett.} {\bfseries 875} no.~1, (2019) L2},
  \href{http://arxiv.org/abs/1906.11239}{{\ttfamily arXiv:1906.11239
  [astro-ph.IM]}}.

\bibitem{EventHorizonTelescope:2019jan}
{\bfseries Event Horizon Telescope} Collaboration, K.~Akiyama {\em et~al.},
  ``{First M87 Event Horizon Telescope Results. III. Data Processing and
  Calibration},'' \href{http://dx.doi.org/10.3847/2041-8213/ab0c57}{{\em
  Astrophys. J. Lett.} {\bfseries 875} no.~1, (2019) L3},
  \href{http://arxiv.org/abs/1906.11240}{{\ttfamily arXiv:1906.11240
  [astro-ph.GA]}}.

\bibitem{EventHorizonTelescope:2019ths}
{\bfseries Event Horizon Telescope} Collaboration, K.~Akiyama {\em et~al.},
  ``{First M87 Event Horizon Telescope Results. IV. Imaging the Central
  Supermassive Black Hole},''
  \href{http://dx.doi.org/10.3847/2041-8213/ab0e85}{{\em Astrophys. J. Lett.}
  {\bfseries 875} no.~1, (2019) L4},
  \href{http://arxiv.org/abs/1906.11241}{{\ttfamily arXiv:1906.11241
  [astro-ph.GA]}}.

\bibitem{EventHorizonTelescope:2019pgp}
{\bfseries Event Horizon Telescope} Collaboration, K.~Akiyama {\em et~al.},
  ``{First M87 Event Horizon Telescope Results. V. Physical Origin of the
  Asymmetric Ring},'' \href{http://dx.doi.org/10.3847/2041-8213/ab0f43}{{\em
  Astrophys. J. Lett.} {\bfseries 875} no.~1, (2019) L5},
  \href{http://arxiv.org/abs/1906.11242}{{\ttfamily arXiv:1906.11242
  [astro-ph.GA]}}.

\bibitem{EventHorizonTelescope:2019ggy}
{\bfseries Event Horizon Telescope} Collaboration, K.~Akiyama {\em et~al.},
  ``{First M87 Event Horizon Telescope Results. VI. The Shadow and Mass of the
  Central Black Hole},'' \href{http://dx.doi.org/10.3847/2041-8213/ab1141}{{\em
  Astrophys. J. Lett.} {\bfseries 875} no.~1, (2019) L6},
  \href{http://arxiv.org/abs/1906.11243}{{\ttfamily arXiv:1906.11243
  [astro-ph.GA]}}.

\bibitem{EventHorizonTelescope:2022wkp}
{\bfseries Event Horizon Telescope} Collaboration, K.~Akiyama {\em et~al.},
  ``{First Sagittarius A* Event Horizon Telescope Results. I. The Shadow of the
  Supermassive Black Hole in the Center of the Milky Way},''
  \href{http://dx.doi.org/10.3847/2041-8213/ac6674}{{\em Astrophys. J. Lett.}
  {\bfseries 930} no.~2, (2022) L12}.

\bibitem{EventHorizonTelescope:2022apq}
{\bfseries Event Horizon Telescope} Collaboration, K.~Akiyama {\em et~al.},
  ``{First Sagittarius A* Event Horizon Telescope Results. II. EHT and
  Multiwavelength Observations, Data Processing, and Calibration},''
  \href{http://dx.doi.org/10.3847/2041-8213/ac6675}{{\em Astrophys. J. Lett.}
  {\bfseries 930} no.~2, (2022) L13}.

\bibitem{EventHorizonTelescope:2022wok}
{\bfseries Event Horizon Telescope} Collaboration, K.~Akiyama {\em et~al.},
  ``{First Sagittarius A* Event Horizon Telescope Results. III. Imaging of the
  Galactic Center Supermassive Black Hole},''
  \href{http://dx.doi.org/10.3847/2041-8213/ac6429}{{\em Astrophys. J. Lett.}
  {\bfseries 930} no.~2, (2022) L14}.

\bibitem{EventHorizonTelescope:2022exc}
{\bfseries Event Horizon Telescope} Collaboration, K.~Akiyama {\em et~al.},
  ``{First Sagittarius A* Event Horizon Telescope Results. IV. Variability,
  Morphology, and Black Hole Mass},''
  \href{http://dx.doi.org/10.3847/2041-8213/ac6736}{{\em Astrophys. J. Lett.}
  {\bfseries 930} no.~2, (2022) L15}.

\bibitem{EventHorizonTelescope:2022urf}
{\bfseries Event Horizon Telescope} Collaboration, K.~Akiyama {\em et~al.},
  ``{First Sagittarius A* Event Horizon Telescope Results. V. Testing
  Astrophysical Models of the Galactic Center Black Hole},''
  \href{http://dx.doi.org/10.3847/2041-8213/ac6672}{{\em Astrophys. J. Lett.}
  {\bfseries 930} no.~2, (2022) L16}.

\bibitem{EventHorizonTelescope:2022xqj}
{\bfseries Event Horizon Telescope} Collaboration, K.~Akiyama {\em et~al.},
  ``{First Sagittarius A* Event Horizon Telescope Results. VI. Testing the
  Black Hole Metric},'' \href{http://dx.doi.org/10.3847/2041-8213/ac6756}{{\em
  Astrophys. J. Lett.} {\bfseries 930} no.~2, (2022) L17}.

\bibitem{Kumar:2018ple}
R.~Kumar and S.~G. Ghosh, ``{Black Hole Parameter Estimation from Its
  Shadow},'' \href{http://dx.doi.org/10.3847/1538-4357/ab77b0}{{\em Astrophys.
  J.} {\bfseries 892} (2020) 78},
  \href{http://arxiv.org/abs/1811.01260}{{\ttfamily arXiv:1811.01260 [gr-qc]}}.

\bibitem{Ghosh:2020spb}
S.~G. Ghosh, R.~Kumar, and S.~U. Islam, ``{Parameters estimation and strong
  gravitational lensing of nonsingular Kerr-Sen black holes},''
  \href{http://dx.doi.org/10.1088/1475-7516/2021/03/056}{{\em JCAP} {\bfseries
  03} (2021) 056}, \href{http://arxiv.org/abs/2011.08023}{{\ttfamily
  arXiv:2011.08023 [gr-qc]}}.

\bibitem{Afrin:2021imp}
M.~Afrin, R.~Kumar, and S.~G. Ghosh, ``{Parameter estimation of hairy Kerr
  black holes from its shadow and constraints from M87*},''
  \href{http://dx.doi.org/10.1093/mnras/stab1260}{{\em Mon. Not. Roy. Astron.
  Soc.} {\bfseries 504} (2021) 5927--5940},
  \href{http://arxiv.org/abs/2103.11417}{{\ttfamily arXiv:2103.11417 [gr-qc]}}.

\bibitem{Ghosh:2022kit}
S.~G. Ghosh and M.~Afrin, ``{An Upper Limit on the Charge of the Black Hole Sgr
  A* from EHT Observations},''
  \href{http://dx.doi.org/10.3847/1538-4357/acb695}{{\em Astrophys. J.}
  {\bfseries 944} no.~2, (2023) 174},
  \href{http://arxiv.org/abs/2206.02488}{{\ttfamily arXiv:2206.02488 [gr-qc]}}.

\bibitem{Vagnozzi:2019apd}
S.~Vagnozzi and L.~Visinelli, ``{Hunting for extra dimensions in the shadow of
  M87*},'' \href{http://dx.doi.org/10.1103/PhysRevD.100.024020}{{\em Phys. Rev.
  D} {\bfseries 100} no.~2, (2019) 024020},
  \href{http://arxiv.org/abs/1905.12421}{{\ttfamily arXiv:1905.12421 [gr-qc]}}.

\bibitem{Banerjee:2019nnj}
I.~Banerjee, S.~Chakraborty, and S.~SenGupta, ``{Silhouette of M87*: A New
  Window to Peek into the World of Hidden Dimensions},''
  \href{http://dx.doi.org/10.1103/PhysRevD.101.041301}{{\em Phys. Rev. D}
  {\bfseries 101} no.~4, (2020) 041301},
  \href{http://arxiv.org/abs/1909.09385}{{\ttfamily arXiv:1909.09385 [gr-qc]}}.

\bibitem{Tang:2022hsu}
Z.-Y. Tang, X.-M. Kuang, B.~Wang, and W.-L. Qian, ``{The length of a compact
  extra dimension from black hole shadow},''
  \href{http://dx.doi.org/10.1016/j.scib.2022.11.002}{{\em Sci. Bull.}
  {\bfseries 67} (2022) 2272--2275},
  \href{http://arxiv.org/abs/2206.08608}{{\ttfamily arXiv:2206.08608 [gr-qc]}}.

\bibitem{Mizuno:2018lxz}
Y.~Mizuno, Z.~Younsi, C.~M. Fromm, O.~Porth, M.~De~Laurentis, H.~Olivares,
  H.~Falcke, M.~Kramer, and L.~Rezzolla, ``{The Current Ability to Test
  Theories of Gravity with Black Hole Shadows},''
  \href{http://dx.doi.org/10.1038/s41550-018-0449-5}{{\em Nature Astron.}
  {\bfseries 2} no.~7, (2018) 585--590},
  \href{http://arxiv.org/abs/1804.05812}{{\ttfamily arXiv:1804.05812
  [astro-ph.GA]}}.

\bibitem{Psaltis:2018xkc}
D.~Psaltis, ``{Testing General Relativity with the Event Horizon Telescope},''
  \href{http://dx.doi.org/10.1007/s10714-019-2611-5}{{\em Gen. Rel. Grav.}
  {\bfseries 51} no.~10, (2019) 137},
  \href{http://arxiv.org/abs/1806.09740}{{\ttfamily arXiv:1806.09740
  [astro-ph.HE]}}.

\bibitem{Stepanian:2021vvk}
A.~Stepanian, S.~Khlghatyan, and V.~G. Gurzadyan, ``{Black hole shadow to probe
  modified gravity},''
  \href{http://dx.doi.org/10.1140/epjp/s13360-021-01119-2}{{\em Eur. Phys. J.
  Plus} {\bfseries 136} no.~1, (2021) 127},
  \href{http://arxiv.org/abs/2101.08261}{{\ttfamily arXiv:2101.08261 [gr-qc]}}.

\bibitem{Younsi:2021dxe}
Z.~Younsi, D.~Psaltis, and F.~\"Ozel, ``{Black Hole Images as Tests of General
  Relativity: Effects of Spacetime Geometry},''
  \href{http://dx.doi.org/10.3847/1538-4357/aca58a}{{\em Astrophys. J.}
  {\bfseries 942} no.~1, (2023) 47},
  \href{http://arxiv.org/abs/2111.01752}{{\ttfamily arXiv:2111.01752
  [astro-ph.HE]}}.

\bibitem{KumarWalia:2022aop}
R.~Kumar~Walia, S.~G. Ghosh, and S.~D. Maharaj, ``{Testing Rotating Regular
  Metrics with EHT Results of Sgr A*},''
  \href{http://dx.doi.org/10.3847/1538-4357/ac9623}{{\em Astrophys. J.}
  {\bfseries 939} no.~2, (2022) 77},
  \href{http://arxiv.org/abs/2207.00078}{{\ttfamily arXiv:2207.00078 [gr-qc]}}.

\bibitem{Vagnozzi:2022moj}
S.~Vagnozzi {\em et~al.}, ``{Horizon-scale tests of gravity theories and
  fundamental physics from the Event Horizon Telescope image of Sagittarius
  A$^*$},'' \href{http://arxiv.org/abs/2205.07787}{{\ttfamily arXiv:2205.07787
  [gr-qc]}}.

\bibitem{Meng:2022kjs}
Y.~Meng, X.-M. Kuang, and Z.-Y. Tang, ``{Photon regions, shadow observables,
  and constraints from M87* of a charged rotating black hole},''
  \href{http://dx.doi.org/10.1103/PhysRevD.106.064006}{{\em Phys. Rev. D}
  {\bfseries 106} no.~6, (2022) 064006},
  \href{http://arxiv.org/abs/2204.00897}{{\ttfamily arXiv:2204.00897 [gr-qc]}}.

\bibitem{Kuang:2022ojj}
X.-M. Kuang, Z.-Y. Tang, B.~Wang, and A.~Wang, ``{Constraining a modified
  gravity theory in strong gravitational lensing and black hole shadow
  observations},'' \href{http://dx.doi.org/10.1103/PhysRevD.106.064012}{{\em
  Phys. Rev. D} {\bfseries 106} no.~6, (2022) 064012},
  \href{http://arxiv.org/abs/2206.05878}{{\ttfamily arXiv:2206.05878 [gr-qc]}}.

\bibitem{Gussmann:2021mjj}
A.~Gu\ss{}mann, ``{Polarimetric signatures of the photon ring of a black hole
  that is pierced by a cosmic axion string},''
  \href{http://dx.doi.org/10.1007/JHEP08(2021)160}{{\em JHEP} {\bfseries 08}
  (2021) 160}, \href{http://arxiv.org/abs/2105.06659}{{\ttfamily
  arXiv:2105.06659 [astro-ph.HE]}}.

\bibitem{Khodadi:2022pqh}
M.~Khodadi and G.~Lambiase, ``{Probing Lorentz symmetry violation using the
  first image of Sagittarius A*: Constraints on standard-model extension
  coefficients},'' \href{http://dx.doi.org/10.1103/PhysRevD.106.104050}{{\em
  Phys. Rev. D} {\bfseries 106} no.~10, (2022) 104050},
  \href{http://arxiv.org/abs/2206.08601}{{\ttfamily arXiv:2206.08601 [gr-qc]}}.

\bibitem{Khodadi:2021gbc}
M.~Khodadi, G.~Lambiase, and D.~F. Mota, ``{No-hair theorem in the wake of
  Event Horizon Telescope},''
  \href{http://dx.doi.org/10.1088/1475-7516/2021/09/028}{{\em JCAP} {\bfseries
  09} (2021) 028}, \href{http://arxiv.org/abs/2107.00834}{{\ttfamily
  arXiv:2107.00834 [gr-qc]}}.

\bibitem{Luminet:1979nyg}
J.~P. Luminet, ``{Image of a spherical black hole with thin accretion disk},''
  {\em Astron. Astrophys.} {\bfseries 75} (1979) 228--235.

\bibitem{Bambi:2013nla}
C.~Bambi, ``{Can the supermassive objects at the centers of galaxies be
  traversable wormholes? The first test of strong gravity for mm/sub-mm very
  long baseline interferometry facilities},''
  \href{http://dx.doi.org/10.1103/PhysRevD.87.107501}{{\em Phys. Rev. D}
  {\bfseries 87} (2013) 107501},
  \href{http://arxiv.org/abs/1304.5691}{{\ttfamily arXiv:1304.5691 [gr-qc]}}.

\bibitem{Gralla:2019xty}
S.~E. Gralla, D.~E. Holz, and R.~M. Wald, ``{Black Hole Shadows, Photon Rings,
  and Lensing Rings},''
  \href{http://dx.doi.org/10.1103/PhysRevD.100.024018}{{\em Phys. Rev. D}
  {\bfseries 100} no.~2, (2019) 024018},
  \href{http://arxiv.org/abs/1906.00873}{{\ttfamily arXiv:1906.00873
  [astro-ph.HE]}}.

\bibitem{Falcke:1999pj}
H.~Falcke, F.~Melia, and E.~Agol, ``{Viewing the shadow of the black hole at
  the galactic center},'' \href{http://dx.doi.org/10.1086/312423}{{\em
  Astrophys. J. Lett.} {\bfseries 528} (2000) L13},
  \href{http://arxiv.org/abs/astro-ph/9912263}{{\ttfamily
  arXiv:astro-ph/9912263}}.

\bibitem{Narayan:2019imo}
R.~Narayan, M.~D. Johnson, and C.~F. Gammie, ``{The Shadow of a Spherically
  Accreting Black Hole},''
  \href{http://dx.doi.org/10.3847/2041-8213/ab518c}{{\em Astrophys. J. Lett.}
  {\bfseries 885} no.~2, (2019) L33},
  \href{http://arxiv.org/abs/1910.02957}{{\ttfamily arXiv:1910.02957
  [astro-ph.HE]}}.

\bibitem{Zeng:2020dco}
X.-X. Zeng, H.-Q. Zhang, and H.~Zhang, ``{Shadows and photon spheres with
  spherical accretions in the four-dimensional Gauss\textendash{}Bonnet black
  hole},'' \href{http://dx.doi.org/10.1140/epjc/s10052-020-08449-y}{{\em Eur.
  Phys. J. C} {\bfseries 80} no.~9, (2020) 872},
  \href{http://arxiv.org/abs/2004.12074}{{\ttfamily arXiv:2004.12074 [gr-qc]}}.

\bibitem{Zeng:2020vsj}
X.-X. Zeng and H.-Q. Zhang, ``{Influence of quintessence dark energy on the
  shadow of black hole},''
  \href{http://dx.doi.org/10.1140/epjc/s10052-020-08656-7}{{\em Eur. Phys. J.
  C} {\bfseries 80} no.~11, (2020) 1058},
  \href{http://arxiv.org/abs/2007.06333}{{\ttfamily arXiv:2007.06333 [gr-qc]}}.

\bibitem{Peng:2020wun}
J.~Peng, M.~Guo, and X.-H. Feng, ``{Influence of quantum correction on black
  hole shadows, photon rings, and lensing rings},''
  \href{http://dx.doi.org/10.1088/1674-1137/ac06bb}{{\em Chin. Phys. C}
  {\bfseries 45} no.~8, (2021) 085103},
  \href{http://arxiv.org/abs/2008.00657}{{\ttfamily arXiv:2008.00657 [gr-qc]}}.

\bibitem{Saurabh:2020zqg}
K.~Saurabh and K.~Jusufi, ``{Imprints of dark matter on black hole shadows
  using spherical accretions},''
  \href{http://dx.doi.org/10.1140/epjc/s10052-021-09280-9}{{\em Eur. Phys. J.
  C} {\bfseries 81} no.~6, (2021) 490},
  \href{http://arxiv.org/abs/2009.10599}{{\ttfamily arXiv:2009.10599 [gr-qc]}}.

\bibitem{Qin:2020xzu}
X.~Qin, S.~Chen, and J.~Jing, ``{Image of a regular phantom compact object and
  its luminosity under spherical accretions},''
  \href{http://dx.doi.org/10.1088/1361-6382/abf712}{{\em Class. Quant. Grav.}
  {\bfseries 38} no.~11, (2021) 115008},
  \href{http://arxiv.org/abs/2011.04310}{{\ttfamily arXiv:2011.04310 [gr-qc]}}.

\bibitem{Gan:2021pwu}
Q.~Gan, P.~Wang, H.~Wu, and H.~Yang, ``{Photon spheres and spherical accretion
  image of a hairy black hole},''
  \href{http://dx.doi.org/10.1103/PhysRevD.104.024003}{{\em Phys. Rev. D}
  {\bfseries 104} no.~2, (2021) 024003},
  \href{http://arxiv.org/abs/2104.08703}{{\ttfamily arXiv:2104.08703 [gr-qc]}}.

\bibitem{Okyay:2021nnh}
M.~Okyay and A.~\"Ovg\"un, ``{Nonlinear electrodynamics effects on the black
  hole shadow, deflection angle, quasinormal modes and greybody factors},''
  \href{http://dx.doi.org/10.1088/1475-7516/2022/01/009}{{\em JCAP} {\bfseries
  01} no.~01, (2022) 009}, \href{http://arxiv.org/abs/2108.07766}{{\ttfamily
  arXiv:2108.07766 [gr-qc]}}.

\bibitem{Li:2021ypw}
G.-P. Li and K.-J. He, ``{Observational appearances of a f(R) global monopole
  black hole illuminated by various accretions},''
  \href{http://dx.doi.org/10.1140/epjc/s10052-021-09817-y}{{\em Eur. Phys. J.
  C} {\bfseries 81} no.~11, (2021) 1018}.

\bibitem{Li:2021riw}
G.-P. Li and K.-J. He, ``{Shadows and rings of the Kehagias-Sfetsos black hole
  surrounded by thin disk accretion},''
  \href{http://dx.doi.org/10.1088/1475-7516/2021/06/037}{{\em JCAP} {\bfseries
  06} (2021) 037}, \href{http://arxiv.org/abs/2105.08521}{{\ttfamily
  arXiv:2105.08521 [gr-qc]}}.

\bibitem{Guo:2021bhr}
S.~Guo, G.-R. Li, and E.-W. Liang, ``{Influence of accretion flow and magnetic
  charge on the observed shadows and rings of the Hayward black hole},''
  \href{http://dx.doi.org/10.1103/PhysRevD.105.023024}{{\em Phys. Rev. D}
  {\bfseries 105} no.~2, (2022) 023024},
  \href{http://arxiv.org/abs/2112.11227}{{\ttfamily arXiv:2112.11227
  [astro-ph.HE]}}.

\bibitem{Hu:2022lek}
S.~Hu, C.~Deng, D.~Li, X.~Wu, and E.~Liang, ``{Observational signatures of
  Schwarzschild-MOG black holes in scalar-tensor-vector gravity: shadows and
  rings with different accretions},''
  \href{http://dx.doi.org/10.1140/epjc/s10052-022-10868-y}{{\em Eur. Phys. J.
  C} {\bfseries 82} no.~10, (2022) 885}.

\bibitem{Guo:2021bwr}
S.~Guo, K.-J. He, G.-R. Li, and G.-P. Li, ``{The shadow and photon sphere of
  the charged black hole in Rastall gravity},''
  \href{http://dx.doi.org/10.1088/1361-6382/ac12e4}{{\em Class. Quant. Grav.}
  {\bfseries 38} no.~16, (2021) 165013},
  \href{http://arxiv.org/abs/2205.07242}{{\ttfamily arXiv:2205.07242 [gr-qc]}}.

\bibitem{Wen:2022hkv}
S.~Wen, W.~Hong, and J.~Tao, ``{Observational Appearances of Magnetically
  Charged Black Holes in Born-Infeld Electrodynamics},''
  \href{http://dx.doi.org/10.1140/epjc/s10052-023-11431-z}{{\em Eur. Phys. J.
  C} {\bfseries 83} (2023) 277},
  \href{http://arxiv.org/abs/2212.03021}{{\ttfamily arXiv:2212.03021 [gr-qc]}}.

\bibitem{Chakhchi:2022fls}
L.~Chakhchi, H.~El~Moumni, and K.~Masmar, ``{Shadows and optical appearance of
  a power-Yang-Mills black hole surrounded by different accretion disk
  profiles},'' \href{http://dx.doi.org/10.1103/PhysRevD.105.064031}{{\em Phys.
  Rev. D} {\bfseries 105} no.~6, (2022) 064031}.

\bibitem{Hou:2022eev}
Y.~Hou, Z.~Zhang, H.~Yan, M.~Guo, and B.~Chen, ``{Image of a Kerr-Melvin black
  hole with a thin accretion disk},''
  \href{http://dx.doi.org/10.1103/PhysRevD.106.064058}{{\em Phys. Rev. D}
  {\bfseries 106} no.~6, (2022) 064058},
  \href{http://arxiv.org/abs/2206.13744}{{\ttfamily arXiv:2206.13744 [gr-qc]}}.

\bibitem{Kuang:2022xjp}
X.-M. Kuang and A.~\"Ovg\"un, ``{Strong gravitational lensing and shadow
  constraint from M87* of slowly rotating Kerr-like black hole},''
  \href{http://dx.doi.org/10.1016/j.aop.2022.169147}{{\em Annals Phys.}
  {\bfseries 447} (2022) 169147},
  \href{http://arxiv.org/abs/2205.11003}{{\ttfamily arXiv:2205.11003 [gr-qc]}}.

\bibitem{Uniyal:2022vdu}
A.~Uniyal, R.~C. Pantig, and A.~\"Ovg\"un, ``{Probing a non-linear
  electrodynamics black hole with thin accretion disk, shadow, and deflection
  angle with M87* and Sgr A* from EHT},''
  \href{http://dx.doi.org/10.1016/j.dark.2023.101178}{{\em Phys. Dark Univ.}
  {\bfseries 40} (2023) 101178},
  \href{http://arxiv.org/abs/2205.11072}{{\ttfamily arXiv:2205.11072 [gr-qc]}}.

\bibitem{Uniyal:2023inx}
A.~Uniyal, S.~Chakrabarti, R.~C. Pantig, and A.~\"Ovg\"un, ``{Nonlinearly
  charged black holes: Shadow and Thin-accretion disk},''
  \href{http://arxiv.org/abs/2303.07174}{{\ttfamily arXiv:2303.07174 [gr-qc]}}.

\bibitem{Kumar:2021cyl}
J.~Kumar, S.~U. Islam, and S.~G. Ghosh, ``{Investigating strong gravitational
  lensing effects by supermassive black holes with Horndeski gravity},''
  \href{http://dx.doi.org/10.1140/epjc/s10052-022-10357-2}{{\em Eur. Phys. J.
  C} {\bfseries 82} no.~5, (2022) 443},
  \href{http://arxiv.org/abs/2109.04450}{{\ttfamily arXiv:2109.04450 [gr-qc]}}.

\bibitem{Atamurotov:2022slw}
F.~Atamurotov, F.~Sarikulov, A.~Abdujabbarov, and B.~Ahmedov, ``{Gravitational
  weak lensing by black hole in Horndeski gravity in presence of plasma},''
  \href{http://dx.doi.org/10.1140/epjp/s13360-022-02548-3}{{\em Eur. Phys. J.
  Plus} {\bfseries 137} no.~3, (2022) 336}.

\bibitem{Afrin:2021wlj}
M.~Afrin and S.~G. Ghosh, ``{Testing Horndeski Gravity from EHT Observational
  Results for Rotating Black Holes},''
  \href{http://dx.doi.org/10.3847/1538-4357/ac6dda}{{\em Astrophys. J.}
  {\bfseries 932} no.~1, (2022) 51},
  \href{http://arxiv.org/abs/2110.05258}{{\ttfamily arXiv:2110.05258 [gr-qc]}}.

\bibitem{Wang:2020jek}
J.~Wang, ``{Multiple rings in the shadow of extremely compact objects},''
  \href{http://dx.doi.org/10.1142/S0218271821501121}{{\em Int. J. Mod. Phys. D}
  {\bfseries 30} no.~15, (2021) 2150112},
  \href{http://arxiv.org/abs/2012.10237}{{\ttfamily arXiv:2012.10237 [gr-qc]}}.

\bibitem{Yuan:2014gma}
F.~Yuan and R.~Narayan, ``{Hot Accretion Flows Around Black Holes},''
  \href{http://dx.doi.org/10.1146/annurev-astro-082812-141003}{{\em Ann. Rev.
  Astron. Astrophys.} {\bfseries 52} (2014) 529--588},
  \href{http://arxiv.org/abs/1401.0586}{{\ttfamily arXiv:1401.0586
  [astro-ph.HE]}}.

\bibitem{Wielgus:2021peu}
M.~Wielgus, ``{Photon rings of spherically symmetric black holes and robust
  tests of non-Kerr metrics},''
  \href{http://dx.doi.org/10.1103/PhysRevD.104.124058}{{\em Phys. Rev. D}
  {\bfseries 104} no.~12, (2021) 124058},
  \href{http://arxiv.org/abs/2109.10840}{{\ttfamily arXiv:2109.10840 [gr-qc]}}.

\bibitem{Bisnovatyi-Kogan:2022ujt}
G.~S. Bisnovatyi-Kogan and O.~Y. Tsupko, ``{Analytical study of higher-order
  ring images of the accretion disk around a black hole},''
  \href{http://dx.doi.org/10.1103/PhysRevD.105.064040}{{\em Phys. Rev. D}
  {\bfseries 105} no.~6, (2022) 064040},
  \href{http://arxiv.org/abs/2201.01716}{{\ttfamily arXiv:2201.01716 [gr-qc]}}.

\bibitem{Wang:2022yvi}
H.-M. Wang, Z.-C. Lin, and S.-W. Wei, ``{Optical appearance of
  Einstein-\AE{}ther black hole surrounded by thin disk},''
  \href{http://dx.doi.org/10.1016/j.nuclphysb.2022.116026}{{\em Nucl. Phys. B}
  {\bfseries 985} (2022) 116026},
  \href{http://arxiv.org/abs/2205.13174}{{\ttfamily arXiv:2205.13174 [gr-qc]}}.

\bibitem{Yang:2022btw}
J.~Yang, C.~Zhang, and Y.~Ma, ``{Loop quantum black hole in a gravitational
  collapse model},'' \href{http://arxiv.org/abs/2211.04263}{{\ttfamily
  arXiv:2211.04263 [gr-qc]}}.

\bibitem{Jaroszynski:1997bw}
M.~Jaroszynski and A.~Kurpiewski, ``{Optics near kerr black holes: spectra of
  advection dominated accretion flows},'' {\em Astron. Astrophys.} {\bfseries
  326} (1997) 419, \href{http://arxiv.org/abs/astro-ph/9705044}{{\ttfamily
  arXiv:astro-ph/9705044}}.

\end{thebibliography}\endgroup

\end{document}